%% file: main.tex
\documentclass[12pt]{article}

\usepackage[T1]{fontenc}
\usepackage[utf8]{inputenc}
\usepackage{textcomp}
\usepackage{amssymb,amsmath,amsfonts,eurosym,geometry,ulem,graphicx,caption,
            color,setspace,sectsty,comment,footmisc,pdflscape,subfigure,
            array,url,booktabs,float,threeparttable}

\usepackage[labelfont=bf, justification=centering]{caption}
\usepackage{threeparttable}
\usepackage{booktabs}

\newcolumntype{L}[1]{>{\raggedright\let\newline\\arraybackslash\hspace{0pt}}m{#1}}
\newcolumntype{C}[1]{>{\centering\let\newline\\arraybackslash\hspace{0pt}}m{#1}}
\newcolumntype{R}[1]{>{\raggedleft\let\newline\\arraybackslash\hspace{0pt}}m{#1}}

\usepackage[natbibapa]{apacite}
\usepackage[colorlinks=true,
            citecolor=blue,
            urlcolor=blue,
            linkcolor=blue]{hyperref}

\begin{document}

\begin{titlepage}
\title{Do Drug Consumption Rooms Reduce Drug-Related Hospitalizations? Evidence from Switzerland}
\author{Ana Armendariz*}
\date{\today}
\maketitle

\begin{abstract}
This paper estimates the causal effect of drug consumption room (DCR) openings on drug-related hospitalizations in Switzerland. I exploit the staggered introduction of DCRs across Swiss cities between 1998 and 2022, using individual-level hospital records and a difference-in-differences estimator. DCR openings reduce drug-related hospitalization rates by roughly 35\% within 5 km of a facility. The reduction is most pronounced for fatal outcomes, which fall by 35\% while narcotics poisoning hospitalizations rise by a comparable amount, consistent with DCRs converting potentially fatal overdoses into non-fatal hospitalizations. Mental and behavioral disorder hospitalizations decline by 32\% while estimates for Hepatitis and HIV are imprecise. The findings support the role of DCRs as an effective harm reduction strategy.

\noindent \\
\vspace{0in}
\vspace{0in}
\end{abstract}

\vfill
\singlespacing

{
\footnotesize
\noindent *University of St.Gallen, Rosenbergstrasse 22, Switzerland.  
anapaula.armendarizpacheco@unisg.ch.\\
I gratefully acknowledge the Swiss Institute for Empirical Economic Research (SEW) at the University of St.\,Gallen for supporting my position throughout the PhD. I further thank the Swiss Society of Health Economics (SGG\"O) for financing the acquisition of the data used in this study through its PhD Project Fund, and the University of St.\,Gallen for the MobiDoc Grant. I am grateful to Michael Lechner, Federica Mascolo, Nora Bearth, Johanna Kurz, and Sofia Sierra for valuable feedback, and to participants in various seminars, workshops, and discussions for their helpful comments. GPT-5 and Claude provided editing and coding assistance. All errors are my own.
}

\setcounter{page}{0}
\thispagestyle{empty}
\end{titlepage}
\pagebreak \newpage

\onehalfspacing

\input{intro}
\input{literature}

\input{background}

\input{data}
\input{method}

\input{results}

\input{conclusion}

\onehalfspacing
\setlength\bibsep{0pt}

\clearpage

\onehalfspacing

% references
\addcontentsline{toc}{section}{References}
\bibliography{references.bib}

\clearpage

\appendix
\renewcommand{\thetable}{A\arabic{table}}
\renewcommand{\thefigure}{A\arabic{figure}}
\setcounter{table}{0}
\setcounter{figure}{0}

\section{Appendix} \label{sec:appendix}
\addcontentsline{toc}{section}{Appendix}

\subsection{Additional Descriptive Tables}
\label{app:tables}

\input{tables/tab_suchtindex_snapshot}

\begin{table}[H]
\caption{Overview of Variables Available in the Hospital Dataset}
\centering
\small
\begin{tabular}{p{5cm}p{10cm}}
\toprule
\textbf{Category} & \textbf{Variables} \\
\hline
\textbf{Identifiers} & Anonymised patient ID, internal technical ID, type of record \\
\hline
\textbf{Demographics} & Place of residence (MedStat region), gender, age group (5-year bands), nationality (Swiss/non-Swiss) \\
\hline
\textbf{Hospital Characteristics} & Canton of hospital \\
\hline
\textbf{Admission Details} & Month of entry, stay before admission, mode of admission, referral decision \\
\hline
\textbf{Diagnoses and Treatments} & Primary and additional diagnoses, main and additional treatments \\
\hline
\textbf{Medical Stay Information} & Length of stay, days until next hospitalization, duration between entry and main treatment, ICU stay \\
\hline
\textbf{Discharge} & Discharge decision, stay after discharge, post-discharge care \\
\bottomrule
\end{tabular}
\label{tab:variables}
\end{table}

\begin{table}[H]
\caption{Distribution of Stay Before Admission}
\centering
\begin{tabular}{lrr}
\toprule
\textbf{Stay Before Admission} & \textbf{Count} & \textbf{Percent} \\
\midrule
1 - Home & 368,646 & 72.9 \\
6 - Other Hospital / Birth Home & 40,172 & 7.9 \\
4 - Non-Med. Non-Hosp. Facility & 24,916 & 4.9 \\
8 - Other & 23,552 & 4.7 \\
5 - Psych. Clinic, Other & 15,093 & 3.0 \\
9 - Unknown & 14,844 & 2.9 \\
7 - Penal Institution & 7,072 & 1.4 \\
66 - Acute Care, Same & 3,678 & 0.7 \\
2 - Home w/ Home Care & 3,396 & 0.7 \\
3 - Med. Non-Hosp. Facility & 3,234 & 0.6 \\
55 - Psych. Clinic, Same & 458 & 0.1 \\
83 - Rehab Clinic, Other & 239 & 0.0 \\
84 - Rehab Clinic, Same & 106 & 0.0 \\
\bottomrule
\end{tabular}
\label{tab:stay_before_adm_full}
\end{table}

\begin{table}[H]
\caption{Distribution of Mode of Admission}
\centering
\begin{tabular}{lrr}
\toprule
\textbf{Mode of Admission} & \textbf{Count} & \textbf{Percent} \\
\midrule
1 - Emergency (within 12h) & 245,143 & 48.5 \\
2 - Scheduled / Planned & 236,488 & 46.8 \\
8 - Other & 10,700 & 2.1 \\
9 - Unknown & 10,105 & 2.0 \\
4 - Internal Transfer & 1,502 & 0.3 \\
5 - Transfer within 24h & 1,468 & 0.3 \\
\bottomrule
\end{tabular}
\label{tab:mode_admission}
\end{table}

\begin{table}[H]
\caption{Distribution of Type of Care}
\centering
\begin{tabular}{lrr}
\toprule
\textbf{Type of Care} & \textbf{Count} & \textbf{Percent} \\
\midrule
3 - Inpatient (Hospitalisation) & 485,941 & 96.1 \\
2 - Semi-hospitalization (Deprecated) & 16,377 & 3.2 \\
1 - Outpatient (Ambulatory) & 2,785 & 0.6 \\
9 - Unknown & 303 & 0.1 \\
\bottomrule
\end{tabular}
\label{tab:type_of_care}
\end{table}

\begin{table}[H]
\caption{Distribution of Cost Coverage Center}
\centering
\begin{tabular}{lrr}
\toprule
\textbf{Cost Coverage Label} & \textbf{Count} & \textbf{Percentage} \\
\midrule
M000 - General Medicine & 8,257 & 1.6 \\
M050 - Intensive Care & 3,173 & 0.6 \\
M100 - Internal Medicine & 121,838 & 24.1 \\
M200 - Surgery & 40,587 & 8.0 \\
M300 - Gynecology/Obstetrics & 7,544 & 1.5 \\
M400 - Pediatrics & 1,419 & 0.3 \\
M500 - Psychiatry/Psychotherapy & 304,685 & 60.3 \\
M600 - Ophthalmology & 307 & 0.1 \\
M700 - ENT (Oto-rhino-laryngology) & 1,824 & 0.4 \\
M800 - Dermatology/Venereology & 951 & 0.2 \\
M850 - Radiology & 302 & 0.1 \\
M900 - Geriatrics & 1,045 & 0.2 \\
M950 - Physical Med/Rehab & 9,306 & 1.8 \\
M960 - Emergency Centers & 2,234 & 0.4 \\
M970 - Emergency Doctor Office & 7 & 0.0 \\
M990 - Other Activities & 1,927 & 0.4 \\
\bottomrule
\end{tabular}
\label{tab:cost_coverage}
\end{table}

\begin{table}[H]
\caption{Distribution of Basic Care Coverage}
\centering
\begin{tabular}{lrr}
\toprule
\textbf{Basic Care Coverage} & \textbf{Count} & \textbf{Percentage} \\
\midrule
1 – Mandatory Health Insurance & 478,765 & 94.7 \\
2 – Disability Insurance & 812 & 0.2 \\
3 – Military Insurance & 287 & 0.1 \\
4 – Accident Insurance & 4,880 & 1.0 \\
5 – Self-Paying (e.g., uninsured foreigners) & 2,950 & 0.6 \\
8 – Other & 7,848 & 1.6 \\
9 – Unknown & 9,864 & 2.0 \\
\bottomrule
\end{tabular}
\label{tab:basic_care_coverage}
\end{table}

\begin{table}[H]
\caption{Distribution of Discharge Decisions}
\centering
\begin{tabular}{lrr}
\toprule
\textbf{Discharge Decision} & \textbf{Count} & \textbf{Percentage} \\
\midrule
1 - By Treating Physician & 417,432 & 82.6 \\
2 - By Patient (Against Advice) & 53,146 & 10.5 \\
9 - Unknown & 20,490 & 4.1 \\
8 - Other & 6,337 & 1.3 \\
5 - Deceased & 3,188 & 0.6 \\
4 - Internal Transfer & 2,515 & 0.5 \\
3 - By Third Party & 2,298 & 0.5 \\
\bottomrule
\end{tabular}
\label{tab:discharge_decision}
\end{table}

\begin{table}[H]
\caption{Distribution of Stay After Discharge}
\centering
\begin{tabular}{lrr}
\toprule
\textbf{Stay After Discharge} & \textbf{Count} & \textbf{Percentage} \\
\midrule
1 - Home & 338,749 & 67.0 \\
4 - Psych. Institution, Other & 40,217 & 8.0 \\
3 - Non-Med. Non-Hosp. Facility & 33,607 & 6.6 \\
9 - Unknown & 28,421 & 5.6 \\
8 - Other & 23,122 & 4.6 \\
6 - Other Hospital / Birth Home & 15,022 & 3.0 \\
5 - Rehab Institution, Other & 8,463 & 1.7 \\
2 - Med. Non-Hosp. Facility & 6,209 & 1.2 \\
7 - Penal Institution & 6,169 & 1.2 \\
0 - Deceased & 3,189 & 0.6 \\
44 - Psych. Division (Same) & 1,525 & 0.3 \\
55 - Rehab Division (Same) & 523 & 0.1 \\
66 - Acute Care Division (Same) & 190 & 0.0 \\
\bottomrule
\end{tabular}
\label{tab:stay_after_discharge}
\end{table}

\begin{table}[H]
\caption{Distribution of Post-Discharge Care}
\centering
\begin{tabular}{lrr}
\toprule
\textbf{Post-Discharge Care} & \textbf{Count} & \textbf{Percentage} \\
\midrule
2 - Outpatient Treatment & 301,794 & 59.7 \\
4 - Inpatient Care & 61,213 & 12.1 \\
1 - Recovered, No Follow-up & 54,610 & 10.8 \\
9 - Unknown & 41,895 & 8.3 \\
8 - Other & 20,914 & 4.1 \\
5 - Rehabilitation & 14,714 & 2.9 \\
3 - Home Care & 7,078 & 1.4 \\
0 - Deceased & 3,188 & 0.6 \\
\bottomrule
\end{tabular}
\label{tab:post_discharge_care}
\end{table}

\begin{table}[H]
\caption{ICD-10-GM Codes for Drug-Related Disorders and Infectious Diseases}
\centering
\small
\begin{tabular}{ll}
\toprule
\textbf{Category} & \textbf{ICD-10-GM Codes} \\
\midrule
\textbf{Drug-Related Disorders (Mental and behavioral)} & \\
\hline
\text{Due to opioids} & F11 \\
\text{Due to cannabinoids} & F12 \\
\text{Due to  sedatives or hypnotics} & F13 \\
\text{Due to cocaine} & F14 \\
\text{Due to other stimulants} & F15 \\
\text{Due to hallucinogens} & F16 \\
\text{Due to volatile solvents} & F18 \\
\text{Due to multiple drug use and other psychoactive substances} & F19 \\
\hline
\textbf{Fatal Drug Overdoses} & \\
\hline
\text{Due to Opium} & T40.0 \\
\text{Due to Heroin} & T40.1 \\
\text{Due to Other opioids} & T40.2 \\
\text{Due to Other synthetic narcotics} & T40.4 \\
\text{Due to Other and unspecified drugs} & T40.6 \\
\hline
\textbf{Accidental Poisoning} & X41–X42 \\
\hline
\textbf{Infectious Diseases} & \\
\hline
\text{HIV} & B20-B24 \\
\text{Viral Hepatitis C} & B15-B19 \\
\bottomrule
\end{tabular}
\label{tab:icd10_codes}
\end{table}

\clearpage

\subsection{Additional Figures}
\label{app:figures}

\begin{figure}[H]
    \centering
    \caption{Persons in Opioid Agonist Treatment (1999--2025)}
    \label{fig:OAT}
    \includegraphics[width=\textwidth]{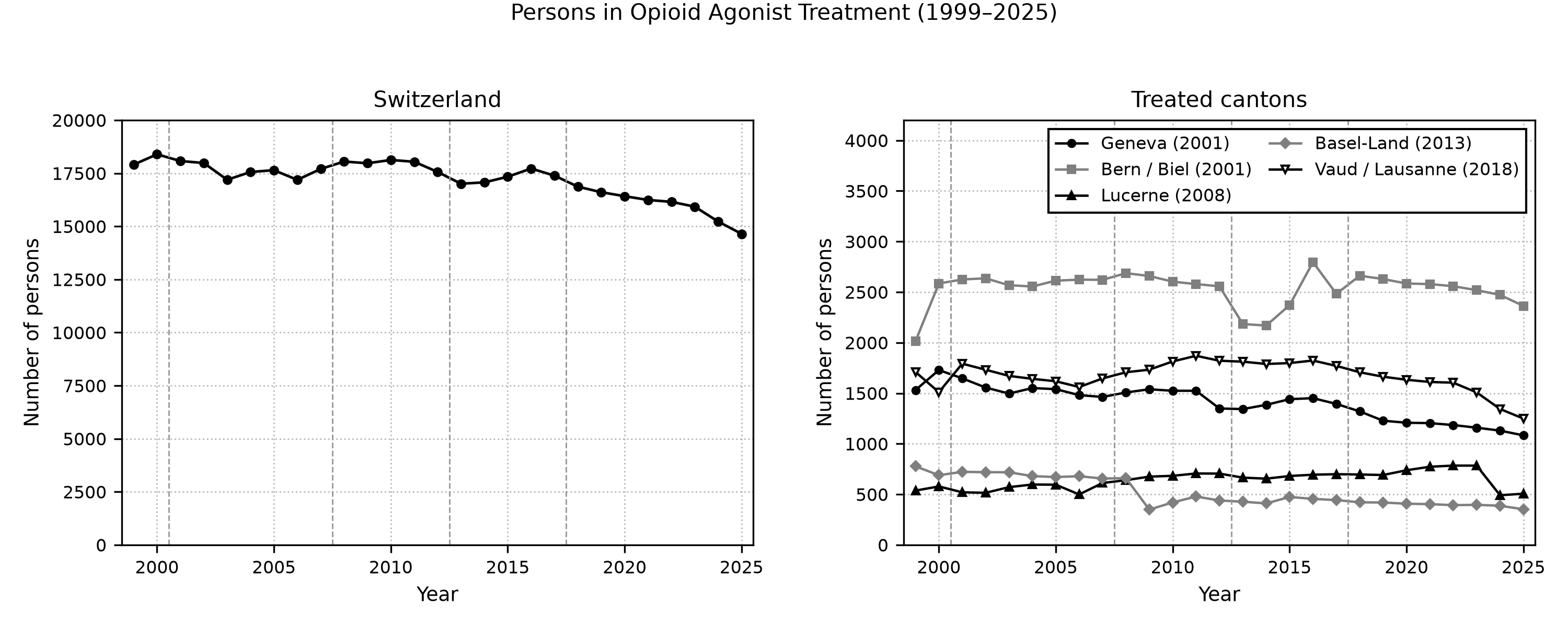}
    \\
    \vspace{0.5cm}
    \setstretch{0.9}
    \raggedright
    \footnotesize \textit{Note:} Number of persons in opioid agonist treatment (methadone, buprenorphine, slow-release oral morphine and other opioid agonists) who were in treatment for at least one day in the year. The left panel shows the national total; the right panel shows the five cantons that host a treatment cohort, with dashed vertical lines at the year the consumption room opened. Persons are attributed to the canton in which the prescribing physician or institution practises, not to the canton of residence. Cantonal physicians have reported these data annually since 1999; figures before 2018 are weighted estimates where cantonal reporting was incomplete, and from 2018 a full census. The drop in Basel-Land in 2009 reflects a change in cantonal reporting rather than in treatment.
    \\
    \footnotesize \textit{Source:} Nationale Statistik der Opioid-Agonisten-Therapie, Sucht Schweiz on behalf of the Federal Office of Public Health (\cite{taooat2026}; \cite{labhart2020}).
\end{figure}

\begin{figure}[H]
    \centering
    \caption{Narcotics Act Convictions (1994--2025)}
    \label{fig:convictions}
    \includegraphics[width=\textwidth]{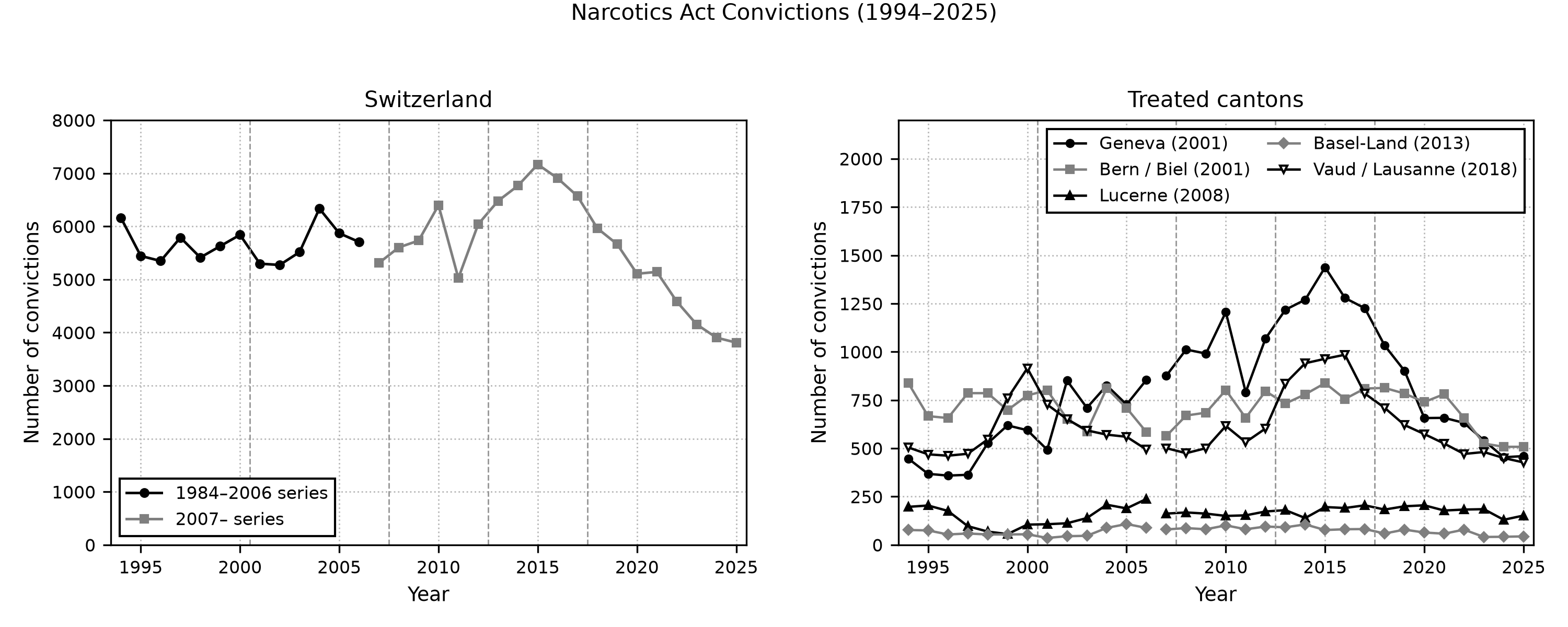}
    \\
    \vspace{0.5cm}
    \setstretch{0.9}
    \raggedright
    \footnotesize \textit{Note:} Convictions of adults for a felony or misdemeanour under the Federal Narcotics Act (BetmG). Consumption offences (Art.~19a BetmG), which are contraventions, are not included, so the series measures enforcement against dealing and larger possession. The left panel shows the national total; the right panel shows the five cantons that host a treatment cohort, with dashed vertical lines at the year the consumption room opened. Convictions are attributed to the canton of the sentencing authority, not to the offender's canton of residence. Filled markers are the 1984--2006 table and open markers the 2007 onward table of the conviction statistics; the two are not joined because the revised Criminal Code of 2007 changed the recording of sanctions.
    \\
    \footnotesize \textit{Source:} Strafurteilsstatistik, Federal Statistical Office (\cite{bfs_sus}; \cite{bfs_sus2007}).
\end{figure}

\begin{figure}[H]
    \centering
    \caption{Population Density by MedStat Region and Location of DCRs}
    \label{fig:medstat_population}
    \includegraphics[width=0.7\textwidth]{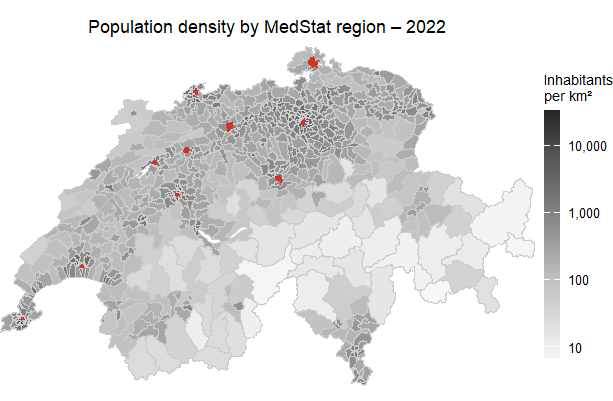}
    \\
    \vspace{0.5cm}
    \setstretch{0.9}
    \raggedright
    \footnotesize \textit{Note:} Each polygon represents a MedStat region, the spatial unit used in the analysis. These regions are defined by the Federal Statistical Office and correspond approximately to hospital catchment areas.
    Shading gives population density in 2022, in inhabitants per square kilometre, on a logarithmic scale. Regions shaded red are the MedStat regions hosting a DCR, listed in Table \ref{tab:dcr_openings}.
    Population figures are constructed from municipal-level data mapped to MedStat regions via postal codes.
    \\
    \footnotesize \textit{Source:} \cite{bfs_medstat}
    \label{fig:medstat_population}
\end{figure}

\begin{figure}[H]
    \centering
    \caption{Overall Hospitalizations}
    \includegraphics[width=0.7\textwidth]{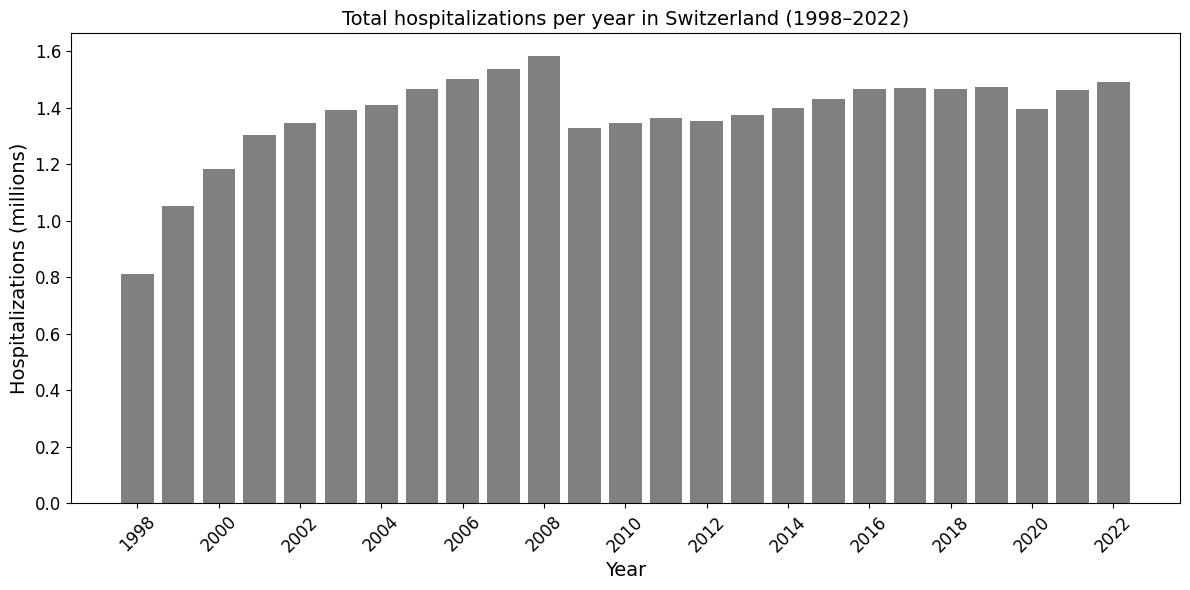}
    \label{fig:hosp_over_time}
\end{figure}

\begin{figure}[H]
    \centering
    \caption{Overall Hospitalizations by Age}
    \includegraphics[width=0.7\textwidth]{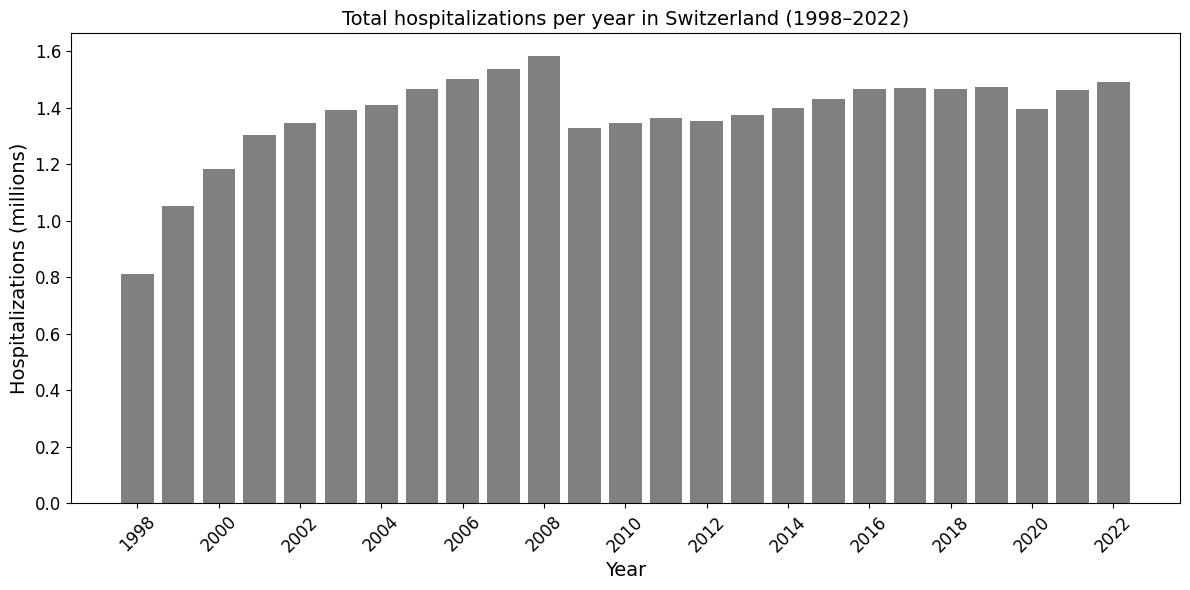}
    \label{fig:hosp_by_age}
\end{figure}

\begin{figure}[H]
    \centering
    \caption{Drug-related Hospitalizations}
    \includegraphics[width=0.7\textwidth]{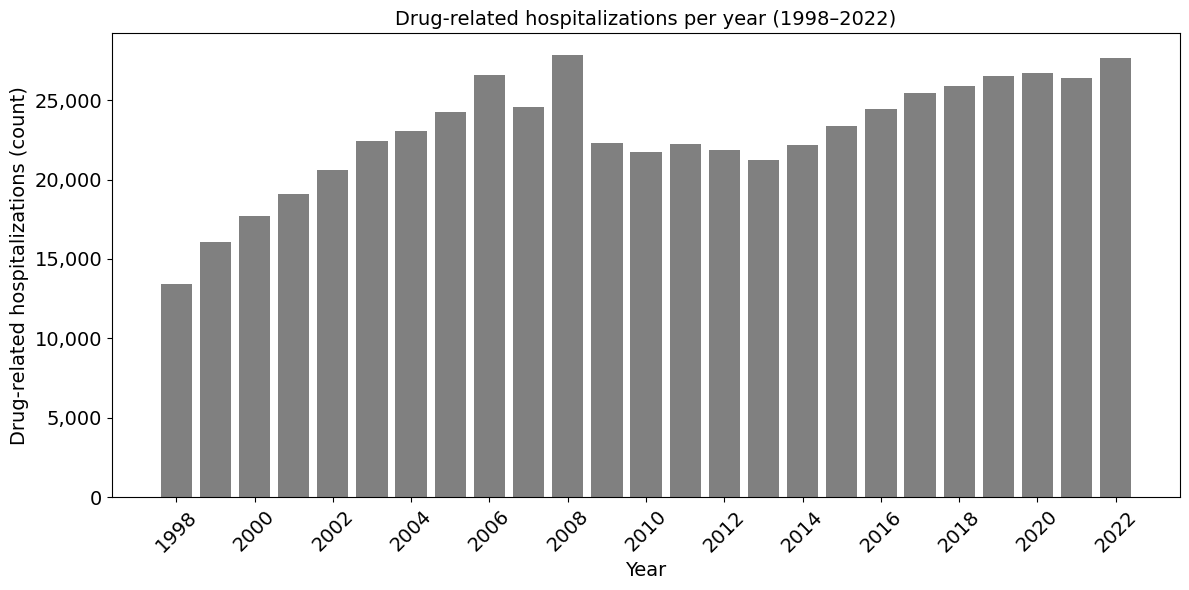}
    \label{fig:drug_hosp_over_time}
\end{figure}

\begin{figure}[H]
    \centering
    \caption{Drug-related Hospitalizations due to Mental/Behavioral disorders}
    \includegraphics[width=0.7\textwidth]{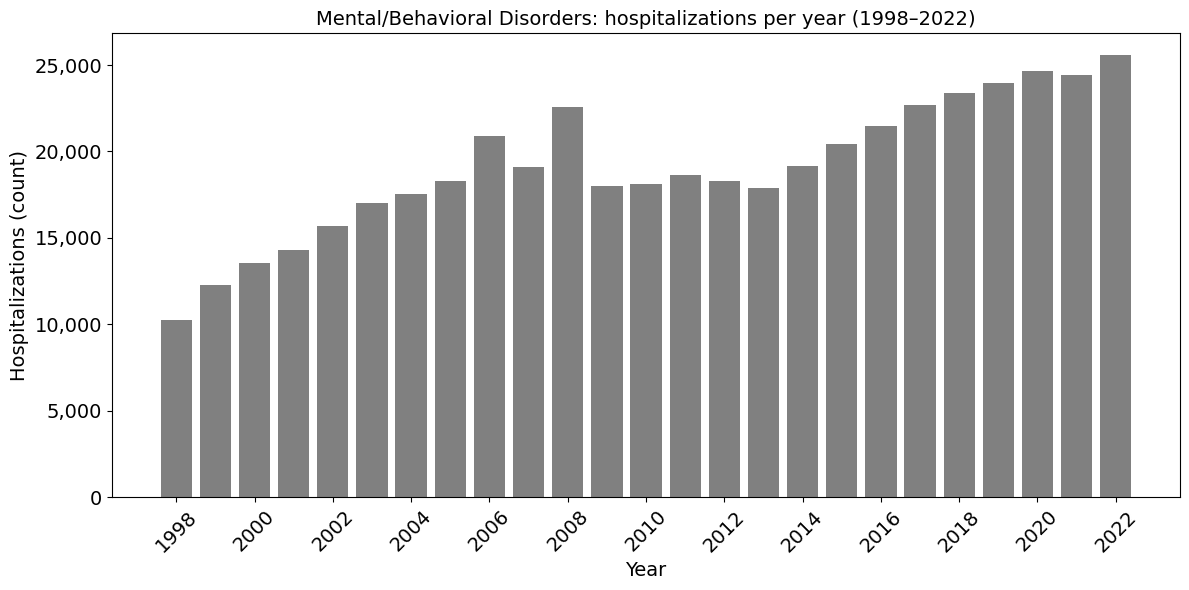}
    \label{fig:mental_hosp_over_time}
\end{figure}

\begin{figure}[H]
    \centering
    \caption{Drug-related Hospitalizations due to Hepatitis}
    \includegraphics[width=0.7\textwidth]{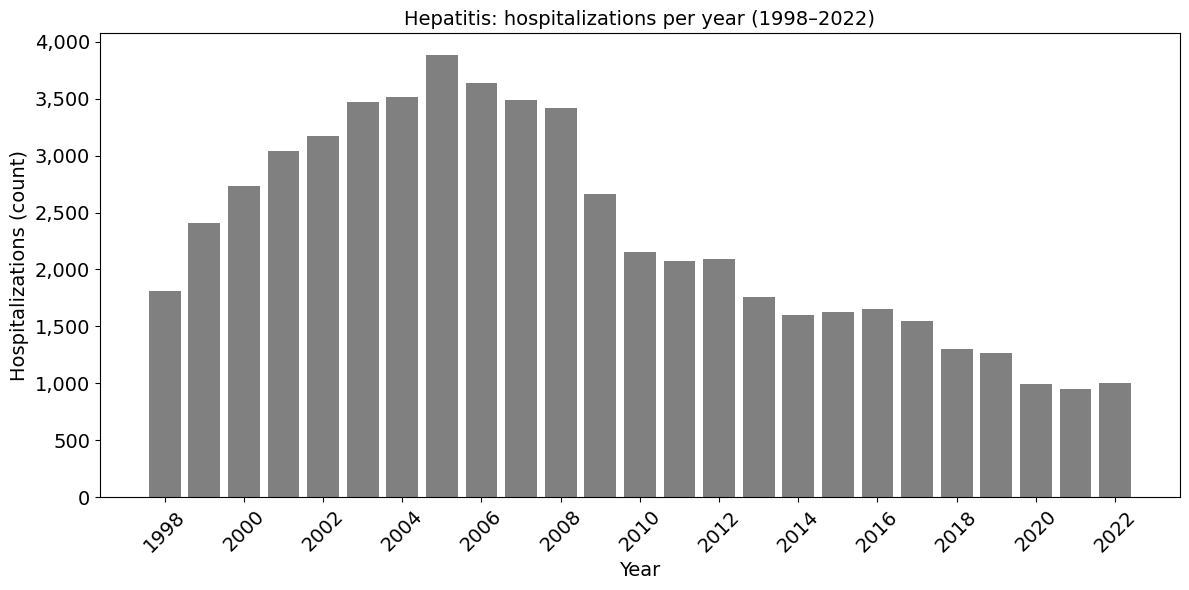}
    \label{fig:hepa_hosp_over_time}
\end{figure}

\begin{figure}[H]
    \centering
    \caption{Drug-related Hospitalizations due to HIV}
    \includegraphics[width=0.7\textwidth]{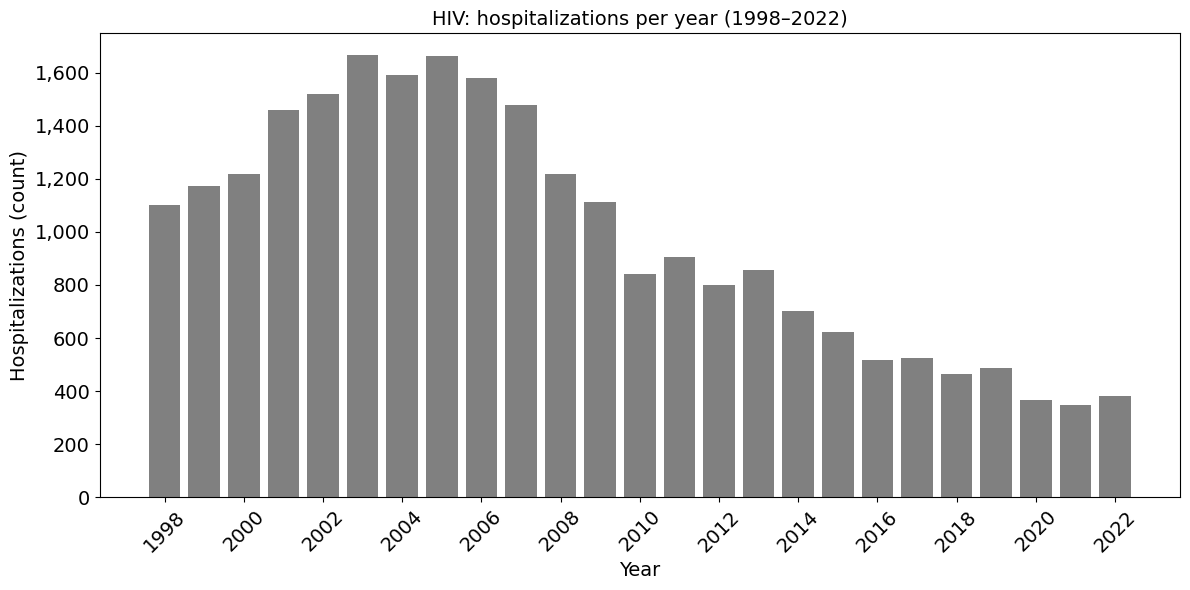}
    \label{fig:hiv_hosp_over_time}
\end{figure}

\begin{figure}[H]
    \centering
    \caption{Drug-related Hospitalizations due Narcotics/Psychodysleptics}
    \includegraphics[width=0.7\textwidth]{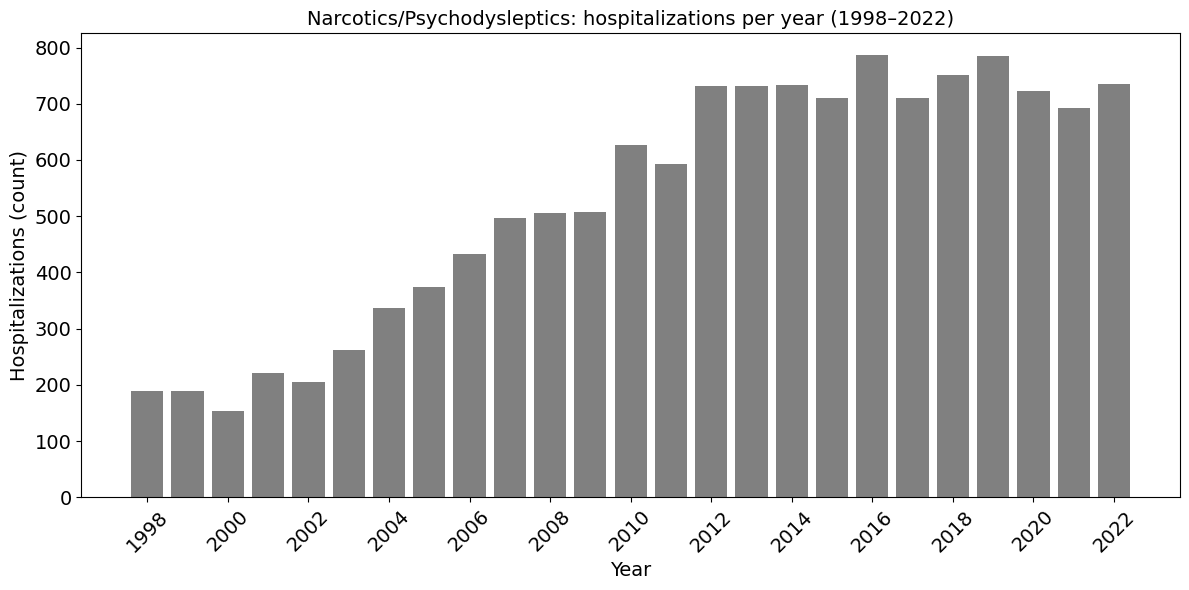}
    \label{fig:narc_hosp_over_time}
\end{figure}

\begin{table}[H]
\input{tables/tab_desc_stats}
\end{table}

\clearpage
\subsection{Validation of Hospital Data with Official Statistics}
\label{app:validation}

This section compares the drug-related outcomes derived from the hospital dataset  with official statistics. The aim is to assess the representativeness of the hospital data. Overall, the hospital-based figures tend to be lower in absolute terms than the official  numbers, but both follow similar trends over time. This supports the use of hospital  records as a proxy for drug-related events in Switzerland. 

\begin{figure}[H]
    \centering
    \caption{Drug-related deaths: Hospital data vs Obsan statistics}
    \includegraphics[width=0.7\textwidth]{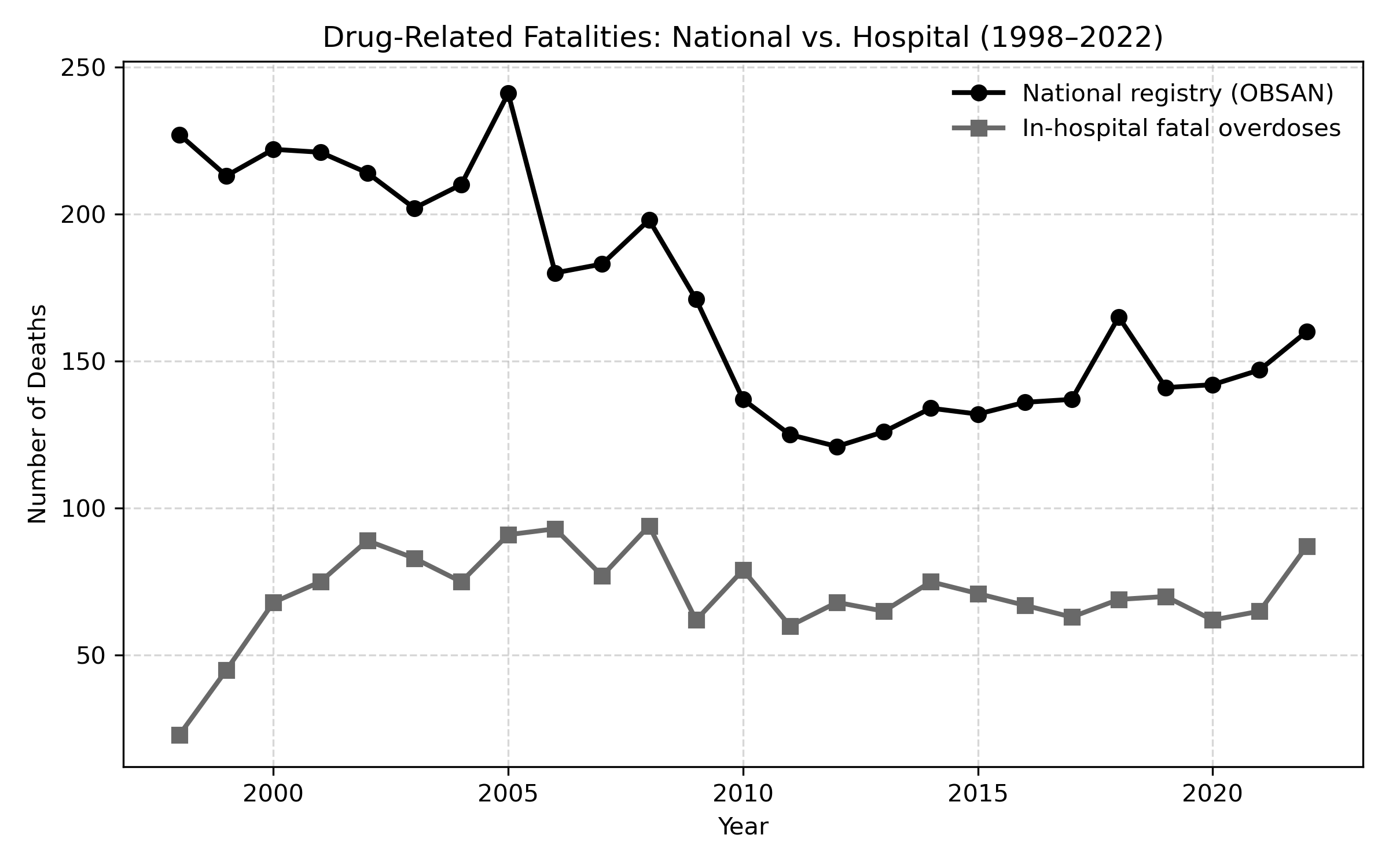}
    \\ 
    \textit{Source: Swiss Health Observatory (Obsan)}
    \label{fig:obsan_trends}
\end{figure}

\begin{figure}[H]
    \centering
    \caption{HIV: Hospital data (primary diagnosis) vs BFS statistics.} 
    \includegraphics[width=0.7\textwidth]{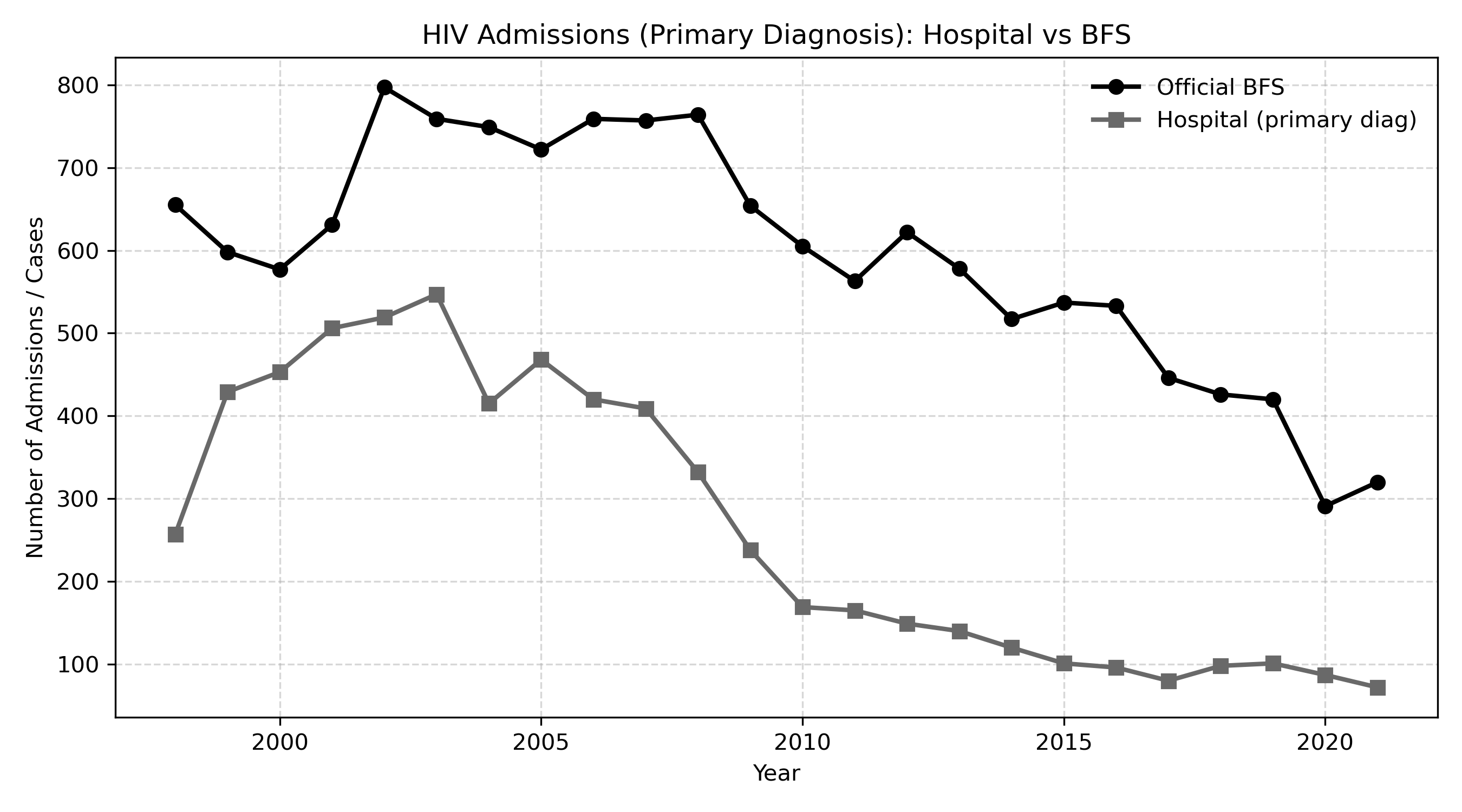}
    \\ 
    \textit{Source: Swiss Federal Office of Public Health}
    \label{fig:hiv_primary_events_vs_bfs}
\end{figure}

\begin{figure}[H]
    \centering
    \caption{HCV: Hospital data (primary diagnosis) vs BFS statistics.}
    \includegraphics[width=0.7\textwidth]{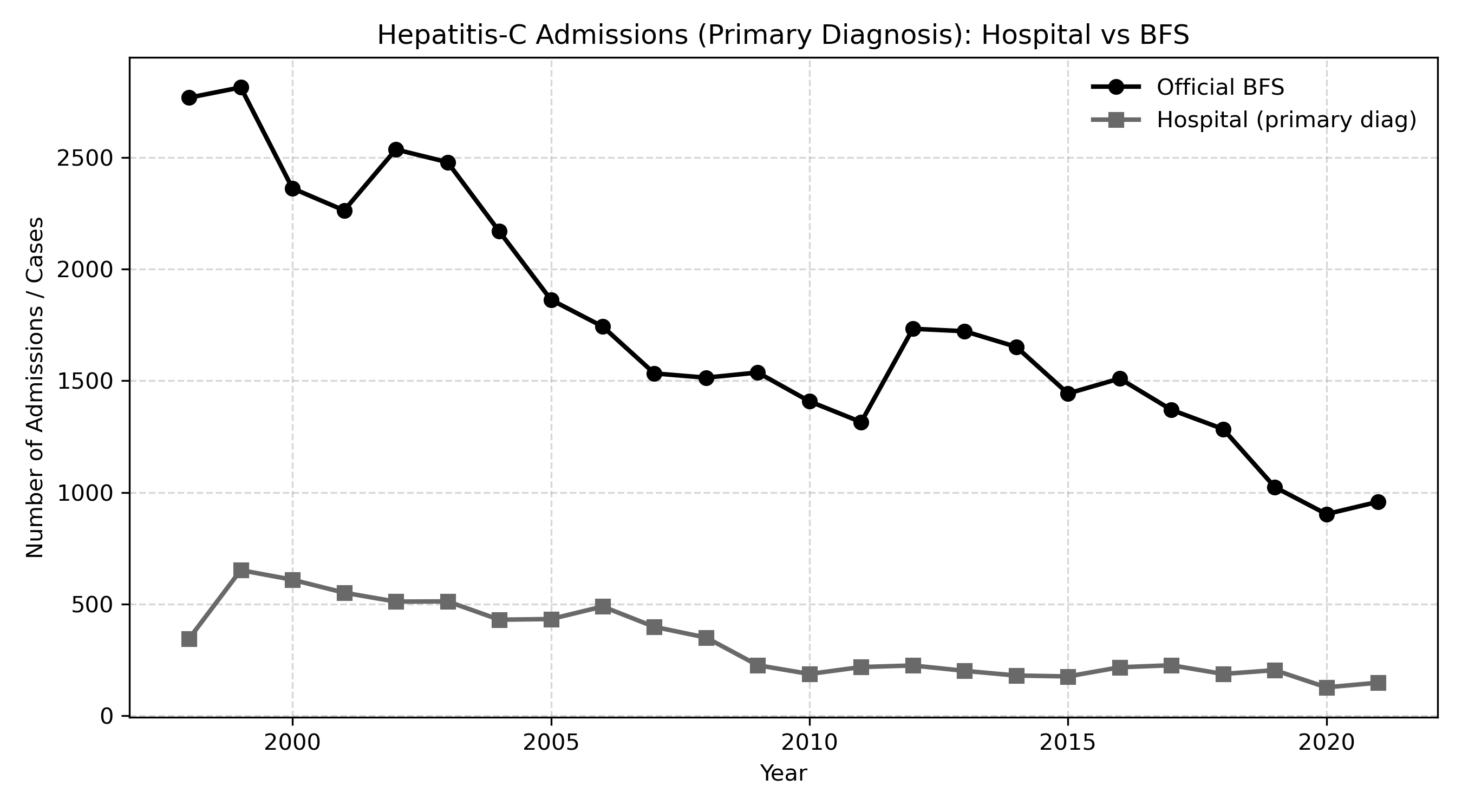}
    \\ 
    \textit{Source: Swiss Federal Office of Public Health}
    \label{fig:hcv_primary_events_vs_bfs}
\end{figure}

\clearpage

\subsection{Balance} 
\label{app:balance}

\begin{table}[H] 
\input{tables/balance_full}
\end{table}

\begin{table}[H] 
\input{tables/balance_drug_pre}
\end{table}

Table \ref{tab:balance_drug_pre} reports the balance of pre-treatment characteristics for the subsample of drug-related hospitalizations. All standardized mean differences are below 20, suggesting that treated and control patients are broadly comparable on observables in this subsample. However, the characteristics of drug-related hospitalizations differ from those of hospitalizations overall. In particular, there is a higher proportion of Swiss patients in treated areas than in control areas. Additionally, approximately 61\% of patients are male, which is considerably higher than the share of male patients in overall hospitalizations (44\%) reported in Table \ref{tab:balance_pre}. The length of hospital stay is longer for drug-related hospitalizations, with a mean of 31 nights, although the median is 9 nights and the mode is 1 night. The number of hours spent in the ICU is higher in control areas. Emergency admissions constitute the most frequent mode of entry for drug-related hospitalizations, and most stays are inpatient. Drug-related hospitalizations are also concentrated among adults aged 20 to 49. These differences reflect how the drug-related subsample differs from the general inpatient population rather than imbalances between the treated and control groups. 

\begin{table}[H] 
\input{tables/balance_full_drug}
\end{table}

For completeness, Table \ref{tab:balance_drug} reports the corresponding balance table for drug-related hospitalizations over the full study period, showing similar patterns and small standardized differences.

\subsection{Parallel Trends} 
\label{app:parallel_trends}

\begin{figure}[H]
    \centering
    \caption{Drug-related hospitalization rate (log) per 100,000 inhabitants}
    \includegraphics[width=0.85\textwidth]{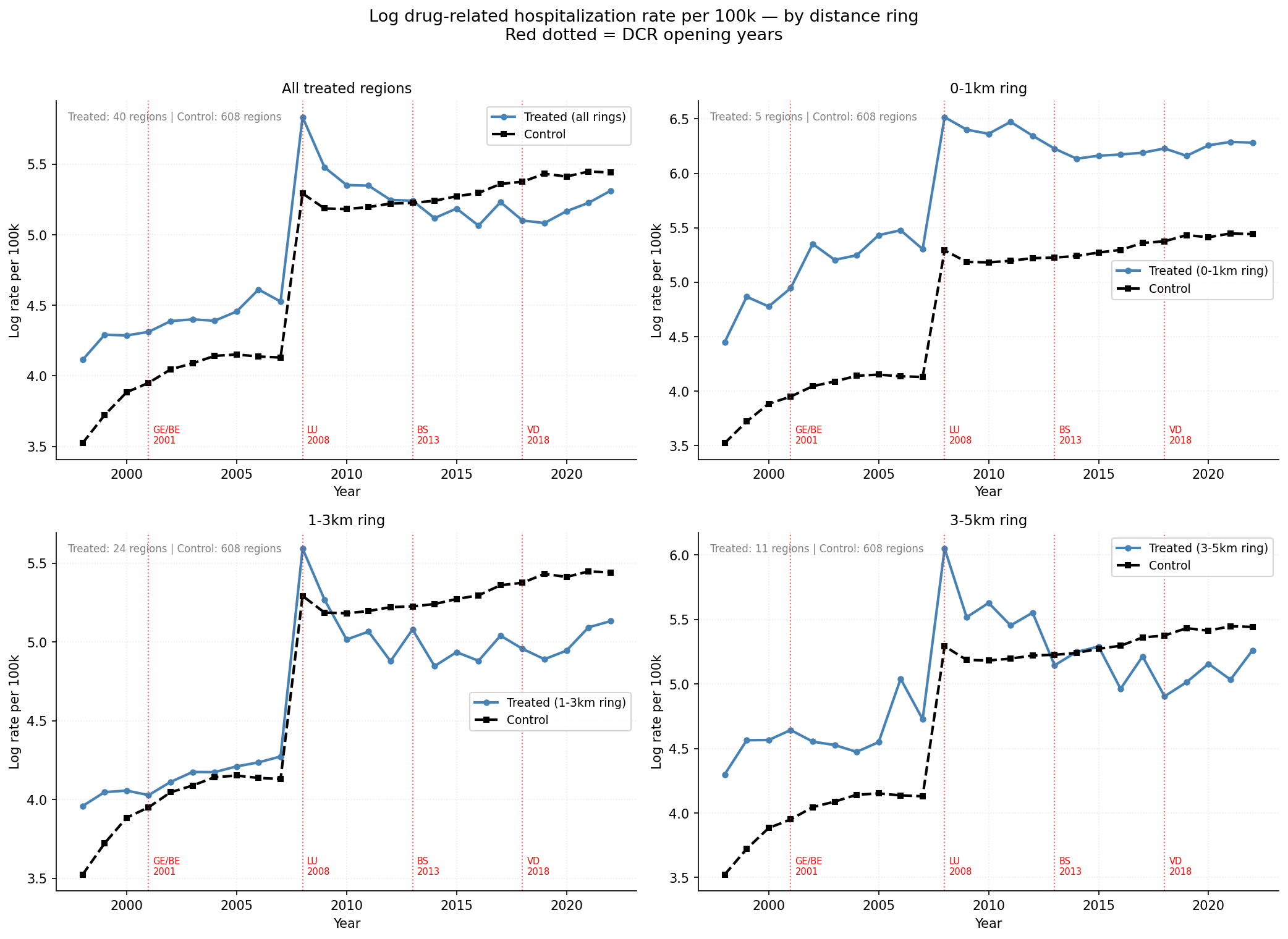}
    \label{fig:log_rate_by-ring}
    \vspace{0.3cm}
    \\
    \setstretch{0.9}
    \raggedright
    \footnotesize{\textit{Note:} The figure plots the average log drug-related hospitalization rate per 100,000 inhabitants for MedStat regions between 1998 and 2022. The top-left panel shows all treated regions within 5\,km of a Drug Consumption Room (DCR), while the remaining panels separate regions by distance rings (0--1\,km, 1--3\,km, and 3--5\,km). Treated regions are compared to MedStat regions with no DCR during the sample period (never-treated). Vertical red dotted lines indicate the opening years of DCRs.}
\end{figure}

Identification requires that treated and never-treated regions would have followed common trends absent a DCR. Figure \ref{fig:log_rate_by-ring} plots average log drug-related hospitalization rates for both groups, pooled and by distance ring.

Treated regions lie above never-treated regions throughout, however, what matters for identification is the co-movement of the two series, not the gap between them. The groups move broadly together up to the first openings in 2001. The treated series is more volatile thereafter, reflecting the few regions per ring (5, 24 and 11 in the 0--1\,km, 1--3\,km and 3--5\,km panels). Both series also shift sharply between 2008 and 2009.

\begin{figure}[H]
    \centering
    \caption{Outcome trends by measure}
    \includegraphics[width=0.95\textwidth]{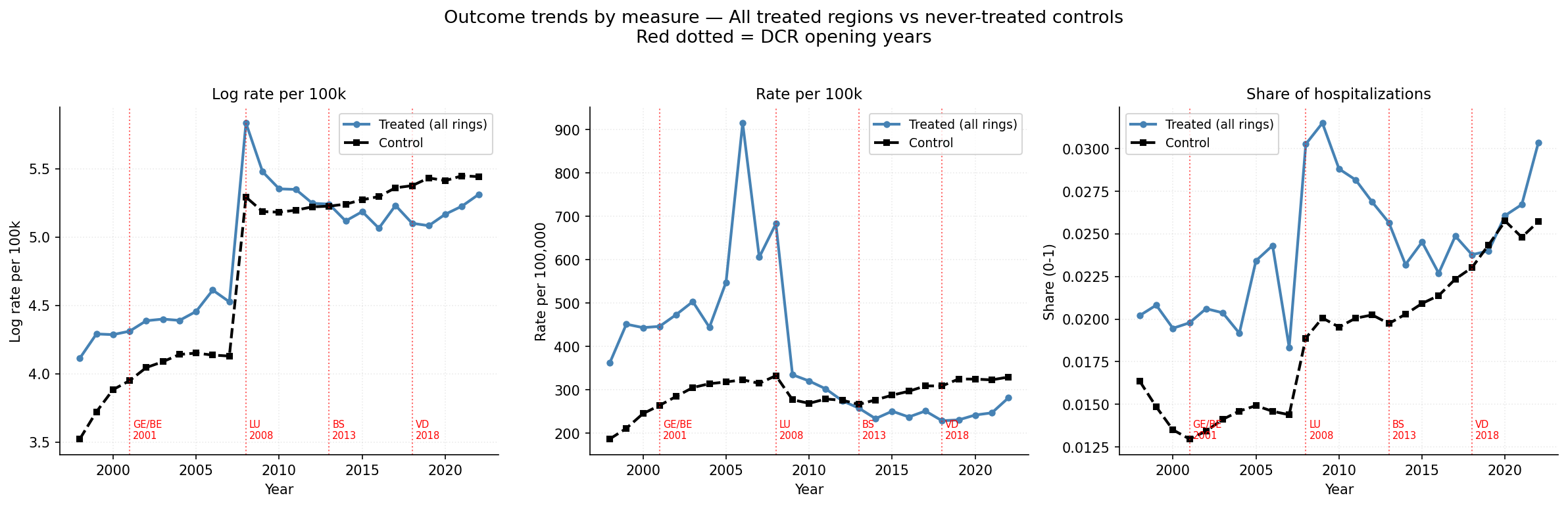}
    \label{fig:all_outcomes_by-ring}
    \vspace{0.3cm}
    \\
    \setstretch{0.9}
    \raggedright
    \footnotesize{\textit{Note:} The figure plots trends for different outcome measures for MedStat regions between 1998 and 2022. The left panel shows the log hospitalization rate per 100,000 inhabitants, the middle panel shows the hospitalization rate per 100,000 inhabitants, and the right panel shows the share of drug-related hospitalizations among all hospitalizations. Treated regions are compared to MedStat regions with no DCR during the sample period (never-treated). Vertical red dotted lines indicate the opening years of DCRs.}
\end{figure}

Figure \ref{fig:all_outcomes_by-ring} repeats the comparison across the three outcome measures. The 2008--2009 shift coincides with the revision of the Hospital Medical Statistics discussed in Section \ref{sec:data} which is common to both groups. 

% Monthly trends:
\begin{figure}[H]
    \centering
    \caption{Monthly drug-related Hospitalizations by Subcategory}
    \includegraphics[width=0.7\textwidth]{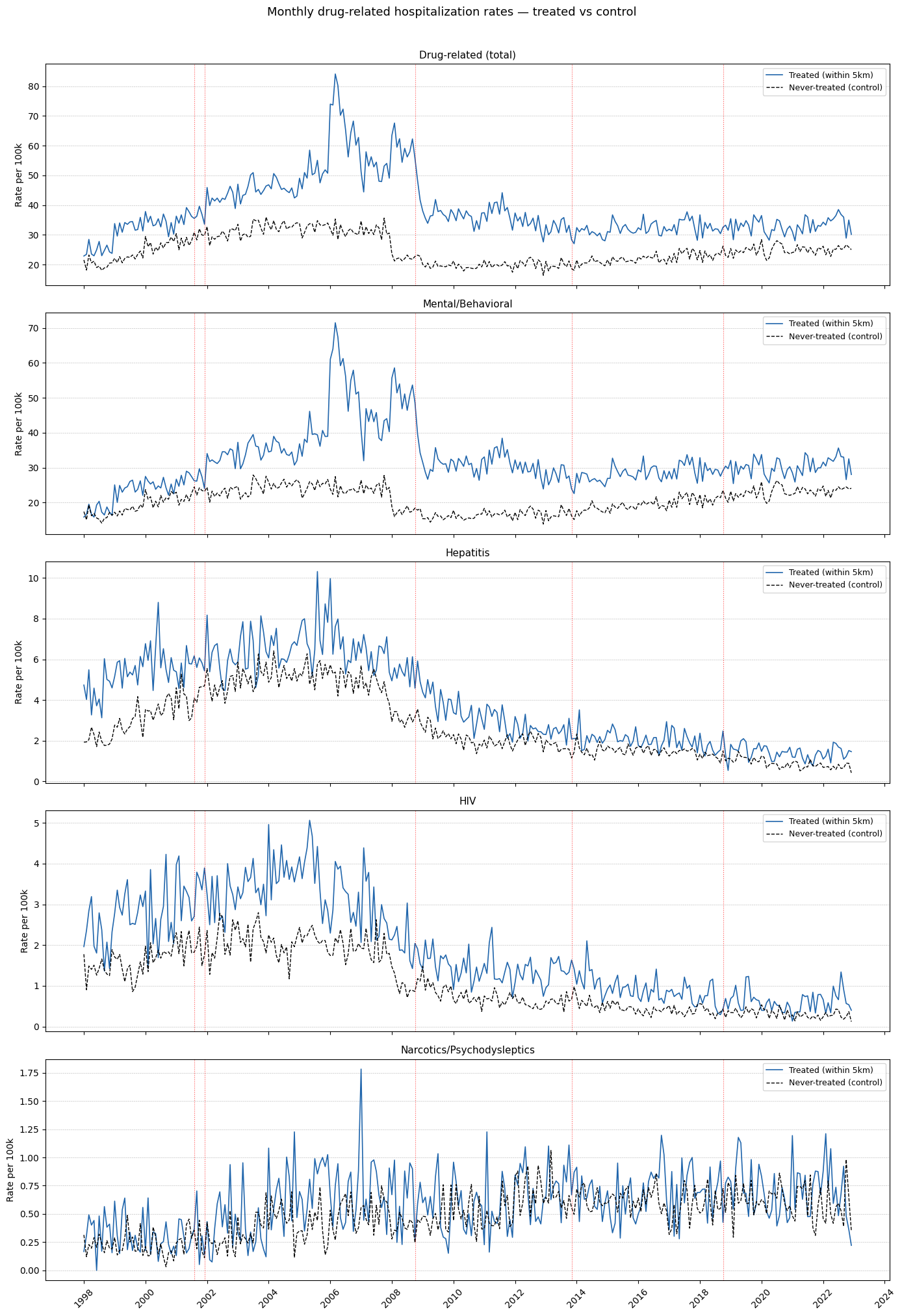}
    \label{fig:lineplot_monthly_outcome}
\end{figure}

\subsection{SUTVA Validation}
\label{app:sutva}

Outcomes are assigned to the patient's MedStat region of residence, and regions
beyond 5\,km of a DCR are treated as unexposed. This requires that patients do not
routinely seek care outside their own area in ways that would blur the boundary
between treated and control regions. Table~\ref{tab:same_canton_by_year} reports, for each
year, the share of hospitalizations that take place in the patient's canton of
residence.

\begin{table}[H]
\centering
\caption{Share of hospitalizations occurring in the patient’s canton of residence}
\small
\begin{tabular}{rrrr}
\toprule
\textbf{Year} & \textbf{Same canton} & \textbf{Valid records} & \textbf{Share same canton (\%)} \\
\midrule
1998 & 614{,}478 & 701{,}998 & 87.53 \\
1999 & 763{,}067 & 867{,}874 & 87.92 \\
2000 & 925{,}838 & 1{,}053{,}279 & 87.90 \\
2001 & 1{,}010{,}817 & 1{,}156{,}617 & 87.39 \\
2002 & 1{,}047{,}693 & 1{,}199{,}927 & 87.31 \\
2003 & 1{,}085{,}331 & 1{,}241{,}167 & 87.44 \\
2004 & 1{,}102{,}313 & 1{,}260{,}477 & 87.45 \\
2005 & 1{,}151{,}872 & 1{,}312{,}613 & 87.75 \\
2006 & 1{,}183{,}320 & 1{,}345{,}387 & 87.95 \\
2007 & 1{,}208{,}539 & 1{,}377{,}289 & 87.75 \\
2008 & 1{,}330{,}669 & 1{,}538{,}921 & 86.47 \\
2009 & 1{,}107{,}437 & 1{,}288{,}432 & 85.95 \\
2010 & 1{,}121{,}673 & 1{,}304{,}454 & 85.99 \\
2011 & 1{,}135{,}999 & 1{,}324{,}288 & 85.78 \\
2012 & 1{,}117{,}158 & 1{,}315{,}132 & 84.95 \\
2013 & 1{,}122{,}626 & 1{,}336{,}296 & 84.01 \\
2014 & 1{,}135{,}151 & 1{,}363{,}212 & 83.27 \\
2015 & 1{,}156{,}687 & 1{,}394{,}135 & 82.97 \\
2016 & 1{,}181{,}862 & 1{,}431{,}726 & 82.55 \\
2017 & 1{,}180{,}354 & 1{,}433{,}681 & 82.33 \\
2018 & 1{,}173{,}396 & 1{,}430{,}541 & 82.02 \\
2019 & 1{,}173{,}560 & 1{,}434{,}735 & 81.80 \\
2020 & 1{,}113{,}501 & 1{,}365{,}321 & 81.56 \\
2021 & 1{,}161{,}172 & 1{,}431{,}774 & 81.10 \\
2022 & 1{,}173{,}312 & 1{,}453{,}982 & 80.70 \\
\bottomrule
\end{tabular}
\label{tab:same_canton_by_year}
\end{table}

The share of same canton hospitalizations is high throughout, but it is not constant. It remains between 87.3\% and 88.0\% until 2007 and then falls to 80.7\% by 2022. The decline is concentrated after 2012. This break coincides with the revision of the Federal Health Insurance Act (KVG) governing hospital financing, which entered into force in 2009 and became fully operative in 2012. During this time, case-based payments (SwissDRG) were introduced and patients gained the right to be treated at any hospital on a cantonal hospital list \cite{lamal_rev2007}. Prior to 2012, treatment outside the canton of residence generally required cantonal authorization. A second, slower channel which operates over the whole period is that the number of hospitals declined as small regional facilities closed or merged, so that residents of cantons without a full-service hospital were increasingly treated across cantonal borders.

Two features of this trend limit its consequences for the analysis. First, it is
driven by national policy rather than by anything specific to regions with a DCR. Second, because outcomes are assigned by MedStat region of residence rather than by the location of the treating hospital, a patient who crosses a cantonal border for care is still counted in their home MedStat region. The trend therefore changes where treatment is delivered but not which region the observation is attributed to.

\subsection{Additional Results \& Robustness}
\label{app:results}

\begin{figure}[H]
    \centering
    \caption{Event study: Callaway \& Sant'Anna estimates, 
    within 1\,km, 3\,km, 5\,km, $\pm$5-year window}
    \includegraphics[width=0.85\textwidth]{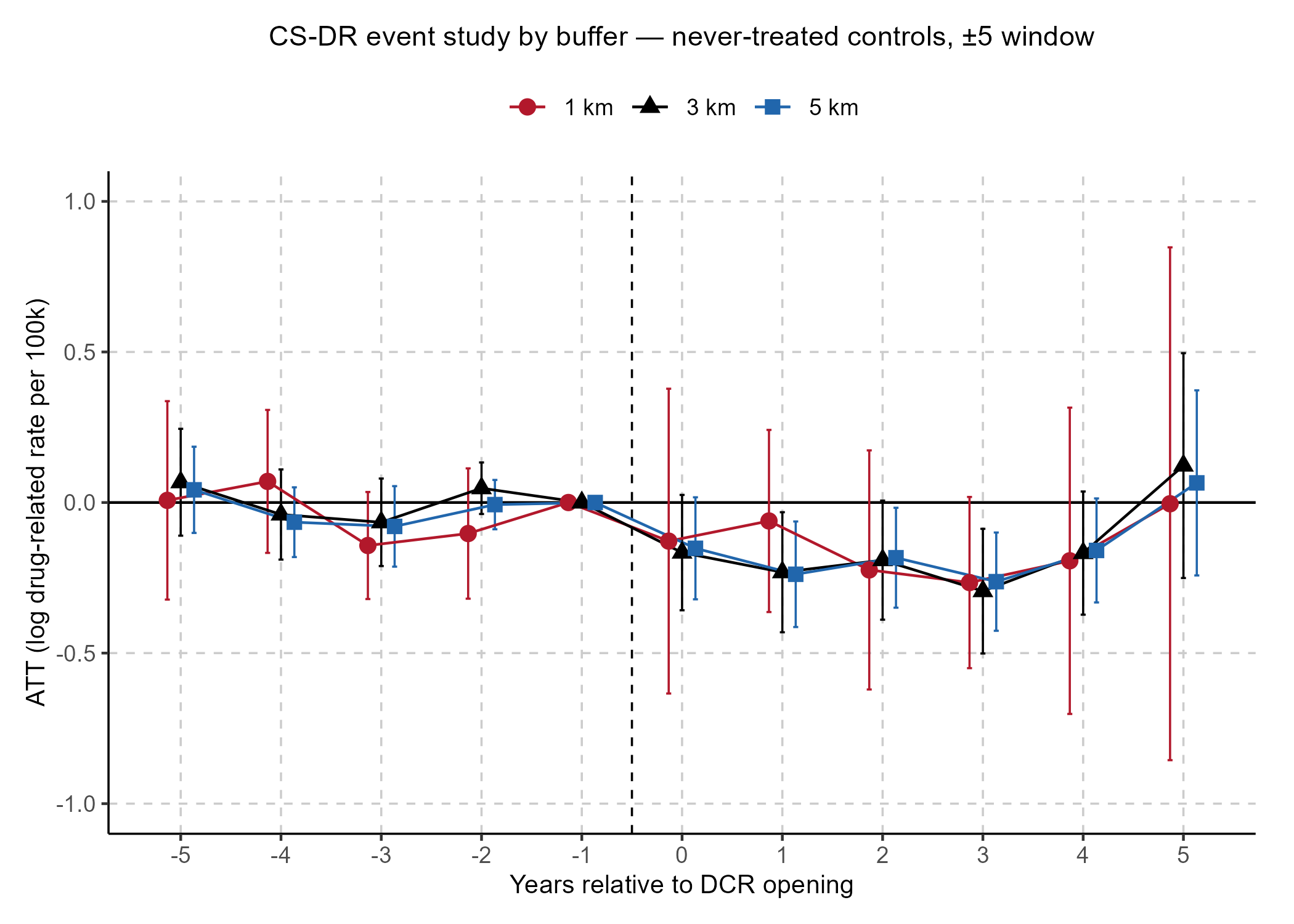}
    \label{fig:event_study_1km_3km_5km_5t_four_estimators}
    \vspace{0.3cm}
    \\
    \setstretch{0.8}
    \raggedright
    \footnotesize{\textit{Note:} The figure plots event-study coefficients from three treatment buffer definitions.}
\end{figure}

The recent difference-in-differences literature has documented that the two-way fixed effects (TWFE) estimator can produce biased and difficult-to-interpret estimates in  settings with staggered treatment adoption and heterogeneous treatment effects \citep{Borusyak_2024, chaisemartin_2020, GOODMANBACON2021254}. In particular, TWFE estimates can be contaminated by ``forbidden comparisons'' in which already-treated units serve as implicit controls for later-treated units, potentially reversing the sign of the estimated effect when treatment effects vary across cohorts or over time \citep{GOODMANBACON2021254}. Given that DCRs opened at different points in 
time across Swiss cantons and that treatment effect heterogeneity is likely in this setting, TWFE estimates should be interpreted with caution. I therefore report the TWFE event study solely as an additional visual diagnostic for pre-treatment parallel trends. The preferred estimates throughout are based on the doubly robust estimator of \citet{CALLAWAY2021200}, which accommodates staggered adoption, allows for cohort-specific treatment effects, and is robust to misspecification of either the propensity score or the outcome model.

\input{tables/robustness_sutva}

Table~\ref{tab:entitlement} assesses whether the results depend on how the
inter-cantonal access agreements are treated. Legal entitlement to enter a DCR
is granted by canton of residence, and two agreements extend it across cantonal
borders, so residents of Basel-Land may use the Basel-Stadt facility and
residents of Nidwalden, Obwalden, Schwyz, Uri and Zug may use the Lucerne
facility. The baseline specification classifies these regions as treated only if
they fall within the 5 km buffer. Column 2 instead classifies every region of
these cantons as treated from the year the corresponding facility opened,
regardless of distance, which is the most expansive reading of entitlement.
Reclassification moves 54 regions from the control group to the treated group
and leaves the estimate essentially unchanged, going from $-0.431$ to $-0.443$,
or from a 35.0\% to a 35.8\% reduction in the drug-related hospitalization rate.
Since the treated group more than doubles, the estimate is also more precisely
estimated, and the standard error falls from 0.216 to 0.189. How the
inter-cantonal agreements are classified therefore does not drive the results.

\input{tables/canton_level}

Table~\ref{tab:cantonlevel} defines treatment at the canton level, which rules
out within-canton spillovers by construction. The estimate falls from $-0.431$
to $-0.276$, or from a 35.0\% to a 24.1\% reduction in the drug-related
hospitalization rate. This attenuation is what averaging exposed and unexposed
regions together mechanically produces. In the three years before each opening,
67.6\% of the drug-related hospitalizations in these cantons occurred in regions
within 5 km of the facility, so if only those regions respond the canton-wide
rate should fall by 23.7\%. The estimate is a 24.1\% reduction. The difference
between the two columns therefore reflects the unit of aggregation and not
spillovers onto untreated regions.

\begin{table}[H] 
\input{tables/tab_combined_att}
\end{table}

\begin{figure}[H]
    \centering
    \caption{Event study: TWFE, Sun \& Abraham, Callaway \& Sant'Anna and, Stacked DiD estimates, 
    within 5\,km, $\pm$5-year window}
    \includegraphics[width=0.85\textwidth]{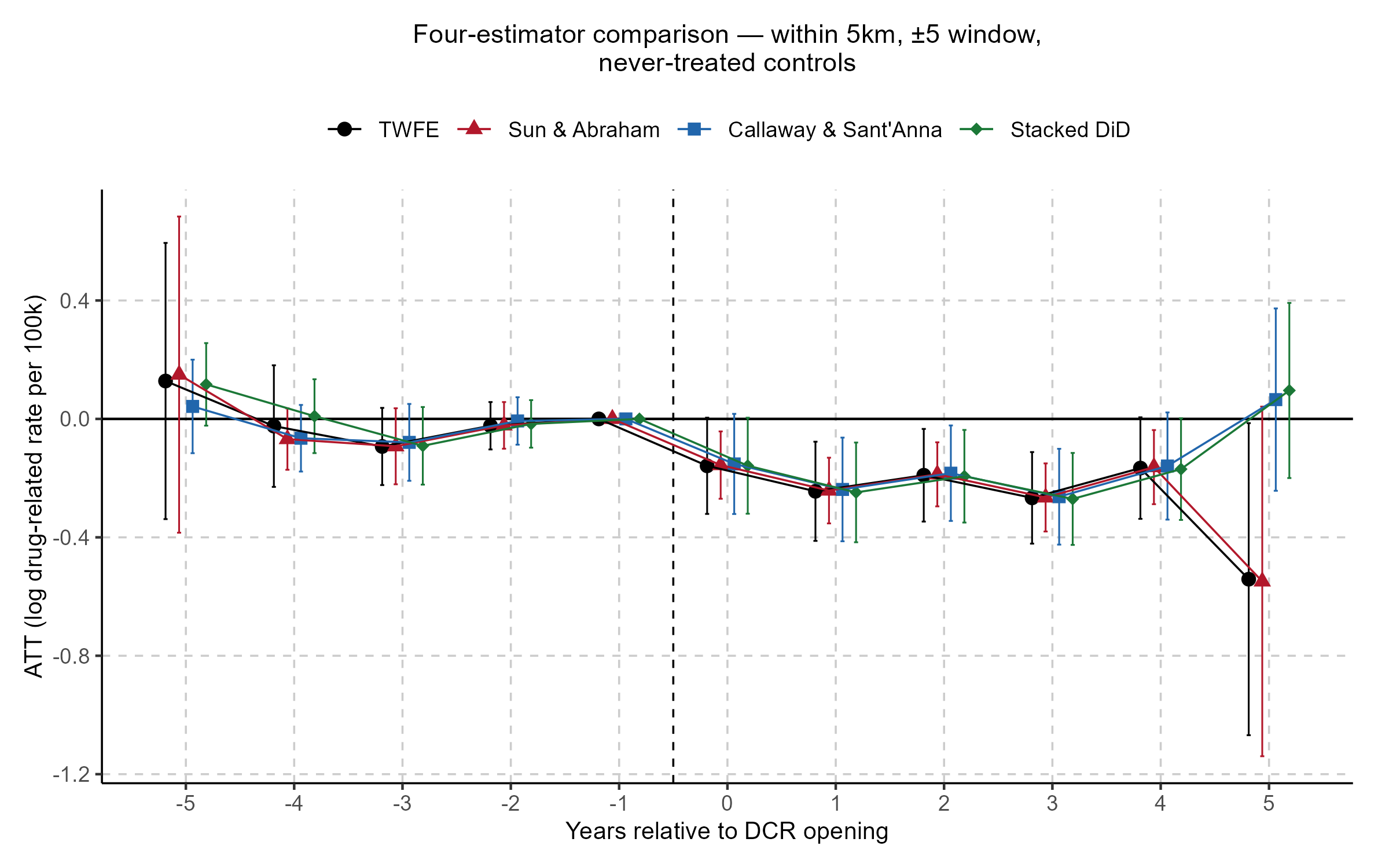}
    \label{fig:event_study_4_estimators}
    \vspace{0.3cm}
    \\
    \setstretch{0.8}
    \raggedright
    \footnotesize{\textit{Note:} The figure plots event-study coefficients from three estimators: the two-way fixed effects model (black circles), the heterogeneity-robust estimator of \cite{SUN2021} (red triangles), the doubly robust estimator of \cite{CALLAWAY2021200} (blue squares) and the stacked DID (green circles) of \cite{stacked_did}. All four are estimated on a $\pm$5-year window relative to DCR opening. The $t = -5$ and $t = +5$ estimates represent pooled endpoint bins that absorb all periods beyond the window. The outcome is the log drug-related hospitalization rate per 100,000 inhabitants. The sample includes all staggered-treated Medstat regions within 5 km of a DCR that opened between 1998 and 2022 (40 treated regions) and all never-treated regions (608 control regions). The reference period is $t = -1$ (the year before DCR opening, normalized to zero). The treated group includes regions across all three distance rings (0--1 km, 1--3 km, and 3--5 km). Regions near DCRs that opened before 1998 (always-treated) are excluded. 95\% confidence intervals are based on standard errors clustered at the MedStat region level.}
\end{figure}

Table~\ref{tab:att_all_estimators} reports the overall ATT across four estimators. All point estimates are negative, ranging from $-0.149$ (stacked 
DiD) to $-0.438$ (Sun \& Abraham). The \cite{CALLAWAY2021200} doubly robust and \cite{SUN2021} estimates are significant at the 5\% level and nearly identical in magnitude ($-35.0\%$ and $-35.5\%$, respectively). This similarity may reflect their shared design for staggered adoption under treatment-effect heterogeneity. The synthetic DiD estimate of $-30.3\%$ which is the most precise estimate, is significant at the 10\% level and consistent in magnitude. The stacked DiD yields a smaller reduction of $-13.9\%$, significant at the 1\% level, the smaller magnitude reflects a more restrictive within-cohort comparison window following \cite{stacked_did}. The effect is consistently negative across all four estimators which suggests that the negative pooled effect is not driven by any particular identification approach. Figure~\ref{fig:event_study_4_estimators} plots the event-study paths for each method and confirms that pre-trends are flat across specifications. However, the pooled ATT masks substantial cohort-level heterogeneity in treatment effects. The \citet{CALLAWAY2021200} framework allows decomposing the pooled estimate into cohort-specific ATTs, which I explore next. 

% ── Urban + intermediate ──────────────────────────────────────

A potential concern with the full sample estimates is that the never treated regions include more rural MedStat regions which may have different health care and drug use dynamics that make them a poor comparison for the treated MedStat units which are predominantly located in urban cantons. To assess the sensitivity of the main results to this concern, I re estimate the \cite{CALLAWAY2021200} estimator restricting the control group to intermediate\footnote{The intermediate comparison was created given that these are regions that were in the border between urban and rural according to the official classification from \cite{bfs_communes_geom_2024}} and rural areas. I first restrict the analysis to urban and intermediate areas (413 control regions)  and then to urban (288 control regions).

\begin{figure}[H]
    \centering
    \caption{CS-DR event study — urban and intermediate regions, $\pm$5-year window}
    \includegraphics[width=0.85\textwidth]{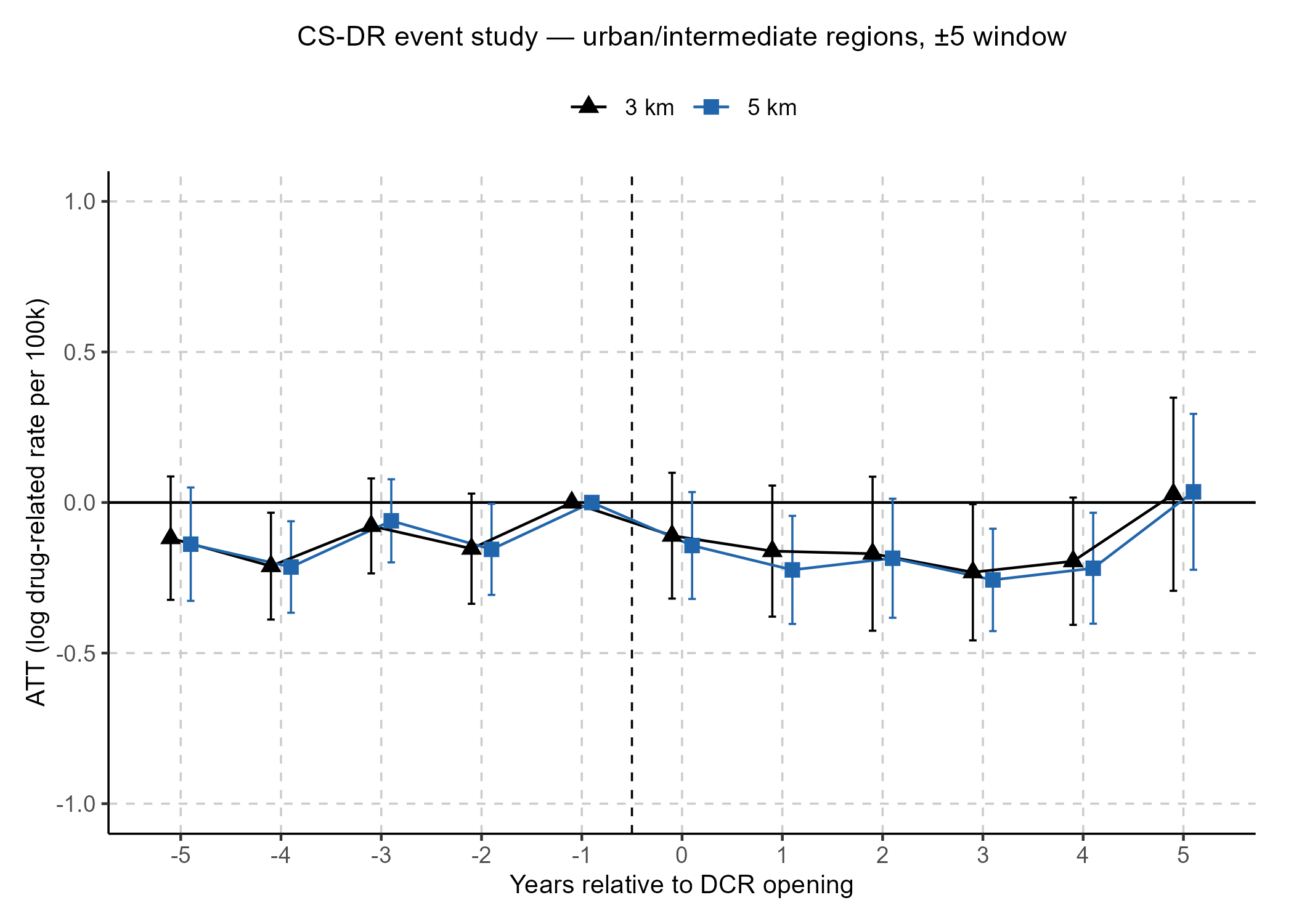}
    \label{fig:cs_eventstudy_urban_inter}
\end{figure}

Figure~\ref{fig:cs_eventstudy_urban_inter} shows consistently negative post-treatment coefficients across most periods, suggesting a reduction in drug-related hospitalizations after DCR opening. The pattern is broadly in line with the full-sample dynamics, with no evidence of pre-trends.

A second set of choices concerns the conditioning set $X$ and the two nuisance functions the doubly robust estimator requires: the propensity score and the outcome regression. The baseline specification leaves $X$ empty, so that parallel trends is imposed unconditionally among regions sharing the same buffer definition. With no covariates the propensity score reduces to a constant and the outcome regression to a cohort-specific mean, and the doubly robust, inverse probability weighted, and outcome regression estimators coincide numerically. The double robustness property therefore has content only once $X$ is non-empty. To assess it, I condition on baseline MedStat-region characteristics measured in 2000, prior to every DCR opening in the estimation window, and re-estimate the ATT three ways: using the propensity score alone, the outcome regression alone, and both together. I do this under two conditioning sets. 

\begin{figure}[H]
    \centering
    \caption{CS event study by nuisance-model specification - geographic controls}
    \includegraphics[width=0.85\textwidth]{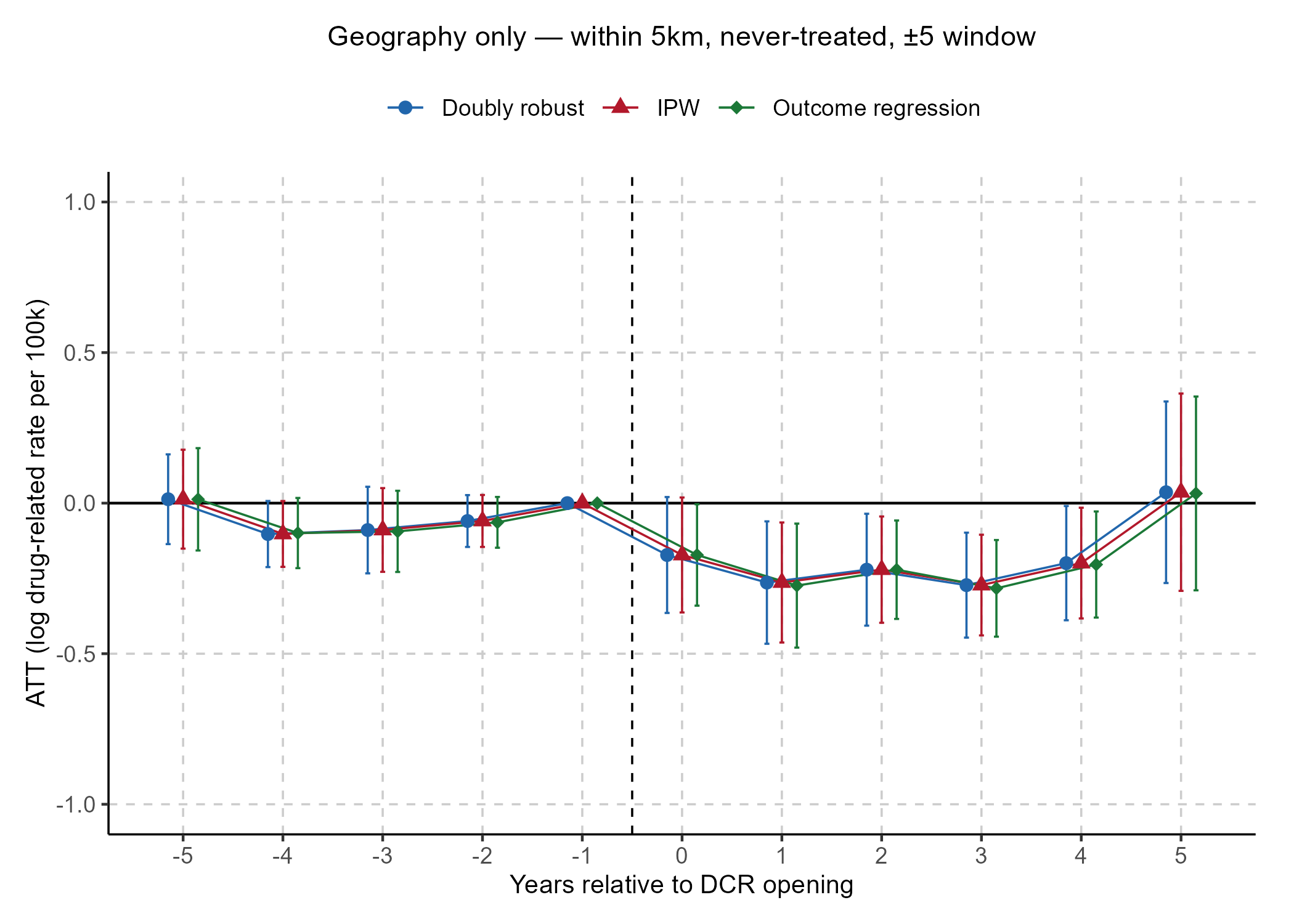}
    \label{fig:cs_nuisance_geo}
    \begin{minipage}{0.85\textwidth}
    \setstretch{0.9}
    \raggedright
    \footnotesize{\textit{Note:} The figure plots event-study coefficients from three specifications of the \citet{CALLAWAY2021200} estimator that differ only in how the nuisance functions are handled, all using the 5\,km buffer and never-treated control regions. The conditioning set is held fixed at urban typology (indicators for urban, intermediate, and rural regions) and log population, both measured in 2000. \textit{Doubly robust} (blue circles): both the propensity score and the outcome regression are estimated. \textit{IPW} (red triangles): the propensity score only, estimated via logit. \textit{Outcome regression} (green diamonds): the outcome regression only, estimated by OLS. 95\% confidence intervals based on standard errors clustered at the MedStat region level.}
    \end{minipage}
\end{figure}

\begin{figure}[H]
    \centering
    \caption{CS event study by nuisance-model specification - geographic and demographic controls}
    \includegraphics[width=0.85\textwidth]{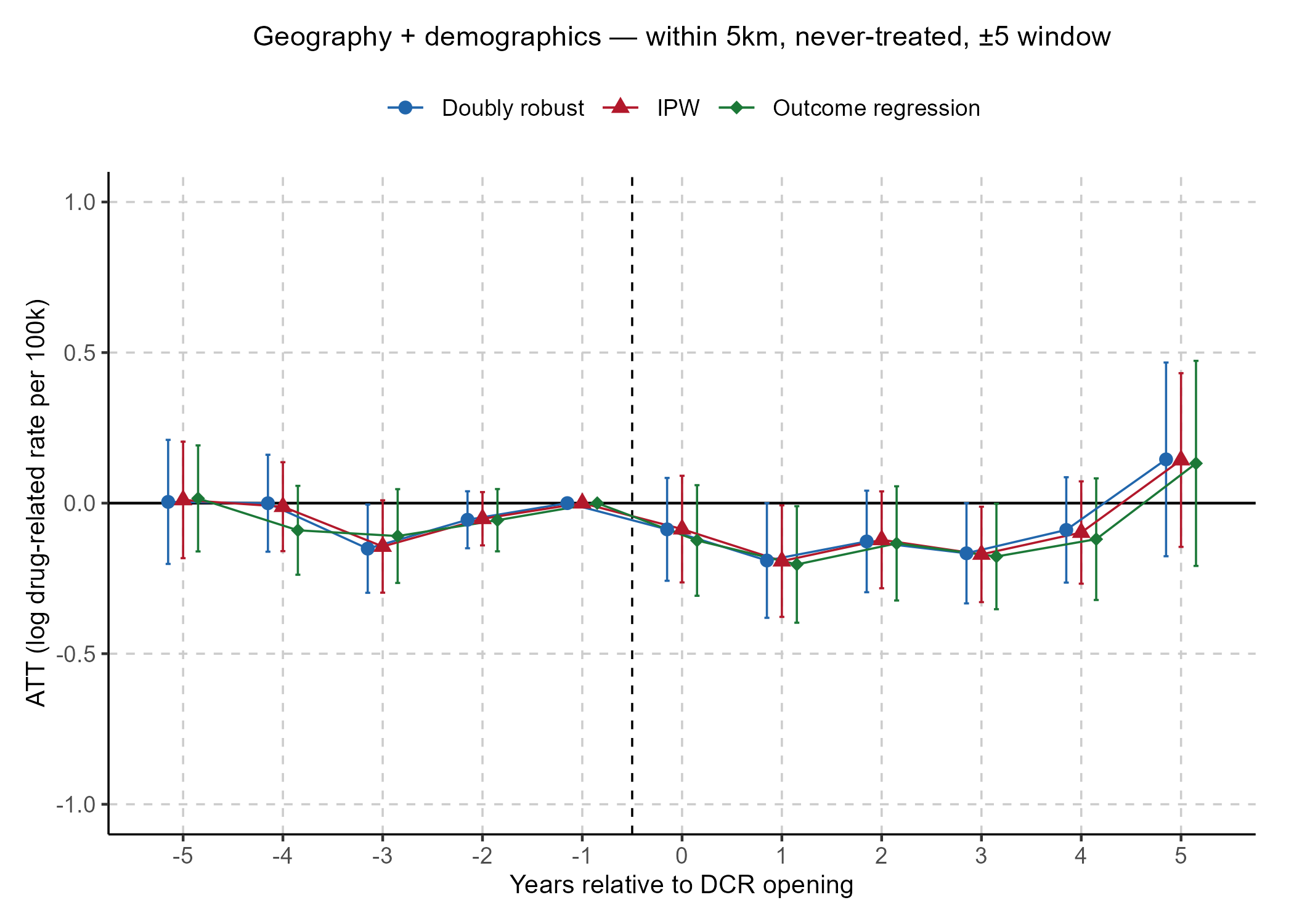}
    \label{fig:cs_nuisance_demo}
    \begin{minipage}{0.85\textwidth}
    \setstretch{0.9}
    \raggedright
    \footnotesize{\textit{Note:} As in Figure~\ref{fig:cs_nuisance_geo}, with the conditioning set extended to demographic structure measured in 2000: the population shares aged 15--24 and 25--44, the share of Swiss nationals, and the share of women. Urban typology and log population are retained. \textit{Doubly robust} (blue circles), \textit{IPW} (red triangles), and \textit{Outcome regression} (green diamonds) again differ only in which nuisance functions are estimated; all other specification choices are unchanged. 95\% confidence intervals based on standard errors clustered at the MedStat region level.}
    \end{minipage}
\end{figure}

Figure~\ref{fig:cs_nuisance_demo} extends the conditioning set to demographic composition. The three nuisance-model specifications again track one another closely, so the choice between them is not consequential in either case. Comparing across the two figures, the results appear to be more sensitive to the conditioning set than to the nuisance model. 

\input{tables/att_ps_dr_results}

Table \ref{tab:cs_est_method} shows that the overall ATT is not sensitive to how the nuisance functions are estimated. The point estimate ranges from $-0.431$ to $-0.512$, and remains negative and statistically significant whether the propensity score or the outcome regression is used alone.

As a robustness check, Figure~\ref{fig:event_study_4_estimators} replicates the main event study using four alternative estimators: TWFE, \citet{SUN2021}, \citet{CALLAWAY2021200}, and stacked difference-in-differences \citep{stacked_did, BAKER2022370}. Despite their methodological differences, pre-treatment coefficients are small and statistically indistinguishable from zero across all four specifications, and post-treatment coefficients are consistently negative, which suggests that the findings are not sensitive to the choice of estimator. The first and last periods have wider confidence intervals because they absorb all periods beyond the window, pooling multiple calendar years into a single coefficient. At $t = -5$, only cohorts with at least 5 pre treatment years contribute. At $t = +5$,, only cohorts with at least 5 post-treatment years contribute. So the effective sample size shrinks as we move away from the event time. The combination of cohort dropout and multi-period pooling at the tails mechanically inflates standard errors.

\end{document}

%% file: intro.tex
\section{Introduction} \label{sec:introduction} 

Illicit drug use remains a serious global public health concern. In 2022, approximately 292 million people used drugs, representing 5.6\% of the global population aged 15 to 64 years \citep{UNODC2024}. This number continues to rise. Of those who used drugs, around 22\% developed drug use disorders. However, more than 80\% do not receive any form of addiction treatment. Drug use is associated with adverse health risks and social consequences. Medically, harmful drug consumption can lead to the development of substance use disorders. In addition, the injection of drugs significantly increases the risk of contracting blood-borne infections such as HIV and hepatitis C. Socially, people who regularly consume drugs often face social marginalization, leading to increased exposure to precarious living conditions, unemployment, homelessness, and are more likely to participate in criminal activities \citep{UNODC2024}. 

In Switzerland in 2023, an estimated 32,000 individuals were living with hepatitis C, a disease that causes around 200 annual deaths \citep{swiss_hepatitis_c_2024}. In addition, 352 new HIV diagnoses were reported \citep{swiss_aids_federation_data_2024} and 192 people died from causes directly related to drug consumption \citep{obsan_drug_deaths}. Furthermore, emerging synthetic drugs, particularly opioids such as fentanyl, pose new challenges to public health due to their high potency that increases the risk of overdose \citep*{POTIER201448, UNODC2024}. The high prevalence and severe consequences of drug use disorders, combined with the treatment gap and scarce evidence on effective interventions, point to a need for further research.

In this paper, I study the effect of Drug Consumption Rooms (DCRs) on drug-related hospitalizations in Switzerland. DCRs are professional-supervised healthcare facilities where people with a drug history can consume their pre-obtained drugs under safer and more hygienic conditions \citep{EMCDDA2024}. DCRs are part of a broader set of harm reduction policies which aim to reduce the health and social consequences of drug use without requiring abstinence, as research has shown that policies focused solely on reducing drug supply are often ineffective \citep{war_on_drugs_2009, war_on_drugs_cunningham,DOBKIN201448}.  DCRs facilities offer clean injection equipment, immediate medical assistance in case of overdose, and access to other health and social services.  

DCRs often emerge in response to public health challenges associated with the use of illicit drugs and serve two main purposes. First, they aim to prevent risky behavior such as overdose, disease transmission through unsafe injecting, morbidity, mortality and provide access to treatment to hard-to-reach, high-risk populations. Second, DCRs seek to improve public order by dissuading drug users from injecting and consuming in public spaces, thus reducing public nuisance and crime \citep{EMCDDA2024}. Despite the expansion of DCRs, their effectiveness remains limited. Existing studies largely rely on correlations and on self reported outcomes. This paper contributes to closing this knowledge gap by investigating whether opening a DCR causally affects drug-related hospitalizations, using a natural experiment and granular hospital data. 

I focus on Switzerland, the first country in the world to introduce a DCRs in 1986 \citep{ZobelMaier2018}. For identification, I exploit the staggered openings of the DCRs across cantons in Switzerland. For estimation of causal effects, I use the staggered difference in difference framework of \cite{CALLAWAY2021200} and complement it with more recent difference in difference methods to assess the sensitivity of the results \citep{SUN2021, stacked_did, BAKER2022370, arkhangelsky2021synthetic}. The main data source of this work relies on individual level administrative records of all inpatient stays in hospitals in Switzerland from 1998 to 2022. The dataset is derived from hospital records containing detailed information on hospitalizations, such as the mode of admission, length of stay, diagnosis, treatment, and discharge. I use this dataset to measure the main outcome of interest, hospitalizations due to drug use, which can be further disaggregated into specific diagnosis categories such as hospitalizations due to mental and behavioral disorders, HIV, hepatitis and also allows to classify fatal and non-fatal hospitalizations. An advantage of studying the Swiss setting is that health care insurance is mandatory. Thus, hospital data provide a more representative coverage of the population than in other countries.

The results suggest that DCR openings are associated with meaningful reductions in drug-related hospitalization rates, driven in part by a shift from fatal to non-fatal overdose events. Effects on mental and behavioral disorder hospitalizations are negative and marginally significant, while the estimates for hepatitis decline gradually in the years following an opening which is consistent with a reduction in needle sharing. HIV effects are imprecise. Treatment effects are heterogeneous across cohorts, with recently opened facilities showing smaller and sometimes an increase in drug-related hospitalizations. 

This study offers three main contributions. First, it provides the first quasi-experimental evidence of the causal effects of DCR opening on hospitalizations due to drug-related events. Second, it expands the focus beyond mortality to include nonfatal overdoses and the incidence of infectious disease, which are outcomes that have received less attention but are central to public health and health care costs. Third, it documents a substitution from fatal to non-fatal drug-related hospital events following DCR openings, providing direct evidence on the mechanism through which supervised consumption reduces mortality. 

The paper is organized as follows. Section \ref{sec:literature} reviews literature on related harm reduction strategies and their effects on health and social outcomes. Section \ref{sec:Background} provides institutional context of drug use and harm reduction policies in Switzerland. It also describes the country’s geographical division and documents the spatial and temporal expansion of DCRs across Swiss cantons. Section \ref{sec:data} describes the sources of data, outcomes and covariates. Section \ref{sec:empirical_strategy} outlines the empirical strategy where identification and estimation are addressed. Section \ref{sec:result} presents the results for the aggregated and disaggregated results, heterogeneity analysis and mechanisms. Section \ref{sec:conclusion} concludes.

%% file: literature.tex
\section{Literature Review} \label{sec:literature}

The conception of DCRs dates back to the 1980s mostly in Europe, Australia and Canada in response to the HIV/AIDS epidemic among people who injected heroin. DCRs have remained one of the more controversial responses \citep{controversial_harmreduction}. Concerns persist that these facilities might create a moral hazard problem and encourage drug use, delay treatment entry, exacerbate local drug market problems \citep{packham2022syringe} or increase crime \citep{TAYLOR2021108397}.

Existing epidemiological literature suggests that DCRs are associated with a reduction in overdose deaths and morbidity \citep{MARSHALL20111429}, improve injection practices \citep{lalanne_2024}, decrease the transmission of blood borne viruses such as HIV and hepatitis C, and increase access to health services \citep{Wood2006Attendance, wood_2007, salmon_2010, POTIER201448, Kerr2017, Kennedy2017, Larson2019, Tran2021}. Evidence from Switzerland further indicates that harm reduction policies, including DCRs, prevented approximately 15,900 HIV infections and 5,400 AIDS deaths among people who inject drugs, using a mathematical model based on HIV cohort data \citep{marzel2018cumulative}. 

However, most of the evidence relies on descriptive methods, pre-post comparisons, or self-reported outcomes, and often lacks a proper control group \citep{Belackova2019, Caulkins2019, Levengood2021}. This is mostly explained by the context in which DCRs operate. First, the number of facilities is small and they are concentrated in a few cities. As a consequence, most of the evidence comes from a single opening at a single point in time. This allows for a before and after comparison, but not for a credible counterfactual. Second, anonymity is a condition for attendance in most facilities and it makes it impossible to link users to administrative records. In addition, there is no accurate sampling for people who consume drugs, so researchers rely on self reported measures. Finally, randomization is not an option, as access to a DCR cannot be withheld for research purposes \citep{Pardo2018}.

In contrast, the economics literature has so far focused on the role of Substance Use Disorder (SUD) treatment facilities, Syringe Exchange Programs (SEPs) or naloxone access laws. SUD facilities offer abstinence-oriented care, such as detoxification, rehabilitation, and counseling, which differ substantially from DCRs in purpose, target population, and expected outcomes. For instance, \citet{SWENSEN201513} shows that a 10\% increase in SUD treatment facilities reduces drug-related deaths by 2\%. Similarly, \citet{CORREDORWALDRON2022102579} find that the opening of a treatment facility in New Jersey reduces drug-related emergency room visits by 6.5\% while closures increase ER visits by 7.4\%. Additionally, \cite{BONDURANT2018124} study crime as an outcome and find that an increase in SUD facilities leads to a reduction in violent and financially motivated crimes. 

Another strand of the literature examines SEPs, which provide access to sterile syringes and allow safe disposal of used needles for injection drug users. \cite{packham2022syringe} studies SEP openings and drug-related health outcomes in the United States. The author finds an increase in opioid-related mortality, with greater effects in rural high-poverty counties. At the same time, SEPs do not appear to increase new HIV diagnoses and may even reduce them. However, the results of the study show that access to SEP also leads to higher rates of emergency room visits and patient stays due to drug-related complications.

Researchers have also examined naloxone access laws, which extend access to an opioid antagonist capable of reversing overdoses. \citet{jennifer_2022} examine the impact of naloxone access laws in the US and report that expanding access may not reduce opioid-related mortality due to potential risk compensation. However, this study was later reviewed by \citet{alexeev2025closer}, who identified problems with the data, timing of the policy and inference. The results of this more recent study show no evidence that access to naloxone increases admission to emergency rooms. Such mixed findings underscore the need for rigorous research to accurately assess the role of harm reduction strategies on public health policy.

The study most closely related to this paper is \citet{franco2024drug}, who examine the effects of DCRs in the Netherlands. Using neighborhood-level survey data, they find that DCR openings reduce self-reported drug use and drug-related crimes, with modest increases in property values in low-income areas. While their findings provide valuable evidence on community-level outcomes, the analysis relies on self-reported data and does not examine individual health outcomes. 

Overall, the current state of the literature emphasizes the potential of harm reduction interventions to shape drug-related outcomes such as health and crime. However, it also highlights that their effects vary across program design and context. Importantly, these findings cannot be generalized to DCRs, which not only provide clean equipment for drug users, but also allow consumption inside the facilities. In addition, existing research focuses on mortality, while other key outcomes associated with drug use, such as nonfatal hospitalizations and infectious diseases, remain understudied. In general, the causal evidence on the impact of DCRs remains scarce. This gap highlights the importance of this study in evaluating the effectiveness of DCRs as a harm reduction policy.

%% file: background.tex
\section{Background} \label{sec:Background}

\subsection{Drug Consumption in Switzerland}

In the 1980s, Switzerland faced a severe opioid crisis that led to an HIV/AIDS epidemic among people who use drugs, increased public disorder, drug-related crime, and mortality. Drug scenes developed, especially in larger cities such as Bern and Zurich. A considerable open drug scene began to emerge at the end of the 1980s in a park in the heart of Zurich. The two key contributing factors to the emergence of drug scenes were the increased drug availability and the emergence of youth movements, which led to the establishment of autonomous youth centers in larger Swiss cities \citep{hedrich2004european}. In response, the Swiss Federal Council implemented a comprehensive four pillar drug policy in February 1991. The four pillars include prevention, treatment, law enforcement, and harm reduction, which were significantly expanded \citep{hedrich2004european, CSETE201282}. DCRs are a strategy which belongs to the harm reduction pillar. 

\begin{figure}[H]
    \centering
    \caption{Drug-related deaths in Switzerland 
    \includegraphics[width=0.7\textwidth]{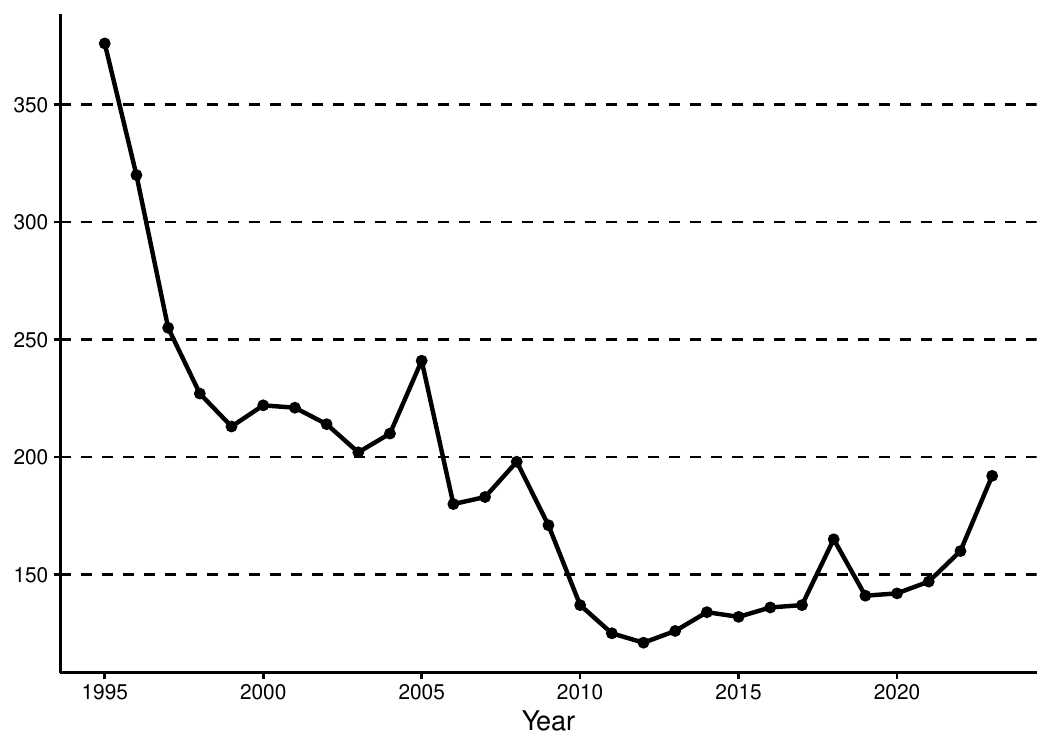}
    \\
    \vspace{0.5cm}
    \setstretch{0.9} % ↓ reduce line spacing in the note
    \raggedright         % ← left-align even inside \centering
    \footnotesize \textit{Note:} Annual number of drug-related deaths in Switzerland, 1995–2023. The sharp decline after the mid-1990s coincides with the introduction of Switzerland’s four-pillar drug policy.
    \\
    \footnotesize \textit{Source:} Federal Office of Public Health (FOPH) \& Swiss Health Observatory (Obsan)}.
    \label{fig:drug_deaths_total}
\end{figure}

Figure~\ref{fig:drug_deaths_total} illustrates the evolution of drug-related deaths in Switzerland since 1995, the earliest year for which consistent data are available \citep{obsan_drug_deaths}. In 1995, there were 376 drug-related deaths, followed by a steep decline, settling at lower levels in the 2000s. However, recent years show an upward trend in drug-related deaths, which highlights the continued relevance of harm reduction interventions to prevent such events.

Drug use in Switzerland is monitored mainly through the Swiss Monitoring System of Addiction and Noncommunicable Diseases (MonAM), coordinated with the Federal Office of Public Health (FOPH) and the Swiss Health Observatory (Obsan). Monitoring relies on national surveys and wastewater analysis which offers the advantage of not being based on self-reported behavior. A recent study by \cite{EawagDroMedArio2025} used wastewater-based epidemiology to measure population-level consumption of illicit drugs, pharmaceuticals, alcohol and tobacco in Switzerland. The study collected wastewater samples from 10 wastewater plants from 2021 to 2024, covering the five largest Swiss cities (Zurich, Geneva, Basel, Lausanne, and Bern) as well as smaller cities (Lugano, Chur, and Neuchâtel) and more rural catchment areas (Schwyz and Zuchwil-Solothurn), together representing about 23\% of the Swiss population. Wastewater-based measures suggest that Zurich, Bern, and Lausanne consistently exhibit drug consumption levels above the national mean for multiple substances. 

These measures can also be compared with other European countries. The European Union
Drugs Agency (EUDA) coordinates an annual wastewater campaign through the Sewage analysis CORe group (SCORE), which applies a common sampling protocol and a common quality control exercise across all participating cities. The 2025 wave covers 115 cities in the European Union and Türkiye, and additionally includes 12 Swiss sampling sites, which allows Swiss consumption to be placed within the European distribution rather than assessed only relative to the national mean \citep{EUDA2026}. The comparison indicates that Switzerland is an outlier for cocaine. The median load across Swiss sites reaches 734 mg per 1 000 inhabitants per day, compared with 386 in the rest of Europe, and Geneva and Zurich rank seventh and eighth among the 129 European cities covered. For cannabis, amphetamine, ketamine and MDMA, Swiss loads remain at or below the European median, while methamphetamine loads lie somewhat above the median but from a very low European base. Overall, these results suggest that drug consumption in Switzerland is moderate by European standards, but that cocaine use is considerably more prevalent than in most other European countries.

\subsection{Drug Consumption Rooms}

In Switzerland, the rollout of DCRs has been gradual and spatially uneven. From 1986 to 2022, Switzerland saw the establishment of 13 DCRs across 9 cantons, primarily in the main cities. The openings of DCRs have been concentrated in urban cantons with higher population densities and higher drug use, such as the above-mentioned Zurich, Bern, Vaud, and Geneva.  
 
Although the adoption of DCRs is decided at the cantonal level, in this study, I use the medical statistical region (MedStat) as the main geographical unit of analysis which is a smaller geographic unit designed to preserve patient anonymity \citep{bfs_medstat}. MedStat regions (706), compared to cantons (26), offer much finer spatial granularity, allowing a more precise treatment definition based on proximity to a DCR and improving the measurement of individual exposure. Figure \ref{fig:medstat_population} in the Appendix illustrates the MedStat regions and their population density\footnote{In total, there are 706 MedStat regions in Switzerland, each of which contains around 11,000 residents. The geographic size of MedStat regions varies substantially. The median MedStat covers about 30.5 km$^2$ while the smallest regions are less than 1 km$^2$, and the largest is nearly 1,000 km$^2$}.

DCRs provide easy access to health and social support for individuals who cannot or do not wish to stop using illegal substances. Users may consume self-obtained psychoactive drugs by injection, smoking, or sniffing in hygienic facilities supervised by trained staff. Beyond supervised consumption, DCRs offer a broad range of services. These include immediate medical assistance in case of overdose or other emergencies; distribution of sterile equipment and education on reducing transmission risks; basic medical care such as wound treatment and hepatitis C testing; psychosocial counseling and crisis interventions; and referral to external health, social, and legal services. Many centers also provide meals, drinks, and low-threshold employment opportunities, and they actively collaborate with local stakeholders, including police, neighbors, and peer organizations \citep{Infodrog2020}.

Access to DCRs in Switzerland is regulated by strict eligibility criteria. Users must be over 18 years old,  demonstrate a drug addiction, and may not be first-time users. Importantly, for this study, access is limited to residents of the canton or city where the facility is located. Regarding the user profile, DCR users are predominantly men (75\%), with an average age of 46. Cocaine is the most frequently used substance, followed by heroin, which are primarily consumed through smoking or injection\footnote{Substances are consumed by smoking (59\%), injection (21\%), or nasal inhalation (20\%).} \citep{Infodrog2020, Infodrog2025}.

Across Switzerland, there are more than 5,400 registered DCR users. All facilities combined record around 1,200 visits per day and more than 3,000 consumptions each day \citep{Infodrog2020}. Each year, around 2 million sterile syringes and 15,000 crack pipes are provided. The use of the DCR varies by city. In Bern, more than 2,000 individuals are registered at the local DCR, with around 600 active clients and 80 to 100 daily users \citep{ContactPointBern}. Zurich operates three DCRs, collectively serving up to 300 people daily, including 160 to 260 frequent users who inject drugs \citep{csak2022harmreduction}. The DCR in Geneva recorded 47,360 visits in 2023  \citep{RapportActivites2023}. However, comprehensive data on capacity and capacity constraints across all DCRs remains unavailable \citep{Infodrog2025}.

\subsection{Treatment Rollout}

Data on the locations and opening years of DCRs are compiled primarily from the \href{https://suchtindex.infodrog.ch/#/}{Swiss National Addiction Index}. The index provides a nationwide directory of harm reduction and addiction services. Additional information is gathered from policy documents, historical archives, and direct communication with facility administrators. To track the evolution of DCRs over time, I record the year of adoption, canton, city, and MedStat where each facility is located. Table~\ref{tab:dcr_openings} presents this information.

\input{tables/adoption_DCR}

As shown in Table \ref{tab:dcr_openings}, the adoption of DCRs in Switzerland has been gradual. The first facility opened in Bern in 1986, making it the first DCR worldwide. Subsequent openings concentrated in urban cantons such as Zurich, and Geneva, with the most recent facility opening in Lausanne in 2018. 

\input{tables/treated_cohorts}

Table~\ref{tab:treatment_cohorts_valid} illustrates the specific treatment cohorts used in the empirical design. I focus on five DCR openings within the period for which hospitalization data is available (1998-2022). As a result, I examine facilities opened in the cities of Biel and Geneva in 2001, Lucerne in 2008\footnote{The Lucerne DCR also admits residents of the neighbouring cantons of Nidwalden, Obwalden, Schwyz, Uri and Zug.}, Basel in 2013\footnote{The treatment cohort for Basel also includes areas of Basel-Land as the DCRs are shared among the cantons of Basel-Stadt and Basel-Land.}, and Lausanne located in the canton of Vaud, in 2018. While the treatment cohorts capture the timing of DCR openings, treatment definition also requires defining the geographic treatment coverage exposure of each DCR. 

\begin{figure}[H]
    \centering
    \caption{Treated Medical Statistical regions with Buffer Zone} 
    \includegraphics[width=0.7\textwidth]{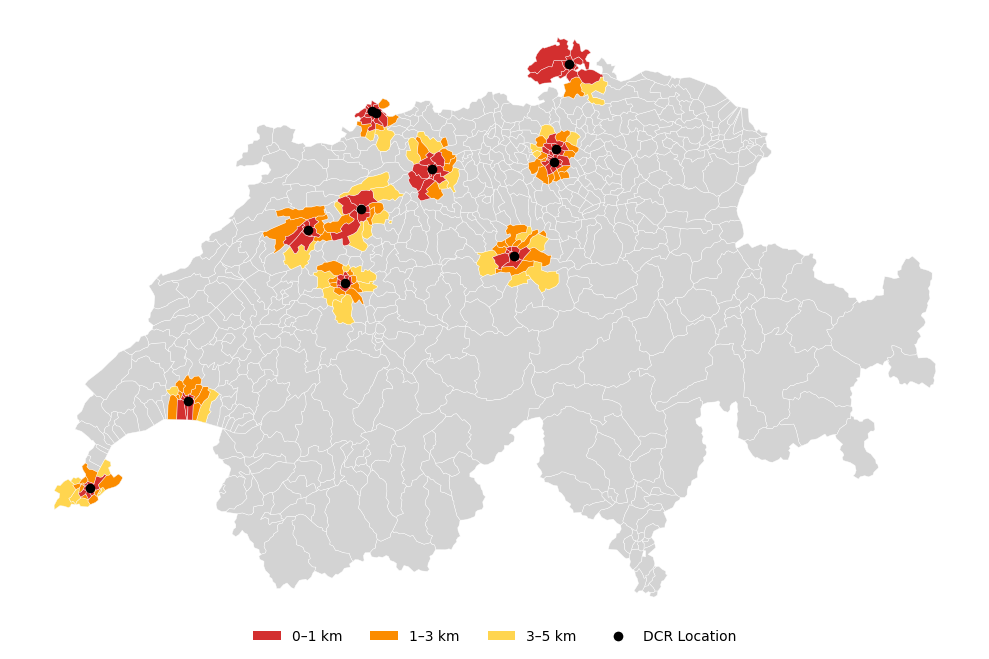}
    \label{fig:map_buffer_treated_medstat}
    \vspace{0.5cm}
    \\
    \setstretch{0.9} % ↓ reduce line spacing in the note
    \raggedright 
    \footnotesize{\textit{Note:} Red areas indicate the 1 km buffer, orange areas the 3 km buffer, and yellow areas the 5 km buffer around the DCR. Grey areas represent never treated regions.}
\end{figure}

Access to treatment is shaped by distance. I assign treatment by proximity to a DCR. I first locate each DCR in the MedStat regions that contains its address, I call this the centroid region and it serves as a reference point for distance. Second, I draw a circular buffer of 1, 3 and 5 km around the centroid. Third, I classify every MedStat region that intersects a buffer entirely as treated at the radius. The buffers are nested, but I also do the analysis with non nested buffers. Figure \ref{fig:map_buffer_treated_medstat} shows the three sets of radius for each DCR, in red, orange and yellow\footnote{Figure \ref{fig:medstat_population} displays centroid MedStat regions within a 1 km buffer from each DCR.}. The assignment is binary at the region level, so a region counts as treated whether the buffer covers all of it or a part of it. This is the finest assignment the data allow, since the MedStat region of residence is the only location the hospital records contain\footnote{The Federal Statistical Office revised the MedStat regions in 2008, and the analysis uses the post-2008 definition throughout. Crucially, these boundary revisions and municipal mergers did not alter regional treatment assignment across the distance buffers.}.

Two features of the geography limit the miss-classification of treated and control areas. First, the MedStat regions are defined to hold comparable populations, so the larger the area, the smaller their density. In the cities where every DCR is located, the regions cover a few square kilometers, whereas regions larger than the buffers exist mostly in rural areas. Second, treatment follows the eligibility at the border. In the cases where the buffer crosses a cantonal border, as in Basel, the foreign country is not treated, because access is restricted to residents of the host canton, and residents of neighboring countries appear neither in the facility nor in the data. Basel is a special case, under a cantonal agreement, residents of Basel-Landschaft are entitled to use the DCR in Basel-Stadt.

The buffer distances are deliberately short. Previous research has shown that when patients must travel more than four miles, approximately 6 kilometers to access care for substance use disorders, the likelihood of treatment initiation or retention decreases significantly \citep{beardsley2003distance, amiri2018increased, CORREDORWALDRON2022102579}. Switzerland’s excellent public transport system could justify defining larger catchment areas. Yet, the compact geography and dense urban structure of its cities suggest that smaller buffers are more appropriate. In many cases, a 6 km radius would encompass almost the entire urban core, leaving little variation in treatment status and raising the risk of misclassifying exposure. Anecdotal reports from Zurich DCR staff suggest that most users live nearby and do not travel long distances to access the facility. For this reason, I focus on 1 km, 3 km, and 5 km thresholds around the DCR. In addition, these thresholds also serve as robustness checks to test the sensitivity of the results to alternative definitions of treatment proximity.

%% file: tables/adoption_DCR.tex
\begin{table}[H]
\centering
\caption{Adoption of Drug Consumption Rooms by Year, Canton, and MedStat Region}
\label{tab:dcr_openings}

\begin{threeparttable}
\begin{tabular}{lllll}
\toprule
\textbf{Year} & \textbf{Canton} & \textbf{City} & \textbf{Acronym} & \textbf{MedStat} \\
\midrule
1986 & Bern         & Bern         & BE & BE02 \\
%1991 & Basel-Stadt  & Basel        & BS & BS01 \\
1992 & Zurich       & Zurich       & ZH & ZH01 \\
%1992 & Luzern       & Luzern       & LU & LU03 \\
1993 & Solothurn    & Solothurn    & SO & SO19 \\
1994 & Schaffhausen & Schaffhausen & SH & SH01 \\
1995 & Solothurn    & Olten        & SO & SO07 \\
2001 & Bern         & Biel         & BE & BE32 \\
2001 & Geneva       & Geneva       & GE & GE01 \\
2008 & Lucerne      & Lucerne      & LU & LU03 \\
2013 & Basel-Stadt   & Basel        & BS & BS02 \\
2018 & Vaud         & Lausanne     & VD & VD03 \\
\bottomrule
\end{tabular}

\begin{tablenotes}[flushleft]
\setstretch{0.9}
\footnotesize
\item \textit{Notes:} The table lists the first year a Drug Consumption Room (DCR) opened in each city during the study period, along with the corresponding canton and central MedStat region code used in the analysis.
\end{tablenotes}

\end{threeparttable}
\end{table}

%% file: tables/treated_cohorts.tex
\begin{table}[H]
\centering
\caption{Treatment Cohorts Based on Year of DCR Adoption}
\label{tab:treatment_cohorts_valid}

\begin{threeparttable}
\begin{tabular}{ll}
\toprule
\textbf{Year of Adoption} & \textbf{Treated MedStat Regions (City)} \\
\midrule
2001 & BE32 (Biel) \\
2001 & GE01 (Geneva) \\
2008 & LU03 (Lucerne) \\
2013 & BS02 (Basel) \\
2018 & VD03 (Lausanne) \\
\bottomrule
\end{tabular}

\begin{tablenotes}[flushleft]
\setstretch{0.9}
\footnotesize\item \textit{Notes:} “Year of adoption” is the first year a DCR opened in the listed city during the study window. Regions that adopted before the window are classified as always treated and are not shown. The second column denotes the central MedStat region closest to the DCR; the city is given in parentheses.
\end{tablenotes}

\end{threeparttable}
\end{table}

%% file: data.tex
\section{Data} \label{sec:data}

This study uses data from the Hospital Medical Statistics dataset obtained from the Swiss \cite{OFS2020_hospitalstats}. This dataset contains individual-level data on all inpatient stays in clinics, hospitals and birth centres in Switzerland from 1998 to 2022\footnote{A limitation of the dataset is that it does not include outpatient cases, which prevents the analysis of care provided outside hospitals.}. The full dataset contains more than 34 million recorded hospitalizations over the period of analysis. 
This corresponds to an average of approximately 1.38 million hospitalizations per year. Annual hospitalizations vary by roughly 170,000 to 180,000 cases. Figure \ref{fig:hosp_over_time} shows the evolution of total hospitalizations over time and by age \ref{fig:hosp_by_age}\footnote{Additional descriptive figures are reported in Appendix \ref{sec:appendix}}.  

The dataset provides individual-level information on patient demographics. These include MedStat region of residence, age in 5-year bands (rather than as a  continuous measure), gender, and nationality. Admission details cover month of entry, stay before admission, mode of admission, and referral decision. Clinical information includes primary and additional diagnoses as well as main and supplementary treatments. For the hospital stay, the dataset reports length of stay, days until the next hospitalization, time from entry to main treatment, and ICU use. Discharge outcomes include the discharge decision, subsequent stay, and post-discharge care. 

An overview of the available variables is reported in Appendix Table~\ref{tab:variables}, with their coding and distribution reported in Tables~\ref{tab:stay_before_adm_full}-\ref{tab:post_discharge_care}. The tables trace the care pathway of drug-related hospitalizations. Most patients are admitted from home (72.9\%), and admissions are split almost evenly between emergency entry (48.5\%) and scheduled admission (46.8\%). Care is almost exclusively inpatient (96.1\%) and financed through mandatory health insurance (94.7\%), consistent with the hospital statistics covering the resident population rather than an insured subgroup. Psychiatry and psychotherapy account for the largest share of cases (60.3\%), followed by internal medicine (24.1\%), reflecting the diagnostic composition of drug-related admissions. Most stays end in discharge by the treating physician (82.6\%), although 10.5\% end in discharge against medical advice. Patients typically return home (67.0\%) and continue in outpatient treatment (59.7\%). 

To evaluate the representativeness of the hospital data, I compare drug-related outcomes recorded in the hospital statistics dataset with official statistics published by the Federal Statistical Office and the Swiss Health Observatory \citep{fso2021_hiv, obsan_drug_deaths}. While the official figures are generally higher in level, both sources display very similar trends over time. This suggests that the hospital data, despite under counting outpatient cases, provide a representative measure of the evolution of drug-related events. Detailed comparisons and figures are provided in Appendix~\ref{app:validation}.

\subsection{Outcomes}

The outcomes of interest are derived from hospital records. A hospitalization is classified as drug-related if the primary diagnosis or any of the first three secondary diagnoses correspond to a drug-related condition. The primary diagnosis is defined as the condition that most justifies the prescribed treatment. If no diagnosis is made, the main condition corresponds to the symptom or disorder with the greatest impact on the patient’s health \citep{ofs2020variables}. 

The diagnosis coding follows the guidelines of the Swiss Federal Statistical Office which are based on ICD-10-GM codes (International Classification of Diseases, Tenth Revision, German Modification), a standard system used by physicians to classify and code all diagnoses, symptoms and procedures for claims processing \citep{bfarm2025icd10, ofs2020variables}. The granularity of ICD-10 codes allows me to classify hospitalizations into detailed subcategories. I group admissions into five categories: (i) drug-related mental and behavioral disorders, (ii) hepatitis, (iii) HIV, (iv) conditions due to narcotics, and (v) accidental drug poisonings\footnote{The appendix Table \ref{tab:icd10_codes} reports the full list of ICD-10 codes used in the analysis.}

\begin{figure}[H]
    \centering
    \caption{Drug-related Hospitalizations} 
    \includegraphics[width=0.85\textwidth]{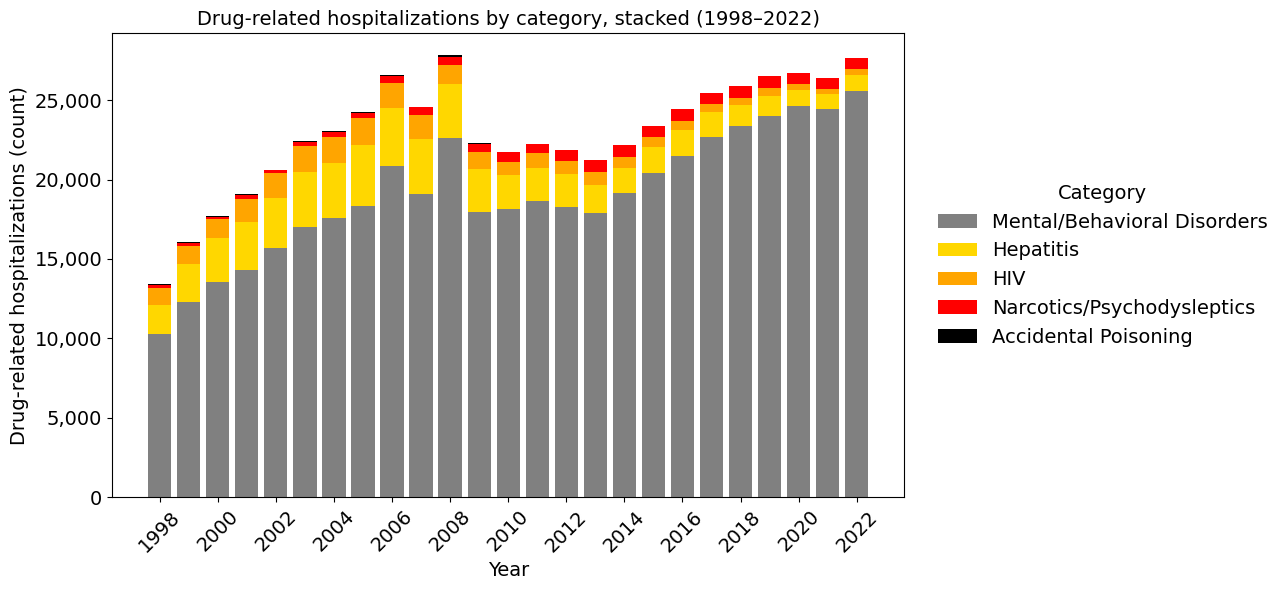}
    \label{fig:stacked_all_drug_related_barplot}
    \vspace{0.5cm}
    \\
    \setstretch{0.9} % ↓ reduce line spacing in the note
    \raggedright
    \footnotesize{\textit{Note:} The figure shows the annual number of drug-related hospitalizations in Switzerland from 1998 to 2022, decomposed into outcome categories. Each hospitalization is assigned to a single category based on the principal diagnosis and the first three secondary diagnoses. 
    \textit{Source:} Swiss Federal Statistical Office (FSO), Hospital Medical Statistics.}
\end{figure}

Figure \ref{fig:stacked_all_drug_related_barplot} presents drug-related hospitalizations from 1998 to 2022 and decomposes the total into the five outcome categories. It appears that the overall number of drug-related hospitalizations increases over time, with breaks around 2006–2008 and again in 2009. The stacked decomposition also shows that there have been shifts in both the level and the composition of drug-related hospitalizations over the period of analysis. Specifically, hospitalizations for mental and behavioral disorders due to psychoactive substance use account for the largest share of drug-related hospitalizations throughout the sample. This category rises steadily over time and drives most of the long-run growth in the aggregate series. Hepatitis and HIV hospitalizations were more frequent up until the mid-2000s and have declined over time, remaining comparatively small in recent years, especially HIV related hospitalizations. Acute outcomes such as accidental poisonings and diagnoses related to narcotics/psychodysleptics represent a small fraction of cases in every year and do not appear to drive the aggregate pattern. In general, Figure \ref{fig:stacked_all_drug_related_barplot}  highlights an increase in substance-related mental and behavioral disorders while infection-related hospitalizations have been decreasing over time. 

Table \ref{tab:desc_stats} reports descriptive statistics for the outcome measures. At the MedStat level, the average drug-related hospitalization rate is 301 per 100,000 inhabitants per year (median 185), with considerable heterogeneity across regions. In absolute terms, the typical MedStat region records approximately 30 drug-related hospitalizations per year, of which around 26 are attributable to mental and behavioral disorders, 4 to hepatitis, 2 to HIV, and fewer than 1 to narcotics-related poisonings. At the canton level, the average canton records approximately 823 drug-related hospitalizations per year (median 574) with hepatitis contributing around 97 cases (median 63) and HIV around 40 (median 19). Figures \ref{fig:hosp_over_time}-\ref{fig:narc_hosp_over_time} show that these categories follow different trajectories. Drug-related hospitalizations roughly double over the period, driven almost entirely by mental and behavioral disorders, which rise from about 10,200 to 25,600 cases. Hepatitis and HIV instead follow an inverted U, peaking in the mid-2000s at roughly 3,900 and 1,700 cases and declining to about 1,000 and 400 by 2022. Narcotics and psychodysleptic poisonings move in the opposite direction, roughly four times from about 190 cases in 1998 to 730 in 2022. These opposing trends motivate the disaggregated outcome analysis.

Two institutional changes contribute to the breaks visible in Figure \ref{fig:stacked_all_drug_related_barplot}. First, Switzerland transitioned from per-diem reimbursement to diagnosis-related group payment (SwissDRG). Parliament adopted the reform in 2007, and hospitals implemented it nationwide by 2012, several cantons adopted it earlier \citep{kutz2019association, boes2021assessment, swissdrg2023outline}. This change may have increased incentives to code diagnoses more completely and altered the case mix. Second, the Hospital Medical Statistics were revised in 2009 to align with SwissDRG. The revision expanded the number of reportable secondary diagnoses (up to 50) and procedures (up to 100), increasing the probability that comorbidities appear in the record even if the underlying morbidity is unchanged \citep{ofs2020variables_interface}. To limit sensitivity to reporting changes, I construct the outcomes using only the main diagnosis and the first three secondary diagnoses. This restriction keeps the information set comparable over time as it does not rely on the additional diagnosis introduced after 2009.  

\subsection{Treatment}

A key variable for treatment definition is the patient's MedStat of residence\footnote{In 2008, there was a change in postal codes where 101 new MedStat regions were created, likely due to splitting existing regions. I harmonize MedStat regions using an official crosswalk linking the pre-2008 and post-2008 classifications via postal codes. For MedStat regions that disappear after the reform, I assign observations to the modern MedStat region that shares the largest number of postal codes with the original region. Regions that remain stable are kept unchanged, which produces a consistent set of MedStat regions across the entire period of analysis.}, which I use to classify individuals as treated or control. The dataset also records the location of the hospital where the patient received treatment. Comparing these two locations, I can identify whether the patient was treated in a canton different from their place of residence. Table \ref{tab:same_canton_by_year} shows that 81\% of all hospitalizations occur in the patient’s home canton, indicating that cross-canton treatment is limited and that residence-based treatment assignment is unlikely to systematically misclassify individuals. The remaining of the cross canton hospitalizations are heavily influenced by regional healthcare infrastructure constraints rather than endogenous patient sorting. Specifically, in small cantons which lack tertiary care or specialized addiction units, care outside of the canton of residence is an institutional necessity driven by inter-cantonal hospital planning. These cases represent a small share of the total, making up 6.8\% of all out of canton hospitalizations and none of the six cantons hosts a DCR, so their residents are control observations. To more accurately define treatment exposure, I construct three distance-based measures using distance buffers of  1 km, 3 km and 5 km around the centroid of the MedStat region where the DCR is located, as illustrated in Figure \ref{fig:map_buffer_treated_medstat}. 

\subsection{Sample Selection}

\begin{figure}[H]
    \centering
    \caption{Drug-related hospitalizations as a share of all hospitalizations, by age group (1998--2022)} 
    \includegraphics[width=0.85\textwidth]{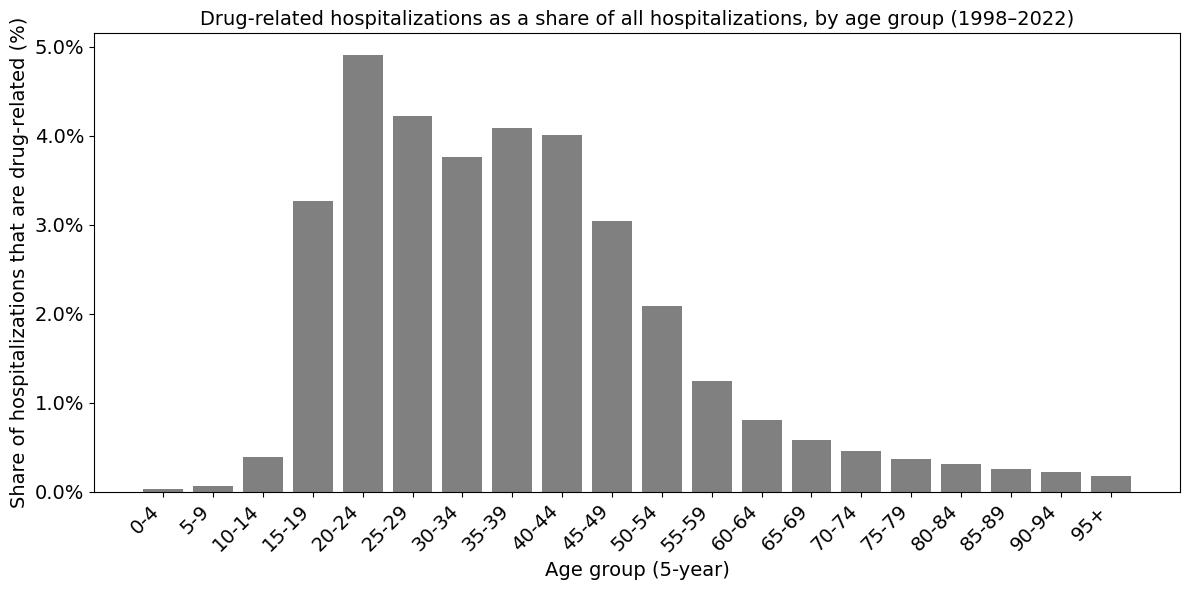}
    \label{fig:drug_related_by_age_share}
    \vspace{0.3cm}
    \\
    \setstretch{0.9} % ↓ reduce line spacing in the note
    \raggedright
    \footnotesize{\textit{Note:} The figure reports, for each 5-year age group, the share of hospitalizations classified as drug-related over the period 1998--2022. Drug-related hospitalizations are identified using ICD codes in the principal diagnosis and the first three secondary diagnoses (see text for the classification).
    \textit{Source:} Swiss Federal Statistical Office (FSO), Hospital Medical Statistics.}
\end{figure}

Of the roughly 34 million recorded hospitalizations available from the Hospital Statistics, as a first step, I exclude records with missing information on the main diagnosis, since diagnosis codes are essential to measure the outcomes. This step removes 737,086 observations (2.14\%), leaving 33,681,628 hospitalizations. The sample is further restricted to hospitalizations among individuals aged 15 to 74 years of age which removes 31.99\% of the sample, leaving 22,460,341 hospitalizations. %This age restriction excludes an additional 10,774,555 observations, corresponding to 31.99\% of the complete sample.  
The motivation behind the age restriction is illustrated in Figure \ref{fig:drug_related_by_age_share} which shows that drug-related hospitalizations among young and middle-aged adults represent the largest share of admissions, peaking in the early twenties. The rate remains relatively stable until the 40-44 years cutoff, followed by a steady decrease for older cohorts. Consistent with this pattern, \citet{Infodrog2020} reports that DCR users are predominantly male and have an average age of 46, which falls within the selected age range.

Moreover, I further restrict the sample to hospitalizations with a valid Swiss MedStat residence code. A small share of records report non-standard residence codes which often come from individuals with a residence abroad. Excluding these observations removes 744,638 hospitalizations (3.32\%) from the age-restricted sample, leaving 21,715,703 hospitalizations. I use the information on the locations and opening years of the DCRs to define indicators for whether a patient's MedStat region of residence falls within 1 km, 3 km and 5 km buffers of a DCR centroid. Table \ref{tab:balance_pre} reports pre-treatment means for key covariates separately for treated and control individuals, along with standardized differences to assess baseline balance. 

\subsection{Balance}

\begin{table}[H] 
\input{tables/balance_pre2001}
\end{table}

Table \ref{tab:balance_pre} reports pre-treatment covariate balance at the level of the individual hospitalization. The variables come from the hospital discharge records which contain hospitalization characteristics of residents in MedStat regions that are eventually treated and never treated. The treated and never treated groups are well balanced on most characteristics. The only notable difference is inpatient status, as hospitalizations among residents in treated MedStat regions are more likely to be inpatient than those in control regions. This pattern can be explained by differences in health care utilization between urban and rural areas as urban MedStat regions have denser hospital networks and ambulatory clinics which are designed for complex procedures that require a inpatient admission \citet{reich2013place, gygli2021regional, giezendanner2020ecology}. Additionally, most hospitalizations are among Swiss citizens, consistent with their larger population share. The sample contains slightly more female than male patients. The average length of stay is around 24 nights in control regions and 27 in treated regions. However, the distribution is highly right-skewed due to a small number of long hospitalizations. The median length of stay is 4 nights and the most frequent stay is 1 night. The mean time spent in the ICU is around 2 hours. About 30\% of admissions are emergencies, and most observations are inpatient stays, which is consistent with the coverage of the hospital medical statistics. Age shares are also similar across groups and stable, especially between ages 25 and 74\footnote{Table \ref{tab:balance_full} in Appendix \ref{app:balance} extends this analysis to the full sample across the entire study period. Also, Tables \ref{tab:balance_drug_pre} and \ref{tab:balance_drug} report covariate balance for the subsample of drug-related hospitalizations only. Overall, hospitalization characteristics of treated and control groups are balanced.}.

\begin{table}[H] 
\input{tables/balance_outcome_pre}
\end{table}

Table \ref{tab:pre_outcome_balance} reports the pretreatment outcome means and standardized difference between treated and control hospitalized patients. The drug-related outcome indicator equals one if a hospitalization is associated with any of the subcategories: mental and behavioral disorders, hepatitis, HIV, narcotics/psychodysleptics and accidental poisoning. Pre-treatment differences in mean outcomes are small. Although balance in levels is not required for difference-in-differences identification, these results suggest no large baseline disparities between the two groups. 

While the balance checks above are conducted at the individual level, all subsequent analysis is done at the MedStat region year level. I aggregate individual-level hospitalizations to annual counts per MedStat region for each of the outcome categories, and combine these with annual MedStat population data to compute hospitalization shares and population-based rates per 100,000 residents. I aggregate for three main reasons. First, treatment variation occurs strictly at the regional level, aggregating to MedStat region-years aligns the unit of analysis with the level of policy intervention. Second, severe drug-related hospitalizations are statistically rare events, with observed individuals appearing once per year on average. Constructing an individual-level panel would result in a very sparse dataset dominated by zeros. Third, evaluating public health impact often requires population-based incidence rates using municipal census data as denominators, analyzing only hospitalized individuals at the micro-level would condition the sample on being admitted.

I next assess pre-treatment balance in region-level characteristics. I obtained data on population density at the cantonal and MedStat region level from the \cite{BFS2025Population}. The data serve as the basis for population weights throughout the empirical analysis\footnote{Since no official population data exist at the MedStat region level, I construct population estimates from municipal-level data spanning 1998–2022, mapping municipalities to their corresponding MedStat regions via postal codes. Where a municipality encompasses multiple postal codes, population figures are divided evenly across them.}. Population figures at the MedStat region level for 2022 are displayed in Figure \ref{fig:medstat_population} in Appendix \ref{app:figures}.  Table \ref{tab:balance_medstat_2000} reports baseline demographic and socioeconomic characteristics for treated and control MedStat regions. The unit of observation is therefore a region, in contrast to Table \ref{tab:balance_pre}, where it is a hospitalization. The characteristics are measured in 2000 using the federal census and include urbanization, educational attainment, socioeconomic profile, unemployment, gender composition and age structure. These are variables which capture important demographic and socioeconomic conditions that can influence both drug-related outcomes and the opening of a DCR.

Mean differences between treated and control regions are generally modest, though the two groups diverge most clearly along the urban–rural dimension. Treated regions are more densely populated and predominantly urban, while control regions are more often rural. Treated regions also exhibit a lower share of Swiss nationals, marginally higher rates of tertiary education and professional employment, and somewhat higher unemployment. Differences in age composition are small across groups. In general, the balance table suggests that the closest DCR regions are systematically more urban than the comparison areas. This motivates the inclusion of urbanicity controls and the use of population weights in the analysis that follows.

\begin{table}[H] 
\input{tables/balance_medstat_pre}
\end{table}

These descriptive comparisons serve two purposes. First, to assess the degree of similarity between treatment and control areas, and second, to verify that the analysis adequately accounts for relevant regional heterogeneity. The individual-level hospitalization records allow for precise measurement of drug-related admissions across diagnostic categories, while the regional covariates help address heterogeneity in both treatment adoption and health outcomes across areas. The data sources available in this study, the universe of hospitalizations, as well as demographic and socioeconomic characteristics at the MedStat-level form an empirical foundation to identify the causal effect of DCRs on health outcomes in Switzerland. 

%% file: tables/balance_pre2001.tex
\begin{table}[H]
\centering
\caption{Pre-treatment covariate balance: Means}
\label{tab:balance_pre}

\begin{threeparttable}
\begin{tabular}{lrrr}
\toprule
\textbf{Variable} & \textbf{Treated} & \textbf{Control} & \textbf{Std Diff} \\
\midrule
Swiss & 0.83 & 0.83 & 1.04 \\
Male & 0.43 & 0.44 & 2.03 \\
Length of stay & 26.81 & 23.60 & 0.80 \\
ICU hours & 2.09 & 2.40 & 0.90 \\
Any ICU stay & 0.03 & 0.04 & 3.33 \\
Emergency admission & 0.28 & 0.30 & 4.76 \\
Inpatient & 0.94 & 0.87 & 23.82 \\
% ---- RECOMPUTE: means aggregated from rounded bin shares (sum to 1.02);
% ---- standardized differences cannot be aggregated from the bin values.
Age 15--24 & 0.08 & 0.10 & 1.16 \\
Age 25--44 & 0.34 & 0.36 & 2.75 \\
Age 45--64 & 0.36 & 0.35 & 1.83 \\
Age 65--74 & 0.21 & 0.21 & 1.72 \\
\midrule
N & 184,346 & 1,556,496 &  \\
\bottomrule
\end{tabular}

\begin{tablenotes}[flushleft]
\setstretch{0.9}
\footnotesize
\item \textit{Notes:} The table reports covariate means in the pre-treatment period $(t<2001)$ for hospitalizations of residents in MedStat regions within 0 to 1km of a DCR centroid (treated) and residents outside 5km (control). Swiss, Male, Any ICU stay, Emergency admission, and Inpatient Stay are indicator variables (means can be read as proportions). Length of stay (SwissDRG) is measured in days. ICU hours is measured in hours (0 for no ICU stay). Age is recorded in five-year classes in the MedStat discharge data and is grouped here to match the categories available in the census aggregates used in Table \ref{tab:balance_medstat_2000}. The sample is restricted to ages 15 to 74, so the highest group is 65 to 74. The standardized difference in means is $\displaystyle \frac{|\bar X_T-\bar X_C|}{\sqrt{(s_T^2+s_C^2)/2}} \times 100$,
where $T$ denotes treated and $C$ control groups. $N$ is the number of hospitalization records used to compute the means. Following \cite{rosenbaum_1985}, a standardized difference higher than 20 is considered large.
\end{tablenotes}

\end{threeparttable}
\end{table}

%% file: tables/balance_outcome_pre.tex
\begin{table}[H]
\centering
\caption{Pre-treatment drug-related outcome shares among all hospitalizations: Means}
\label{tab:pre_outcome_balance}

\begin{threeparttable}
\begin{tabular}{lrrr}
\toprule
\textbf{Outcome} & \textbf{Treated} & \textbf{Control} & \textbf{Std Diff} \\
\midrule
Drug-related (any) & 2.59 & 2.27 & 2.07 \\
Mental/Behavioral disorders & 2.19 & 1.80 & 2.79 \\
Hepatitis & 0.34 & 0.45 & 1.76 \\
HIV & 0.15 & 0.19 & 0.94 \\
Narcotics/Psychodysleptics & 0.03 & 0.02 & 0.28 \\
Accidental poisoning & 0.01 & 0.01 & 0.13 \\
\midrule
N & 184,346 & 1,556,496 &  \\
\bottomrule
\end{tabular}

\begin{tablenotes}[flushleft]
\setstretch{0.9}
\footnotesize
\item \textit{Notes:} The table reports pre-treatment outcome means (in percent) for hospitalizations in MedStat regions within 0--1km of a DCR centroid (treated) and outside 5km (control), for the period $t<2001$. The last column reports absolute standardized differences in means multiplied by 100, computed as in Table \ref{tab:balance_pre}.
\end{tablenotes}

\end{threeparttable}
\end{table}

%% file: tables/balance_medstat_pre.tex
\begin{table}[H]
\centering
\caption{Baseline MedStat covariate balance: Means}
\label{tab:balance_medstat_2000}

\begin{threeparttable}
\begin{tabular}{lrrr}
\toprule
\textbf{Variable} & \textbf{Treated} & \textbf{Control} & \textbf{Std Diff} \\
\midrule
Female & 0.52 & 0.51 & 14.60 \\
Swiss & 0.75 & 0.83 & 10.11 \\
Rural & 0.06 & 0.35 & 7.72 \\
Urban & 0.88 & 0.41 & 11.17 \\
Intermediate & 0.06 & 0.23 & 5.09 \\
Age 15--24 & 0.11 & 0.12 & 4.52 \\
Age 25--44 & 0.32 & 0.31 & 6.57 \\
Age 45--64 & 0.24 & 0.24 & 0.85 \\
Age 65+ & 0.18 & 0.15 & 11.69 \\
Low education & 0.21 & 0.24 & 7.74 \\
Middle education & 0.34 & 0.34 & 0.64 \\
High education & 0.08 & 0.04 & 12.65 \\
Professionals & 0.05 & 0.04 & 11.28 \\
Intermediate occupations & 0.09 & 0.08 & 3.55 \\
Low-skill manual & 0.06 & 0.06 & 6.05 \\
Unemployed & 0.03 & 0.02 & 14.14 \\
Population & 12897.65 & 9217.28 & 5.89 \\
\midrule
N (MedStat regions) & 50 & 608 &  \\
\bottomrule
\end{tabular}

\begin{tablenotes}[flushleft]
\setstretch{0.9}
\footnotesize
\item \textit{Notes:} The table reports baseline MedStat-region covariate means measured in the year 2000 for regions within 0--1km of a DCR centroid (treated) and regions outside 5km (control). Urban, Intermediate and Rural are indicators for the urbanization class of the region and sum to one. All other variables are population shares within the region, computed over the total resident population. Age groups therefore do not sum to one because the population under 15 is not reported. Age groups follow the categories available in the 2000 census aggregates and match those in Table \ref{tab:balance_pre}. Population is the resident population of the region in 2000. The last column reports absolute standardized differences in means multiplied by 100, computed as in Table \ref{tab:balance_pre}.
\end{tablenotes}

\end{threeparttable}
\end{table}

%% file: method.tex
\section{Empirical Strategy}
\label{sec:empirical_strategy}

\subsection{Identification}

In an ideal scenario, to estimate the causal effect of DCRs on health outcomes, one would randomly assign access to DCRs. However, in practice, such randomization is not feasible due to practical and ethical concerns. Instead, this study leverages a ``natural experiment'' by exploiting the staggered adoption of harm reduction policies across Swiss cantons and cities. The main analysis compares drug-related hospitalizations of patients living in treated MedStat regions with those living in not-yet-treated MedStat regions, before and after the opening of a DCR. 

To identify the effect, I use the staggered difference-in-differences (DiD) design. Specifically, I use the estimator proposed by \cite{CALLAWAY2021200}, which is particularly well suited for the analysis. The method accommodates the staggered timing of DCR adoption across cantons by comparing each treated group to not-yet-treated or never treated units. This approach allows me to estimate the impact of DCRs while accounting for relevant trends and regional differences beyond the timing of DCR openings. It also allows for dynamic effect estimation and event-study analyses, cohort level treatment effect estimation and further heterogeneity analysis. 

The parameter of interest is the group--time average treatment effect on the treated:

\begin{equation}
ATT(g,t) = \mathbb{E}\left[ Y_t(1) - Y_t(0) \mid G = g \right], \quad t \ge g
\end{equation}

where $G=g$ denotes the group of MedStat regions that first gain access to a DCR in year $g$, and $t$ denotes the calendar year. $Y_t(1)$ represents the potential outcome in year $t$ if a MedStat region has access to a DCR, while $Y_t(0)$ represents the potential outcome in the absence of a DCR. In this context, $ATT(g,t)$ measures the average change in drug-related hospitalization outcomes in year $t$ for patients living in MedStat regions that gained access to a DCR in year $g$, relative to the counterfactual outcome that would have been observed in the absence of the policy.

The key identifying assumption needed for this study is that changes in health outcomes, in the control group (or not yet treated), provide a good counterfactual for the changes that would have been observed in the treated groups in the absence of harm reduction policies. Accordingly, before estimating the DiD model, I must validate the identifying assumptions of the model. 

\subsubsection{Parallel Trends}

The validity of the DiD design rests on the parallel trends assumption, which requires that, in the absence of the DCR opening, drug-related hospitalizations in treated MedStat regions would have evolved as they did in control regions. Conceptually, this assumption is supported by three features of the Swiss institutional context, supply and demand of drugs, and access to health care services.

First, DCRs were established primarily to address visible open drug scenes in high-density urban hotspots \citet{EMCDDA2024}. Consequently, baseline levels of drug-related hospitalizations are higher in treated areas than in control areas. Crucially, DiD accommodates baseline level differences provided that underlying growth rates do not systematically diverge. Local governments introduced DCRs based on intensity of drug use, not because hospitalizations trends were accelerating faster in these regions relative to neighboring areas prior to treatment. 

Second, trends in drug related hospitalizations are heavily influenced by broader shifts in illicit drug supply and demand. On the supply side, because Switzerland is geographically compact and highly interconnected, changes in drug purity, price fluctuations, supply chain disruptions, and the emergence of synthetic opioids impact both urban and semi-urban regions. The argument for market integration is supported by empirical evidence from a cohort study on substance use risk factors, which demonstrates that access to hard illegal substances is surprisingly widespread and easy to purchase in rural areas \citet{quednow2022high}. On the demand side, treated regions are denser and differ in composition. Therefore, to ensure comparability, the analysis excludes highly rural regions with low common support and condition for regional socioeconomic characteristics and population density. What remains compares urban with semi-urban regions, facing common market shocks.

Third, institutional drivers of hospital admissions are highly standardized in Switzerland. Basic health insurance has been compulsory for every resident since 1996, and the benefits it covers are defined federally. Emergency care carries a treatment obligation, and hospitals code admissions and diagnoses are encoded under uniform guidelines issued by the  Federal Statistical Office for all facilities in all 26 cantons \citet{depietro2015switzerland, OFS2020_hospitalstats}. 

Although the parallel trends assumption cannot be tested directly, one can examine pre-treatment dynamics to evaluate its plausibility. I assess this using visual inspection of pre-treatment trends and more formal event study plots presented in Section~\ref{sec:result}. Before turning to the figures, I note that all outcome variables are expressed in logarithms throughout the analysis. Drug-related hospitalization rates are heavily right-skewed and exhibit substantial variation across MedStat regions, so the log transformation gives a more symmetric distribution and stabilizes variance. It also carries a natural interpretation, as estimated coefficients can be read as approximate percentage changes. 

\begin{figure}[H]
    \centering
    \caption{Drug-related hospitalization rate (log) per 100,000 inhabitants (1998--2022)}
    \includegraphics[width=0.85\textwidth]{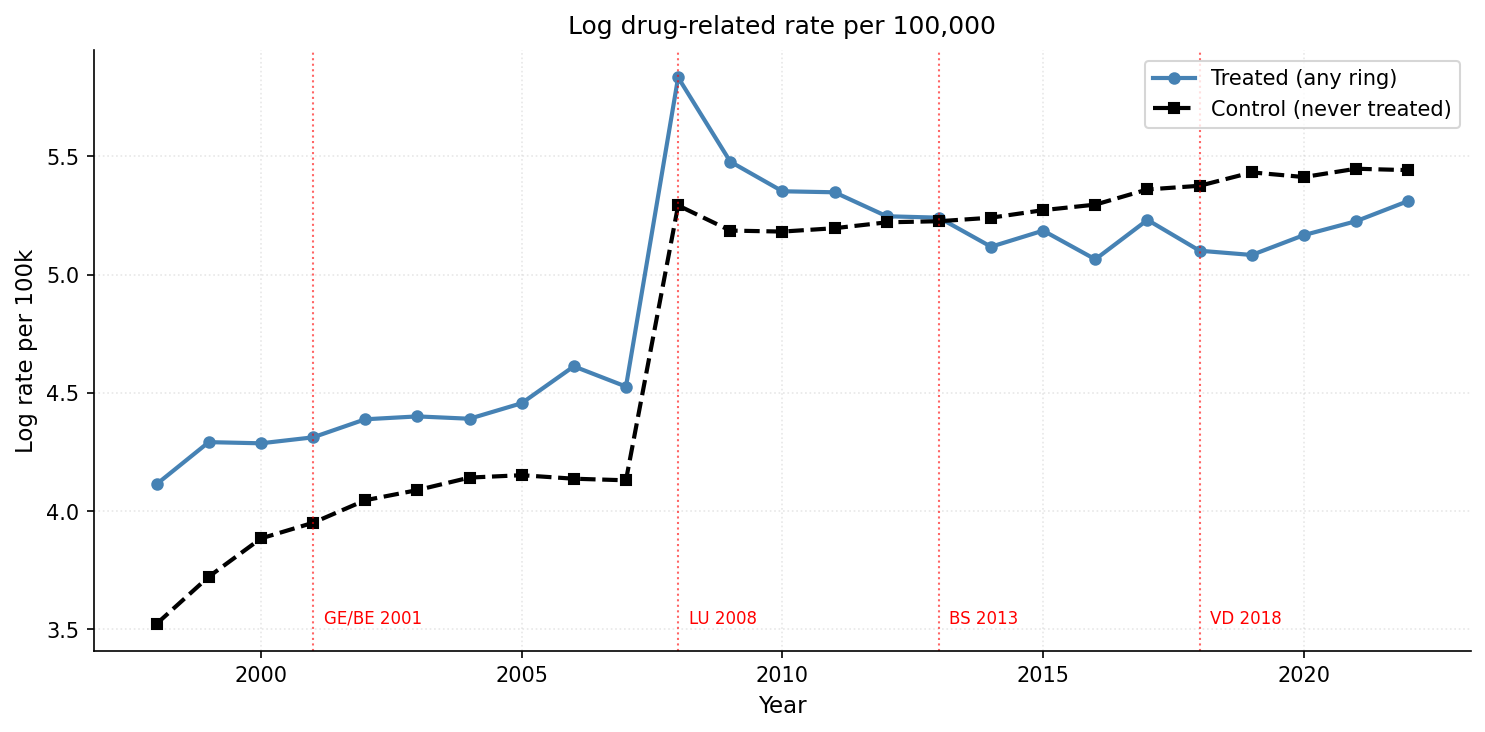}
    \label{fig:pretrends_log_rate}
    \vspace{0.3cm}
    \\
    \setstretch{0.9}
    \raggedright
    \footnotesize{\textit{Note:} The figure plots the average log drug-related hospitalization rate per 100,000 inhabitants for MedStat regions over 1998--2022. The \textit{treated} group (blue solid line) comprises all staggered-treated MedStat regions within 5\,km of a DCR that opened between 1998 and 2022, across all three 
    distance rings (0--1\,km, 1--3\,km, and 3--5\,km). The \textit{control} group (black dashed line) includes all MedStat regions with no DCR over the sample period (\textit{never-treated}). Regions near DCRs that opened before 1998 (\textit{always-treated}) are excluded.  Vertical red dotted lines indicate the opening years of DCRs in each treatment cohort.}
\end{figure}

Figure \ref{fig:pretrends_log_rate} presents the evolution of the logarithm of drug-related hospitalizations rates for treated and never treated MedStat regions. This visual inspection shows that treated and never treated regions follow similar trajectories, particularly prior to the first DCR opening in 2001. There is a noticeable jump in 2008 due to coding changes in the Hospital Medical Statistics as well as a reclassification of MedStat regions due to updates to postal codes and municipality codes. Importantly, these changes affect both treated and control regions in the same direction and therefore does not threaten identification. 

Additional trend figures covering alternative outcome measures, including raw counts, hospitalization shares, and the log of counts, as well as alternative comparison groups and treatment definitions, are reported in Appendix \ref{app:parallel_trends} in Figures \ref{fig:log_rate_by-ring},  \ref{fig:all_outcomes_by-ring} \& \ref{fig:lineplot_monthly_outcome}. For example, Figure \ref{fig:lineplot_monthly_outcome} plots monthly trends separately for each diagnostic subcategory, distinguishing treated units from control units. Across all specifications and outcome measures, the figures reveal broadly similar pre-treatment dynamics between treated and control regions. Overall, the graphical evidence provides informal support for the parallel trends assumption underlying the difference-in-differences identification strategy. However, since visual inspection alone does not constitute a formal test, I further evaluate the parallel trends assumption through event study estimates, which are presented in Section~\ref{sec:result}.

\subsubsection{Four Pillar Drug Policy}

In response to the opioid crisis, the Swiss Federal Council adopted a four pillar drug policy in 1991 that includes prevention, treatment, law enforcement and harm reduction \citep{CSETE201282}. A potential challenge to identification is that reductions in hospitalizations might reflect concurrent changes in the other three pillars rather than DCR availability alone. This would threaten identification if these pillars moved differentially in treated regions at the opening dates. I therefore examine, for each pillar, how it evolved and whether it changed when the DCRs opened.

Prevention is set nationally and delivered through campaigns and cantonal statutory bodies that run school and media programs. They have been established decades before DCRs opened \citep{CSETE201282}. As they operate at the cantonal level, it reaches treated and control regions equally within a canton, although prevention intensity may differ across cantons. Its mandates cover alcohol, tobacco, gambling and young people rather than dependent opioid users \citep{foph2017addiction}. Prevention works through schools, employers and parents on entry into dependence, acts with a delay of years, and thus cannot produce a break at a DCR opening date in the population my outcome measures.

Treatment consists of opioid agonist therapy (OAT) and heroin-assisted treatment (HAT). OAT replaces illicit heroin with a prescribed long-acting opioid, most often methadone, buprenorphine or slow-release oral morphine, taken daily under medical supervision. It stabilizes consumption, removes the need to inject street heroin and is the standard treatment for opioid dependence in Switzerland. HAT prescribes pharmaceutical heroin (diacetylmorphine), injected or taken orally in a specialized center several times a day, and is reserved for long-term users for whom at least two other treatments have failed. Both act directly on overdose, infection and injury, and thus on drug-related hospitalizations, so this is the pillar where co-timing would matter most. In most cantons, however, the outpatient centers providing OAT and HAT were established between six and nineteen years before the DCR opened\footnote{In Biel, Suprax has provided both treatments since 1995 \citep{infodrogsuprax}. In Geneva, methadone prescription expanded from the early 1990s and the heroin-prescription program PEPS opened in September 1995 \citep{premiereligne}. In Lucerne, the Drop-In center has dispensed substitution treatment since 1992 \citep{stadtluzern2008}. In Basel, methadone dispensing was established in the early 1990s and the HAT centre Janus opened in 1994 \citep{steffen2002development}.}. HAT expanded from 387 patients in 1994 to 1,598 in 2003 and has remained between 1,500 and 1,800 since \citep{gmel2020diacetylmorphine, amos2025diacetylmorphine}. For opioid agonist therapy, I observe the number of persons in treatment by canton and year from 1999 to 2025, from the national substitution statistic that cantonal physicians report \citep{labhart2020}. Figure~\ref{fig:OAT} shows the series of patients taking OAT. Switzerland counts between 17,000 and 18,000 people each year until 2012 and the number slowly declines afterward, and no treated canton shows a break at its opening.

Law enforcement is the pillar most often coupled with the opening of a DCR, since a DCR typically opens alongside a negotiated policing agreement for the surrounding area \citep{hedrich2004european}. I proxy enforcement with adult convictions for a felony or misdemeanor under the Narcotics Act, which the Federal Statistical Office reports for each canton from 1984. These convictions capture enforcement against dealing and larger possession.\footnote{Unfortunately, statistics on consumption offenses are available only from 2009.} Figure~\ref{fig:convictions} shows the series. Nationally, convictions move between 5,300 and 6,300 a year from 1994 to 2012, rise to 7,200 in 2015 and fall to 3,800 by 2025. It appears that enforcement did not co-evolve with the opening of DCRs. Two further features of the design limit what local policing agreements can do to the estimates. First, hospitalizations are attributed to the patient's MedStat region of residence and not to the location of the incident, so displacement of the drug scene, the dominant channel through which policing acts, cannot move a case from one region to another except through residential relocation, which operates far more slowly than the annual frequency of my data. What remains is that policing raises the riskiness of consumption for the users who stay, leading them to rush injections, to inject in less safe settings and to avoid carrying syringes \citep{KERR2005210, wood2004displacement}. This pushes measured hospitalizations up in treated regions and therefore makes an estimated decline conservative. 

The remaining harm-reduction services, contact centers, needle exchange and outreach, have no time series at the cantonal level \citep{duboisarber2008trends}. As a cross-sectional check, I use the current national directory of addiction services, which lists 997 facilities with coordinates and service type. Table~\ref{tab:suchtindex_snapshot} shows that treated cantons have no more treatment capacity per resident than cantons without a DCR.

\subsubsection{No Anticipation}

The no-anticipation assumption requires that potential users do not adjust their behavior prior to a DCR opening in response to its anticipated availability. Several features of the Swiss institutional setting and the target population make this assumption credible. At the institutional level, DCR opening dates are often uncertain due to long administrative and licensing procedures that introduce uncertainty over both timing and location. Importantly, the decisions are made by cantonal authorities in response to public health needs rather than driven by individual preferences of the prospective users, prospective users have limited ability to form reliable expectations about when or where a facility will open. 

At the individual level, the behavioral and socioeconomic characteristics of the target population further constrain the likelihood of anticipatory responses. Drug-dependent individuals are unlikely to adjust their consumption behavior in anticipation of a facility as addiction reduces the scope for strategic waiting or behavioral adjustment. Even if information about a planned facility were available, the ability to act on it is limited. In addition, relocating to areas where DCRs may open is constrained by high housing costs in urban centers where DCRs tend to be concentrated. 

While some word-of-mouth information may circulate, there is little reason to expect such information would meaningfully influence health outcomes before a DCR facility becomes operational. Overall, the centralized nature of policy decisions, the uncertainty of location and opening dates, the behavioral constraints imposed by addiction, and the limited housing mobility opportunities of DCR users suggest that anticipation effects are unlikely to bias the estimates.

\subsubsection{Irreversibility of Treatment}

The irreversibility assumption requires that once a unit becomes treated, it remains treated. In the context of this study, the assumption is satisfied. All DCRs that opened in Biel, Geneva, Lucerne, Basel, and Lausanne during the study period have remained in continuous operation without interruption. The facilities did not close even during the Covid-19 pandemic, a period of widespread disruption to health and social services, though some temporarily reduced their operating hours in response to public health restrictions. 

\subsubsection{Overlap}

Identification in the staggered difference in differences design requires that the overlap assumption holds. This condition requires that for each period, units have a probability of treatment that is strictly between zero and one. This ensures that each newly treated cohort can be compared against a set of suitable not-yet-treated or never-treated units with similar characteristics.

Several pieces of evidence support the plausibility of this assumption in the present setting. First, the staggered rollout of DCRs across Switzerland means that, for each treatment cohort, a large pool of never treated (or not-yet-treated) regions is available as a comparison group at the time of treatment. Second, I assess covariate balance between treated and control units prior to treatment. Tables \ref{tab:balance_pre} and \ref{tab:balance_drug_pre} report pre-treatment balance in hospitalization characteristics at the individual level, with standardized mean differences remaining small across all variables. In addition, Table \ref{tab:balance_medstat_2000} complements this by documenting baseline balance in demographic and socioeconomic characteristics at the MedStat region level, where differences in means are also modest.

\subsubsection{Stable Unit Treatment Value Assumption (SUTVA)}

SUTVA requires that the treatment status of one unit does not affect the outcomes in other units and that the treatment is of the same type for all treated units. The second requirement is not a concern since in this setting, DCRs follow a common service model across cantons and the intervention is therefore comparable wherever it is available. The first requirement is the one that needs to be further checked. In this context, it means that opening a DCR in one MedStat region should not alter drug-related outcomes in untreated MedStat regions. Several institutional features of the Swiss context support these assumptions. 

First, access to DCRs is restricted by canton of residence, and in some cases by city, which means that only cantonal residents can enter the facility. However, I define treatment by exposure to a DCR and not by the legal entitlement to use one. The reason behind this is that after speaking with managers in two of the largest DCRs in Switzerland, they explained that the regular DCR users do not travel more than half an hour to reach the DCR. This was there further assessed by the annual reports, where there was information available about where the users lived. For example, for the DCR in Lausanne, around 70\% of the clients lived in the city itself while the rest lived in areas more outside of the city. The facility attributes the non-Lausanne registrations to family or employment ties to the city and not to the journey being undertaken to reach the service \citep{abs2024}. This shows that the recurrent individuals for whom the DCR is often used are the ones living in the city itself, while the rest come rarely and have some other reason to be in the city. 

Although legal eligibility is decided at the canton of residence level, assigning treatment at that level would give identical exposure to a region bordering a facility and to one forty kilometers away in the same canton, and it would reduce the design to 26 units. Defining treatment at the MedStat region measures who is actually within reach of a facility and retains the variation needed to estimate cohort-specific effects. Entitlement is not irrelevant and there are some exceptions depending on the canton. For example, in the case of smaller cantons such as Basel-Stadt and Basel-Land, the two cantons have an agreement in which residents of Basel-Land can enter the DCR in Basel-Stadt. A similar arrangement applies to the DCR in Lucerne, which welcomes residents from the cantons of Nidwalden, Obwalden, Schwyz, Uri and Zug. I re-estimate the effects classifying Basel-Land as well as Nidwalden, Obwalden, Schwyz, Uri and Zug as treated. Reclassifying these 54 regions leaves the estimate essentially unchanged. For details please refer to Table~\ref{tab:entitlement}.

Additionally, hospitalizations are attributed to the MedStat region of residence of the patient and not to the location of the hospital. As a descriptive check of the spillover across cantons, I show that throughout the study period, the share of hospitalizations occurring in a patients same canton of residence ranges from 88\% in 1998 to 80\% in 2022 and remains above 81\% in every year.(A year-by-year series is reported in Appendix~\ref{tab:same_canton_by_year}.) This pattern suggests that patients typically seek care close to home, leaving little scope for treatment in one canton to affect outcomes in another. However, I am aware that the exposure outside the buffer is smaller, but not exactly zero. Thus, I  re-estimate the effects at the canton level, which removed the within-canton spillovers by construction and quantifies how much of the effect is diluted when exposed and unexposed regions are averaged together. The results are in Table \ref{tab:cantonlevel}.

\subsection{Estimation}

After assessing the identifying assumptions, I estimate the average treatment effect on the treated (ATT) using the doubly robust estimator proposed by \cite{CALLAWAY2021200} as it accommodates staggered treatment adoption, allows for cohort-specific treatment effects, and is robust to misspecifications of the nuisance functions. The estimator is specified as follows:

\begin{equation}
\label{eq:att_estimation}
\widehat{\text{ATT}}^{\text{nev}}_{\text{dr}}(g, t)
= \mathbb{E}_n\!\left[
\left(\widehat{w}^{\text{treat}}_g - \widehat{w}^{\text{comp,nev}}_g\right)
\left(Y_t - Y_{g-1} - \widehat{m}^{\text{nev}}_{g,t}\!\left(X;\widehat{\beta}^{\text{nev}}_{g,t}\right)\right)
\right]
\end{equation}
where
\[
\widehat{w}^{\text{treat}}_g = \frac{G_g}{\mathbb{E}_n[G_g]},
\qquad
\widehat{w}^{\text{comp,nev}}_g =
\frac{\dfrac{\widehat{p}_g(X;\widehat{\pi}_g)\,C}{1 - \widehat{p}_g(X;\widehat{\pi}_g)}}
     {\mathbb{E}_n\!\left[\dfrac{\widehat{p}_g(X;\widehat{\pi}_g)\,C}{1 - \widehat{p}_g(X;\widehat{\pi}_g)}\right]}.
\]

Here $Y_t$ denotes the log drug-related hospitalization rate per 100,000 inhabitants in year $t$. $G_g$ is an indicator equal to one for units first treated in cohort $g$, and $C$ is an indicator equal to one for never-treated or not-yet-treated units. The function $\widehat{p}_g(X;\widehat{\pi}_g)$ is an estimator of the propensity score $p_g(X) = \Pr(G_g = 1 \mid X,\, G_g + C = 1)$, and $\widehat{m}^{\text{nev}}_{g,t}(X;\widehat{\beta}^{\text{nev}}_{g,t})$ is an estimator of the outcome regression $m^{\text{nev}}_{g,t}(X) = \mathbb{E}[Y_t - Y_{g-1} \mid X, C = 1]$. 

Implementing the estimator requires specifying the conditioning set $X$ and the models used for the two nuisance functions, the propensity score and the outcome regression. I condition on baseline MedStat-region characteristics measured in 2000, prior to any DCR opening. This covariate set include educational attainment, socio-professional composition, demographic structure (the share of individuals aged 15--24 and 25--44, the share of Swiss nationals, and the share of women), urban typology (indicators for urban, intermediate, and rural regions), and log population. Following \cite{CALLAWAY2021200}, I estimate the propensity score using a logit model and the outcome regression via ordinary least squares (OLS). Finally, to ensure the results are not driven by these specific functional forms, I conduct robustness checks by varying these nuisance estimators, specifically by using inverse probability weighting and outcome regression individually, the results are in Figure \ref{fig:cs_nuisance_demo}, \ref{fig:cs_nuisance_geo} and Table \ref{tab:cs_est_method}.

The estimator is doubly robust in that it remains consistent if either the propensity model $\widehat{\pi}_g$ or the outcome regression $\widehat{\beta}^{\text{nev}}_{g,t}$ is correctly specified \citep{CALLAWAY2021200}. The estimator $\widehat{\text{ATT}}^{\text{nev}}_{\text{dr}}(g,t)$ therefore measures the average difference at year $t$ between drug-related hospitalization rates in cohort-$g$ regions and never-treated or not-yet-treated regions, with both sides expressed as deviations from the pre-treatment reference year $g-1$. A negative estimate implies that hospitalization rates fell in regions near a DCR relative to what would have been observed had no DCR opened nearby. 

The baseline specification throughout is the doubly robust \citet{CALLAWAY2021200} estimator with never-treated controls and a universal base period.\footnote{The universal base period normalizes pre-treatment coefficients relative to the average pre-treatment period rather than to a single reference year, which allows pre-trends to be compared across cohorts with different treatment timing \citep{Roth2026_eventstudy}.} Each of the specification choices described above are then varied. Estimates under alternative buffer distances of 1\,km and 3\,km are reported in Figure~\ref{fig:event_study_1km_3km_5km_5t_four_estimators}. Finally, I re-estimate the effects at the 5\,km buffer using three alternative estimators: the interaction-weighted estimator of \citet{SUN2021}, stacked difference-in-differences \citep{stacked_did, BAKER2022370}, and synthetic difference-in-differences \citep{arkhangelsky2021synthetic}, reported in Table~\ref{tab:att_all_estimators} with the corresponding event studies in Figure~\ref{fig:event_study_4_estimators}. Results are consistent across all specifications.

Additionally, \cite{CALLAWAY2021200} propose different ways of aggregating the $\text{ATT}(g,t)$'s to answer different empirical questions. First, event study estimates, aggregate across cohorts at a common length of exposure and trace the dynamic profile of effects following DCR opening. Second, an overall pooled ATT averages the cohort specific effects weighted by cohort size to produce a single summary. Finally, I estimate the $\widehat{\text{ATT}}^{\text{nev}}_{\text{dr}}(g,t)$ separately for the five diagnostic subcategories, since DCRs are expected to move infection-related and poisoning outcomes in different directions.

%% file: results.tex
\section{Results} \label{sec:result}

This section reports the effect of DCR openings on the rate of drug-related hospitalizations per 100,000 inhabitants. I begin with the event-study estimates and then present the overall pooled ATT. I then disaggregate the overall outcome into its sub-outcome categories and report the same estimates for each. 

\begin{figure}[H]
    \centering
    \caption{Event study estimates by buffer zone, $\pm$5-year window}
    \includegraphics[width=0.85\textwidth]{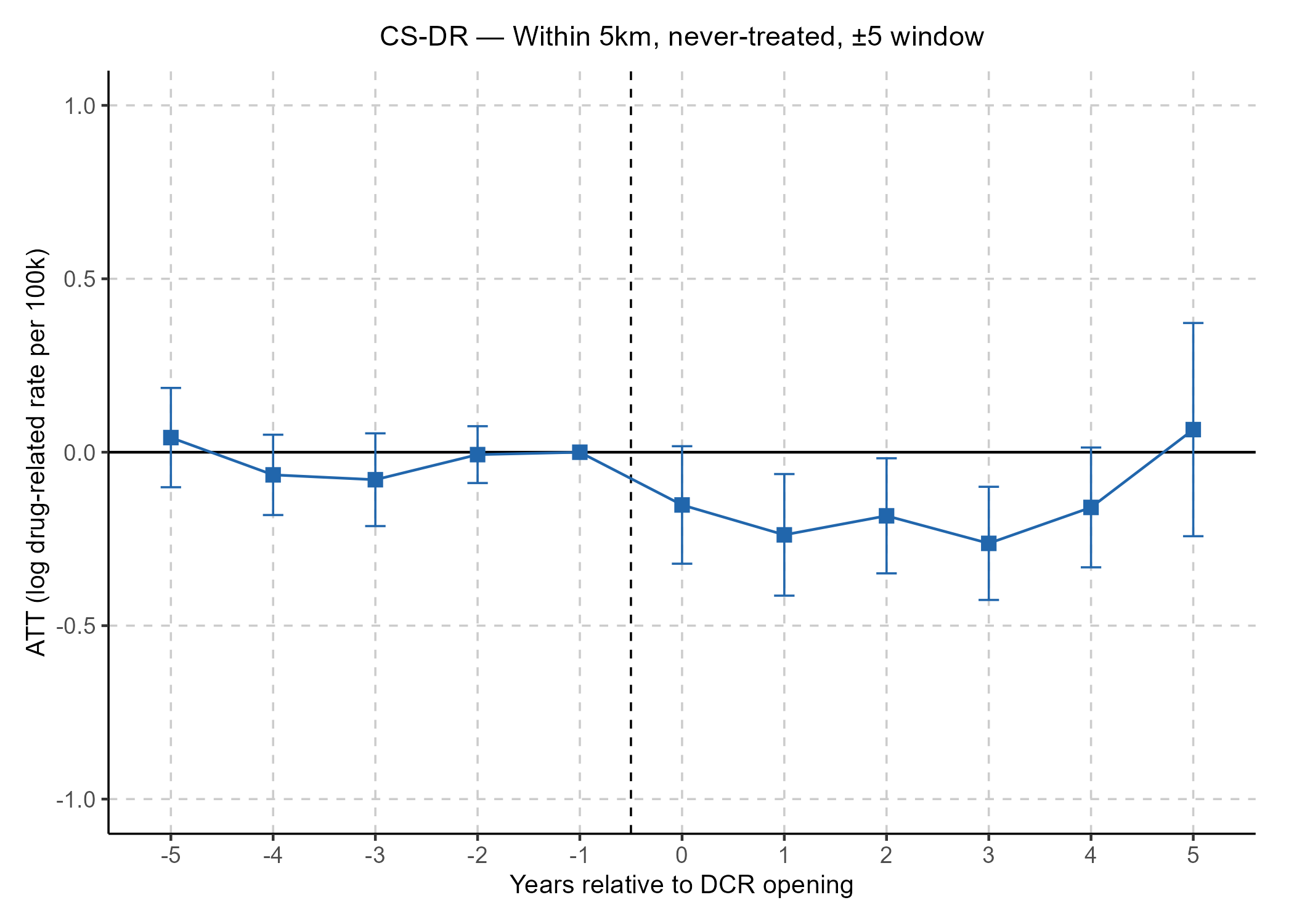}
    \label{fig:cs_eventstudy_buffers_5t}
    \vspace{0.3cm}
    \\
    \setstretch{0.9}
    \raggedright
    \footnotesize{\textit{Note:} The figure plots event-study coefficients from the doubly robust estimator of \cite{CALLAWAY2021200} for regions within 5\,km of a DCR opening. The comparison group consists of never-treated regions (608 control regions). The outcome is the log drug-related hospitalization rate per 100,000 inhabitants. The sample includes all staggered-treated MedStat regions within 5\,km of a DCR that opened between 1998 and 2022 and all never-treated regions. The estimation uses a $\pm$5-year window relative to DCR opening, with $t = -5$ and $t = +5$ representing pooled endpoint bins that absorb all periods beyond the window. The reference period is a universal base period. Regions near DCRs that opened before 1998 (\textit{always-treated}) are excluded through out. 95\% confidence intervals are based on standard errors clustered at the MedStat region level. Results for the 1\,km and 3\,km buffer specifications are reported in Appendix Figure~\ref{fig:event_study_1km_3km_5km_5t_four_estimators}.}
\end{figure}

Figure \ref{fig:cs_eventstudy_buffers_5t} presents the event study estimates for the 5\,km buffer specifications. I focus on the 5 km buffer specification, as it includes the largest number of treated regions and yields the most precise estimates. I summarize dynamic effects over a symmetric five year window around DCR openings. This window avoids relying on distant leads and lags that are informed by only a small subset of adopters. Pre-treatment coefficients are small and statistically insignificant, supporting the parallel trends assumption. The post-treatment coefficients are consistently negative, indicating a reduction in drug-related hospitalizations following DCR openings. The estimates of the overall summary of ATTs based on calendar time aggregation of \cite{CALLAWAY2021200} is around -35\%  reduction in hospitalizations due to drugs. 

\begin{table}[H] 
\input{tables/tab_overall_att_BL}
\end{table}

Table~\ref{tab:overall_att_BL} reports the pooled ATT for the 1\,km, 3\,km and 5\,km buffer specification, aggregating the dynamic treatment effects into a single estimate averaged across all post-treatment periods and cohorts. The 5\,km specification, yields an ATT of -0.431
corresponding to a 35\% reduction in the drug-related hospitalization rate. This estimate is statistically significant at the 5\% level. The 3\,km and 1\,km estimates point in the same direction but are less precise, due to the smaller number of treated regions. 

To further explore the mechanisms through which DCR openings affect drug-related hospitalizations, I disaggregate the main outcome into five diagnostic 
sub-categories: (i) mental and behavioral disorders due to psychoactive substance use, (ii) viral hepatitis, (iii) HIV, (iv) poisoning by narcotics and psychodysleptics, and (v) fatal drug-related hospitalizations. Accidental poisoning is excluded due to the very small number of cases. 

\begin{figure}[H]
    \centering
    \caption{Event study: Sub-outcome estimates,
    within 5\,km, $\pm$5-year window}
    \includegraphics[width=0.85\textwidth]{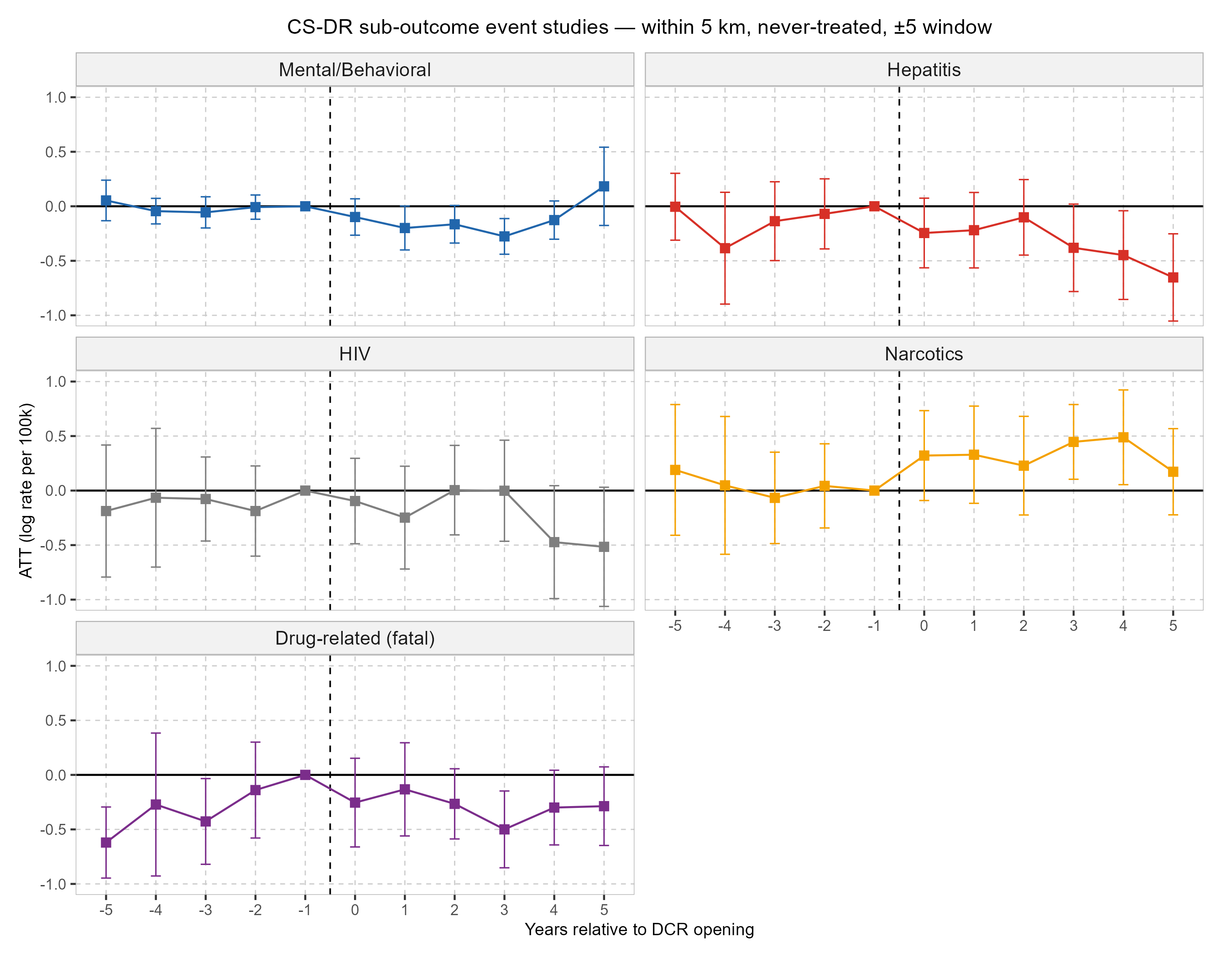}
    \label{fig:sub_outcomes_5km}
    \vspace{0.3cm}
    \\
    \setstretch{0.9}
    \raggedright
    \footnotesize{\textit{Note:} The figure plots event-study coefficients from the doubly robust Callaway \& Sant'Anna (2021) estimator for four diagnostic sub-categories of drug-related hospitalizations. Each panel uses a separate $y$-axis to accommodate differences in scale across outcomes. The sample includes all staggered-treated MedStat regions within 5\,km of a
    DCR that opened between 1998 and 2022. The reference period is a universal base period. Regions near DCRs that opened before 1998 (\textit{always-treated}) are excluded. 95\% confidence intervals are based on standard errors clustered at the MedStat region level.}
\end{figure}

\begin{table}[H] 
\input{tables/tab_sub_overall_att}
\end{table}

Figure~\ref{fig:sub_outcomes_5km} and Table~\ref{tab:sub_overall_att} repeat the estimation for the five diagnostic categories. From  Table~\ref{tab:sub_overall_att}, the ATT estimates for hospitalizations due to mental and behavioral disorders are consistently negative and marginally significant at the 5\,km buffer specification. Moreover, the estimates of hepatitis are also negative and of comparable magnitude but insignificant. However, the event-study path in Figure~\ref{fig:sub_outcomes_5km} shows a gradual decline that builds from year 3 onward, consistent with a slow-acting reduction in needle sharing. HIV estimates are insignificant and close to zero. Narcotics poisoning hospitalizations rise after DCR opening and remain positive across all buffers. Lastly, fatal drug-related hospitalizations move in the opposite direction, falling after opening the DCR. The opposing patterns in narcotics and fatal hospitalizations suggest that DCRs convert events that would otherwise be fatal into survivable hospital contacts.  

\subsection{Mechanisms}

The sub-outcome results point to two channels through which DCRs reduce drug-related hospitalizations. The first is overdose supervision. DCRs provide a medically supervised environment where staff can control the dose consumed, perform resuscitation, and call emergency services when an overdose occurs. When supervision prevents a death that would otherwise happen outside a hospital, the consequence is a simultaneous decline in fatal drug-related hospitalizations and a rise in non-fatal poisoning contacts, as events that would have ended in death instead end in a hospital visit. The results are consistent with this prediction. Drug-related fatal hospitalizations fall near DCRs, by approximately 30\% at 5\,km, while narcotics and psychodysleptics hospitalizations rise by 47\% at 5\,km. The opposing signs and spatial concentration of these two effects are consistent with a 
substitution from out-of-hospital mortality to in-hospital survival. Since I cannot observe overdose events that do not result in a hospital contact, this interpretation is suggestive rather than identified, but it is consistent with the clinical evidence on DCR-based overdose reversals 
\citep{Kerr2017, hedrich2004european}.

The second channel is harm reduction for blood-borne diseases. DCRs distribute sterile injection equipment, which reduces needle sharing and the transmission of viral hepatitis and HIV. Unlike overdose prevention, these effects are expected to emerge slowly, since disease progression and hospitalization take years to follow from initial infection. The event-study path for hepatitis is consistent with this lag. The estimates are close to zero  in the first two post-treatment years and become progressively more negative from year 3 onward, reaching approximately 40\% below baseline by year 5. HIV estimates are near zero through out the analysis, with wide confidence intervals. A longer panel and additional out-patient data would be needed to determine whether these trends continue and to get more precise estimates.

A third potential channel is referral to treatment and social services. DCRs routinely connect users to addiction treatment, primary care, and housing support, and a sustained reduction in hospitalizations could partly reflect better long-term management of drug dependence. The regional hospitalization panel used in this paper does not contain information on treatment entry or service use, so I cannot separate this channel from the two described above. It remains a plausible additional mechanism that future work with information on treatment take up data could examine.

%% file: tables/tab_overall_att_BL.tex
\begin{table}[H]
\centering
\caption{Overall Average Treatment Effect on the Treated (ATT)}
\label{tab:overall_att_BL}
\begin{tabular}{lccc}
\toprule
 & 1\,km & 3\,km & 5\,km \\
\midrule
ATT & -0.399 & -0.283 & -0.431** \\
    & (0.412) & (0.279) & (0.210) \\
\% change & -32.9\% & -24.7\% & -35.0\% \\
\midrule
Treated regions & 7      & 35     & 50     \\
Control regions & 608    & 608    & 608    \\
Observations    & 16,175 & 16,875 & 17,250 \\
\bottomrule
\end{tabular}
\vspace{0.3cm}
\begin{minipage}{\linewidth}
\setstretch{0.9}
\raggedright
\footnotesize{\textit{Note:} The table reports the overall ATT estimated using the doubly robust \cite{CALLAWAY2021200} estimator with never-treated controls and a universal base period. The outcome is the log drug-related hospitalization rate per 100,000 inhabitants. Columns differ in the geographic buffer used to define treated Medstat regions. The control group consists of 608 never-treated Medstat regions
throughout. Regions near DCRs that opened before 1998
(\textit{always-treated}) are excluded. Standard errors clustered at the Medstat region level in
parentheses. Significance: $^{***}p<0.01$, $^{**}p<0.05$,
$^{*}p<0.10$.}
\end{minipage}
\end{table}

%% file: tables/tab_sub_overall_att.tex
\begin{table}[H]
\centering
\caption{Overall ATT by Outcome Category and Buffer Zone}
\label{tab:sub_overall_att}
\begin{tabular}{lccc}
\toprule
Outcome & 1\,km & 3\,km & 5\,km \\
\midrule
Mental/Behavioral   & -0.354    & -0.202   & -0.381*  \\
                    & (0.382)   & (0.244)  & (0.220)  \\
\addlinespace
Hepatitis           & -0.270    & -0.413   & -0.406   \\
                    & (0.486)   & (0.270)  & (0.250)  \\
\addlinespace
HIV                 & 0.063     & -0.239   & -0.340   \\
                    & (0.801)   & (0.337)  & (0.294)  \\
\addlinespace
Narcotics           & 1.186***  & 0.549*** & 0.390**  \\
                    & (0.288)   & (0.187)  & (0.165)  \\
\addlinespace
Drug-related (fatal)& -1.207*** & -0.290   & -0.354** \\
                    & (0.137)   & (0.215)  & (0.178)  \\
\midrule
Treated regions & 7   & 35  & 50  \\
Control regions & \multicolumn{3}{c}{608 (never-treated)} \\
\bottomrule
\end{tabular}
\vspace{0.3cm}
\begin{minipage}{\linewidth}
\setstretch{0.9}
\raggedright
\footnotesize{\textit{Note:} Each cell reports the overall ATT from
the doubly robust Callaway \& Sant'Anna (2021) estimator with
never-treated controls and a universal base period. Outcomes are
log hospitalization rates per 100,000 inhabitants. Sub-outcomes:
mental and behavioral disorders (F11--F19), viral hepatitis (B15--B19),
HIV disease (B20--B24), narcotics/psychodysleptics (T40). The 1\,km
estimates are identified from a single treated region per cohort and
should be interpreted with caution. Regions near DCRs opened before
1998 excluded. Standard errors clustered at the Medstat region level.
Significance: $^{***}p<0.01$, $^{**}p<0.05$, $^{*}p<0.10$.}
\end{minipage}
\end{table}

%% file: conclusion.tex
\section{Conclusion} \label{sec:conclusion}

The results suggest that DCRs reduce drug-related hospitalizations in nearby regions. The pooled ATT implies a 35\% reduction in hospitalization rates within 5 km of a DCR opening, and this estimate is consistent in sign across all buffer distances and robust across four alternative estimators. The sub-outcome results help clarify the channels behind this average. Mental and behavioral disorder hospitalizations, which form the largest category, decline by around 32\% at 5 km. Hepatitis estimates are negative and gradually build up from about three years after opening, consistent with a slow-acting reduction in needle sharing, while HIV estimates remain near zero.

The clearest mechanism comes from the overdose outcomes. Fatal drug-related hospitalizations fall by roughly 35\% within 5 km, while non-fatal narcotics poisoning hospitalizations rise by a comparable magnitude. This offsetting pattern is consistent with a substitution effect, in which on-site supervision allows staff to intervene in overdoses that would otherwise become fatal and convert them into non-fatal hospital contacts. 

These reductions in drug-related hospitalizations carry economic significance beyond their health value. Fewer drug-related hospitalizations imply lower inpatient costs, and the shift from fatal to non-fatal overdose events represents a larger welfare gain through avoided mortality. A limitation from the current analysis is that it relies only on inpatient hospitalization records. Incorporating outpatient data, insurance claims, emergency call records, and drug-related deaths occurring outside hospital settings would provide a more complete picture of how DCRs affect health, including fatal overdoses that never reach a hospital.

%% file: tables/tab_suchtindex_snapshot.tex
\begin{table}[htbp]
\centering
\caption{Addiction-care facilities by type: Switzerland, treated cantons and never-DCR cantons (Suchtindex, 2026 snapshot)}
\label{tab:suchtindex_snapshot}
\footnotesize
\setlength{\tabcolsep}{4pt}
% Fallback if the table still overflows: \resizebox{\textwidth}{!}{ ...tabular... }
\begin{tabular}{lcccccccc}
\toprule
 & Switzerland & GE & BE & LU & BS & BL & VD & Never-DCR \\
\midrule
\multicolumn{9}{l}{\textit{Number of facilities}} \\
OAT site & 161 & 7 & 22 & 5 & 5 & 7 & 15 & 65 \\
\quad of which HAT & 31 & 1 & 6 & 2 & 1 & 0 & 1 & 7 \\
Consumption room (DCR) & 18 & 1 & 3 & 1 & 2 & 0 & 2 & 2 \\
Contact centre, no room & 46 & 2 & 7 & 1 & 7 & 0 & 8 & 13 \\
Needle exchange / automat & 110 & 2 & 16 & 5 & 3 & 0 & 10 & 49 \\
Street outreach & 35 & 1 & 8 & 2 & 3 & 0 & 3 & 9 \\
Residential therapy & 83 & 3 & 9 & 5 & 4 & 2 & 11 & 27 \\
Outpatient therapy & 40 & 1 & 10 & 2 & 2 & 0 & 1 & 10 \\
Inpatient withdrawal & 68 & 3 & 11 & 3 & 3 & 1 & 3 & 32 \\
On-site counselling & 443 & 24 & 80 & 16 & 19 & 9 & 39 & 171 \\
Any entry & 994 & 53 & 186 & 37 & 43 & 22 & 80 & 354 \\
\midrule
\multicolumn{9}{l}{\textit{Per 100,000 residents}} \\
OAT site & 1.76 & 1.30 & 2.04 & 1.13 & 2.45 & 2.30 & 1.74 & 1.76 \\
\quad of which HAT & 0.34 & 0.19 & 0.56 & 0.45 & 0.49 & 0.00 & 0.12 & 0.19 \\
Contact centre, no room & 0.50 & 0.37 & 0.65 & 0.23 & 3.43 & 0.00 & 0.93 & 0.35 \\
Needle exchange / automat & 1.21 & 0.37 & 1.49 & 1.13 & 1.47 & 0.00 & 1.16 & 1.33 \\
Street outreach & 0.38 & 0.19 & 0.74 & 0.45 & 1.47 & 0.00 & 0.35 & 0.24 \\
Residential therapy & 0.91 & 0.56 & 0.84 & 1.13 & 1.96 & 0.66 & 1.27 & 0.73 \\
Outpatient therapy & 0.44 & 0.19 & 0.93 & 0.45 & 0.98 & 0.00 & 0.12 & 0.27 \\
Inpatient withdrawal & 0.75 & 0.56 & 1.02 & 0.68 & 1.47 & 0.33 & 0.35 & 0.87 \\
On-site counselling & 4.85 & 4.47 & 7.43 & 3.62 & 9.31 & 2.96 & 4.52 & 4.64 \\
Any entry & 10.89 & 9.87 & 17.28 & 8.38 & 21.06 & 7.24 & 9.26 & 9.60 \\
\midrule
Population 2025 (thousands) & 9,127 & 537 & 1,076 & 442 & 204 & 304 & 864 & 3,686 \\
\bottomrule
\end{tabular}
\begin{minipage}{\linewidth}\footnotesize
\textit{Notes:} Entries in the Infodrog Suchtindex (suchtindex.infodrog.ch) as of 11 September 2026, classified by the directory's own service tags; an entry is counted once per type it lists, so rows are not exclusive. Liechtenstein excluded. The directory has no opening dates, so counts describe the current service environment only. Never-DCR: the 17 cantons with no consumption room in the 1998--2022 treatment definition; the two rooms listed there (Chur, Fribourg) are not part of it. Population: permanent residents, 31 December 2025 (BFS).
\end{minipage}
\end{table}

%% file: tables/tab_desc_stats.tex
\begin{table}[H]
\centering
\caption{Descriptive Statistics: Drug-Related Hospitalization Outcomes}
\label{tab:desc_stats}
\begin{threeparttable}
\begin{tabular}{lrr}
\toprule
 & \multicolumn{1}{c}{Mean} & \multicolumn{1}{c}{Median} \\
\midrule
\multicolumn{3}{l}{\textit{Panel A: Medstat region level (count per region per year)}} \\
\midrule
Drug-related (total) & 30.4 & 19.0 \\
Mental/Behavioral    & 26.0 & 16.0 \\
Hepatitis            &  3.6 &  1.0 \\
HIV                  &  1.5 &  0.0 \\
Narcotics            &  0.6 &  0.0 \\
\midrule
\multicolumn{3}{l}{\textit{Panel B: Medstat region level (rate per 100,000 per year)}} \\
\midrule
Drug-related (total) & 301.1 & 184.7 \\
Mental/Behavioral    & 256.0 & 157.1 \\
Hepatitis            &  37.7 &  14.2 \\
HIV                  &  14.7 &   0.0 \\
Narcotics            &   5.8 &   0.0 \\
\midrule
\multicolumn{3}{l}{\textit{Panel C: Canton level (count per canton per year)}} \\
\midrule
Drug-related (total) & 823.4 & 573.7 \\
Mental/Behavioral    & 705.0 & 488.1 \\
Hepatitis            &  97.2 &  63.2 \\
HIV                  &  39.9 &  18.8 \\
Narcotics            &  15.8 &  11.9 \\
\midrule
\multicolumn{3}{l}{\textit{Panel D: Canton level (rate per 100,000 per year)}} \\
\midrule
Drug-related (total) & 8165.1 & 6536.4 \\
Mental/Behavioral    & 6942.3 & 5521.1 \\
Hepatitis            & 1021.3 &  657.8 \\
HIV                  &  399.2 &  225.0 \\
Narcotics            &  156.9 &  119.9 \\
\bottomrule
\end{tabular}
\begin{tablenotes}[flushleft]
\setstretch{0.9}
\footnotesize
\item \textit{Notes:} Panels A and B are computed at the Medstat
region $\times$ year level (705 regions, 1998--2022). Panels C and D
are computed at the canton $\times$ year level (26 cantons). The rate
is expressed per 100,000 inhabitants. Sub-outcomes: mental and
behavioral disorders (F11--F19), hepatitis (B15--B19), HIV
(B20--B24), narcotics/psychodysleptics (T40). The median of HIV and
narcotics at the Medstat level is zero, reflecting that more than
half of region--year observations record no hospitalization in these
categories.
\end{tablenotes}
\end{threeparttable}
\end{table}

%% file: tables/balance_full.tex
\begin{table}[H]
\centering
\caption{Covariate balance (full sample): Means}
\label{tab:balance_full}

\begin{threeparttable}
\begin{tabular}{lrrr}
\toprule
\textbf{Variable} & \textbf{Treated} & \textbf{Control} & \textbf{Std Diff} \\
\midrule
Swiss & 0.74 & 0.81 & 16.23 \\
Male & 0.44 & 0.46 & 3.96 \\
Length of stay & 13.42 & 12.29 & 0.55 \\
ICU hours & 3.03 & 3.29 & 0.71 \\
Any ICU stay & 0.04 & 0.05 & 2.22 \\
Emergency admission & 0.37 & 0.37 & 1.01 \\
Inpatient Stay & 0.93 & 0.93 & 0.44 \\
Age 15--19 & 0.03 & 0.04 & 3.36 \\
Age 20--24 & 0.05 & 0.05 & 0.99 \\
Age 25--29 & 0.07 & 0.07 & 0.44 \\
Age 30--34 & 0.10 & 0.09 & 3.33 \\
Age 35--39 & 0.09 & 0.08 & 4.21 \\
Age 40--44 & 0.08 & 0.07 & 1.14 \\
Age 45--49 & 0.08 & 0.08 & 0.27 \\
Age 50--54 & 0.09 & 0.09 & 1.38 \\
Age 55--59 & 0.09 & 0.10 & 1.68 \\
Age 60--64 & 0.10 & 0.11 & 1.61 \\
Age 65--69 & 0.11 & 0.11 & 1.21 \\
Age 70--74 & 0.12 & 0.12 & 0.37 \\
\midrule
N & 3,557,995 & 15,171,913 &  \\
\bottomrule
\end{tabular}

\begin{tablenotes}[flushleft]
\setstretch{0.9}
\footnotesize
\item \textit{Notes:} The table reports covariate means in the full sample for hospitalizations of residents in MedStat regions within 0 to 1km of a DCR centroid (treated) and residents outside 5km (control). Swiss, Male, Any ICU stay, Emergency admission, and Inpatient Stay are indicator variables (means can be read as proportions). Length of stay (SwissDRG) is measured in days. ICU hours is measured in hours (0 for no ICU stay). Age rows are shares of hospitalizations in each 5-year age bin. The standardized difference in means is
$\displaystyle \frac{|\bar X_T-\bar X_C|}{\sqrt{(s_T^2+s_C^2)/2}} \times 100$,
where $T$ denotes treated and $C$ control groups. $N$ is the number of hospitalization records used to compute the means. Following \cite{rosenbaum_1985}, a standardized difference higher than 20 is considered large.
\end{tablenotes}

\end{threeparttable}
\end{table}

%% file: tables/balance_drug_pre.tex
\begin{table}[H]
\centering
\caption{Pre-treatment covariate balance among drug-related hospitalizations: Means}
\label{tab:balance_drug_pre}

\begin{threeparttable}
\begin{tabular}{lrrr}
\toprule
\textbf{Variable} & \textbf{Treated} & \textbf{Control} & \textbf{Std Diff} \\
\midrule
Swiss & 0.83 & 0.81 & 5.44 \\
Male & 0.60 & 0.61 & 1.70 \\
Length of stay & 30.57 & 30.93 & 0.16 \\
ICU hours & 0.91 & 2.62 & 6.87 \\
Any ICU stay & 0.02 & 0.04 & 9.49 \\
Emergency admission & 0.44 & 0.42 & 2.57 \\
Inpatient Stay & 0.98 & 0.94 & 18.60 \\
Age 15--19 & 0.05 & 0.05 & 0.61 \\
Age 20--24 & 0.13 & 0.13 & 1.01 \\
Age 25--29 & 0.17 & 0.18 & 1.71 \\
Age 30--34 & 0.20 & 0.19 & 0.70 \\
Age 35--39 & 0.15 & 0.15 & 0.11 \\
Age 40--44 & 0.10 & 0.10 & 1.54 \\
Age 45--49 & 0.05 & 0.06 & 2.03 \\
Age 50--54 & 0.05 & 0.04 & 4.14 \\
Age 55--59 & 0.03 & 0.03 & 0.17 \\
Age 60--64 & 0.03 & 0.02 & 2.00 \\
Age 65--69 & 0.02 & 0.02 & 0.86 \\
Age 70--74 & 0.02 & 0.02 & 1.75 \\
\midrule
N & 4,775 & 35,353 &  \\
\bottomrule
\end{tabular}

\begin{tablenotes}[flushleft]
\setstretch{0.9}
\footnotesize
\item \textit{Notes:} The table reports covariate means in the pre-treatment period ($t<2001$) among drug-related hospitalizations for residents in MedStat regions within 0--1km of a DCR centroid (treated) and residents outside 5km (control). Variable definitions and standardized differences are computed as in Table \ref{tab:balance_full}. Following \cite{rosenbaum_1985}, a standardized difference above 20 is considered large.
\end{tablenotes}

\end{threeparttable}
\end{table}

%% file: tables/balance_full_drug.tex
\begin{table}[H]
\centering
\caption{Covariate balance among drug-related hospitalizations: Means}
\label{tab:balance_drug}

\begin{threeparttable}
\begin{tabular}{lrrr}
\toprule
\textbf{Variable} & \textbf{Treated} & \textbf{Control} & \textbf{Std Diff} \\
\midrule
Swiss & 0.74 & 0.78 & 8.42 \\
Male & 0.62 & 0.63 & 1.81 \\
Length of stay & 25.61 & 25.67 & 0.05 \\
ICU hours & 1.83 & 2.59 & 3.08 \\
Any ICU stay & 0.04 & 0.05 & 6.38 \\
Emergency admission & 0.49 & 0.48 & 1.78 \\
Inpatient Stay & 0.96 & 0.97 & 8.84 \\
Age 15--19 & 0.04 & 0.06 & 8.93 \\
Age 20--24 & 0.09 & 0.11 & 9.11 \\
Age 25--29 & 0.12 & 0.13 & 4.91 \\
Age 30--34 & 0.13 & 0.14 & 2.95 \\
Age 35--39 & 0.14 & 0.14 & 1.44 \\
Age 40--44 & 0.13 & 0.12 & 4.72 \\
Age 45--49 & 0.12 & 0.09 & 7.98 \\
Age 50--54 & 0.09 & 0.07 & 6.78 \\
Age 55--59 & 0.06 & 0.05 & 4.68 \\
Age 60--64 & 0.04 & 0.03 & 1.98 \\
Age 65--69 & 0.03 & 0.03 & 1.42 \\
Age 70--74 & 0.02 & 0.02 & 2.28 \\
\midrule
N & 128,948 & 331,660 &  \\
\bottomrule
\end{tabular}

\begin{tablenotes}[flushleft]
\setstretch{0.9}
\footnotesize
\item \textit{Notes:} The table reports covariate means among drug-related hospitalizations for residents in MedStat regions within 0--1km of a DCR centroid (treated) and residents outside 5km (control). Variable definitions and standardized differences are computed as in Table \ref{tab:balance_pre}.
\end{tablenotes}

\end{threeparttable}
\end{table}

%% file: tables/robustness_sutva.tex
\begin{table}[htbp]
\centering
\caption{Reclassifying Cantons with Inter-Cantonal Access
Agreements --- 5 km buffer}
\label{tab:entitlement}
\begin{tabular}{lcc}
\toprule
 & Baseline & Agreement cantons treated \\
\midrule
ATT              & $-0.431^{**}$ & $-0.443^{**}$ \\
                 & (0.218)       & (0.189)       \\
\% change        & $-35.0\%$     & $-35.8\%$     \\
Treated regions  & 50            & 104           \\
Control regions  & 608           & 586           \\
Observations     & 17{,}250      & 17{,}250      \\
\bottomrule
\end{tabular}

\vspace{0.5em}
\begin{minipage}{0.95\textwidth}\footnotesize
Note: Both columns use the 5 km buffer, never-treated comparison regions, and
the doubly robust estimator of \cite{CALLAWAY2021200} with a universal
base period and simple cohort-size weighting. The outcome is the log
drug-related hospitalization rate per 100,000 inhabitants. Column 1 is the
baseline specification reported in Table A16. Column 2 classifies all regions of
Basel-Land as treated from 2013 and all regions of Nidwalden, Obwalden, Schwyz,
Uri and Zug as treated from 2008, following the inter-cantonal agreements that
allow their residents to enter the Basel-Stadt and Lucerne facilities. This
moves 54 regions from the control group to the treated group. The estimation
sample is identical across columns and only the classification of these regions
changes. Standard errors are clustered at the Medstat region level in
parentheses. Significance: $^{***}p<0.01$, $^{**}p<0.05$, $^{*}p<0.10$.
\end{minipage}
\end{table}

%% file: tables/canton_level.tex
\begin{table}[htbp]
\centering
\caption{Robustness: Canton-Level Treatment Definition}
\label{tab:cantonlevel}
\begin{tabular}{lcc}
\toprule
 & Within 5 km of a DCR & Canton hosts a DCR \\
 & (MedStat region)     & (Canton)           \\
\midrule
ATT             & $-0.431^{**}$ & $-0.276^{**}$ \\
                & (0.216)       & (0.121)       \\
\% change       & $-35.0\%$     & $-24.1\%$     \\
Treated units   & 50            & 4             \\
Control units   & 608           & 18            \\
Observations    & 17{,}250      & 550           \\
\bottomrule
\end{tabular}

\vspace{0.5em}
\begin{minipage}{0.95\textwidth}\footnotesize
Note: Both columns use the doubly robust estimator of \cite{CALLAWAY2021200} with never-treated comparison units and a universal base period. The
outcome is the log drug-related hospitalization rate per 100,000 inhabitants.
Column 2 aggregates counts and population to the canton by year, with cohorts
Geneva 2001, Lucerne 2008, Basel 2013 and Vaud 2018. Bern, Zurich, Solothurn and
Schaffhausen opened their first facility before 1998 and are excluded, since
they carry no in-panel variation. Standard errors clustered at the unit of
treatment in parentheses. With four treated cantons the canton-level standard
error should be interpreted with caution. Significance: $^{***}p<0.01$,
$^{**}p<0.05$, $^{*}p<0.10$.
\end{minipage}
\end{table}

%% file: tables/tab_combined_att.tex
\begin{table}[H]
\centering
\caption{Robustness: Overall ATT Across Estimators --- 5\,km buffer, $\pm$5-year window}
\label{tab:att_all_estimators}
\begin{tabular}{lcccc}
\toprule
 & CS-DR & Sun \& Abraham & Synthetic DiD & Stacked DiD \\
\midrule
ATT       & -0.431** & -0.438** & -0.362* & -0.149*** \\
          & (0.218)  & (0.208)  & (0.205) & (0.038)   \\
\% change & -35.0\%  & -35.5\%  & -30.3\% & -13.9\%   \\
\midrule
Treated regions & \multicolumn{4}{c}{50} \\
Control regions & \multicolumn{4}{c}{608} \\
Observations    & \multicolumn{4}{c}{17,250} \\
\bottomrule
\end{tabular}
\vspace{0.3cm}
\begin{minipage}{\linewidth}
\setstretch{0.9}
\raggedright
\footnotesize{\textit{Note:} All specifications use the 5\,km buffer,
never-treated comparison regions (640 regions), and a $\pm$5-year window
relative to DCR opening. The outcome is the log drug-related hospitalization
rate per 100,000 inhabitants. CS-DR follows the doubly robust estimator of
\cite{CALLAWAY2021200} with a universal base period and simple cohort-size
weighting. Sun \& Abraham uses the interaction-weighted estimator of
\cite{SUN2021}. Synthetic DiD follows \cite{arkhangelsky2021synthetic}. The
overall ATT is a cohort-size-weighted average of cohort-specific estimates.
Stacked DiD constructs a separate clean comparison group for each treatment
cohort following \cite{stacked_did}. Standard errors are clustered at the
Medstat region level in parentheses. Significance: $^{***}p<0.01$,
$^{**}p<0.05$, $^{*}p<0.10$.}
\end{minipage}
\end{table}

%% file: tables/att_ps_dr_results.tex
\begin{table}[H]
\centering
\caption{Overall ATT by estimation of the nuisance functions}
\label{tab:cs_est_method}
\begin{tabular}{lccc}
\toprule
 & Doubly robust & IPW & Outcome reg \\
\midrule
ATT & -0.431* & -0.512* & -0.485* \\
    & (0.229) & (0.285) & (0.277) \\
\% change & -35.0\% & -40.1\% & -38.5\% \\
\midrule
Treated regions & 50 & 50 & 50 \\
Control regions & 608 & 608 & 608 \\
Observations    & 17,250 & 17,250 & 17,250 \\
\bottomrule
\end{tabular}
\vspace{0.3cm}
\begin{minipage}{\linewidth}
\setstretch{0.9}
\raggedright
\footnotesize{\textit{Note:} The table reports the overall ATT
\cite{CALLAWAY2021200} at the 5\,km buffer with never-treated controls and a
universal base period. The outcome is the log drug-related hospitalization rate
per 100,000 inhabitants. The first column is the doubly robust specification reported in the main result. The second and third columns condition on log population, the Swiss share, the share aged 25 to 44 and the female share, all measured in 2000, and estimate the nuisance functions by inverse probability weighting and by outcome regression respectively. Standard errors clustered at the Medstat region level in parentheses. Significance: $^{***}p<0.01$, $^{**}p<0.05$, $^{*}p<0.10$.}
\end{minipage}
\end{table}